\documentclass[nofootinbib,a4paper,11pt]{article}
\usepackage{jheppub}
\usepackage[T1]{fontenc}
\usepackage[utf8]{inputenc}
\usepackage[compat=1.1.0]{tikz-feynman}
\usepackage{float}
\usepackage{multicol,multirow}
\usepackage{amsmath}
\usepackage{subcaption}
\usepackage{array}
\usepackage{color}
\usepackage{hyperref}
\hypersetup{colorlinks=true}
\usepackage{longtable}
\usepackage{graphics,graphicx,epsfig,ulem,booktabs}
\makeatletter
\newcommand\footnoteref[1]{\protected@xdef\@thefnmark{\ref{#1}}\@footnotemark}
\makeatother

\newcolumntype{C}[1]{>{\centering\let\newline\\\arraybackslash\hspace{0pt}}m{#1}}

\newcommand{\odc}{\mathcal{O}(\delta^3)}
\newcommand{\GMt}{\Gamma^{-2}_\beta}
\newcommand{\GMo}{\Gamma^{-1}_\beta}
\newcommand{\Gz}{\Gamma^0_\beta}
\newcommand{\lt}[1]{\Tilde{\lambda}_{#1}}
\newcommand{\Sb}{s_\beta}
\newcommand{\Cb}{c_\beta}
\newcommand{\Stb}{s_{2\beta}}
\newcommand{\Ctb}{c_{2\beta}}

\definecolor{green}{rgb}{0.1,0.5,0.0}

\newcommand{\red}{\color{red}}

\newcommand\inputpgf[2]{{
\let\pgfimageWithoutPath\pgfimage
\renewcommand{\pgfimage}[2][]{\pgfimageWithoutPath[##1]{#1/##2}}
\input{#1/#2}
}}

\title{Cornering the Two Higgs Doublet Model Type II}

\author[a]{Oliver Atkinson,}
\author[b]{Matthew Black,}
\author[b]{Alexander Lenz,}
\author[b]{Aleksey Rusov,}
\author[c]{James Wynne}

\affiliation[a]{SUPA, School of Physics \& Astronomy, University of Glasgow, Glasgow G12 8QQ, UK}

\affiliation[b]{Physik Department, Universit\"{a}t Siegen, Walter-Flex-Str. 3, 57068 Siegen, Germany}

\affiliation[c]{IPPP, Department of Physics, University of Durham, DH1 3LE, UK}

\emailAdd{o.atkinson.1@research.gla.ac.uk}
\emailAdd{Matthew.Black@uni-siegen.de}
\emailAdd{Alexander.Lenz@uni-siegen.de}
\emailAdd{rusov@physik.uni-siegen.de}
\emailAdd{jameswynne39@gmail.com}

\abstract{We perform a comprehensive study of the allowed parameter space of the Two Higgs Doublet Model of Type II (2HDM-II). Using the theoretical framework {\bf flavio} we combine the most recent flavour, collider and electroweak precision observables with theoretical constraints to obtain bounds on the mass spectrum of the theory. In particular we find that
the 2HDM-II fits the data slightly better than the Standard Model (SM) with best fit values of the heavy Higgs 
masses  around 2 TeV and a value of $\tan \beta \approx 4$. Moreover, we conclude that
the wrong-sign limit is disfavoured by Higgs signal strengths and excluded by the global fit by more than five standard deviations and potential deviations from the alignment limit can only be tiny. 
Finally we test the consequences of our study on electroweak baryogenesis via the program 
package {\bf BSMPT} and we find that the  allowed parameter space strongly discourages a strong first order phase transition within the 2HDM-II. 
}

\begin{document}
\hfill SI-HEP-2021-20, SFB-257-P3H-21-049

\maketitle
\flushbottom

\section{Introduction}
\label{sec:introduction}
Two Higgs Doublet Models (2HDM) \cite{Lee:1973iz}
(see Ref.~\cite{Gunion:1989we} for a textbook and Ref.~\cite{Branco:2011iw} 
for a more recent review) are very well-studied extensions of the Standard Model (SM) and probably one of the simplest possibilities for beyond the SM (BSM) physics. Besides the SM particles, four new scalars arise in the 2HDM: two charged ones ($H^+$ and $H^-$), a pseudoscalar ($A^0$, CP-odd) and a scalar ($H^0$, CP-even; when denoting the SM Higgs with $h^0$). 
Such an extension might solve some of the conceptual problems that arise within the SM, see e.g.~Ref.~\cite{Trodden:1998qg}. In particular, 2HDMs could provide the missing amount of CP violation to explain the matter-antimatter asymmetry in the Universe and they could also lead to a first order phase transition in the early Universe, as required by the Sakharov criteria \cite{Sakharov:1967dj}.
2HDMs further appear as subsets of more elaborate BSM models, such as the minimal supersymmetric SM (MSSM).
Depending on the structure of the Yukawa couplings one distinguishes different types of 2HDMs~--~in this work we will
consider the 2HDM-II. 

Since we have no direct experimental 
evidence so far for an extended Higgs 
sector, we investigate indirect 
constraints on the 2HDM-II by
comparing the most recent measurements 
with high precision calculations.
We perform a comprehensive study of
more than 250 observables, where the new Higgs
particles could appear as virtual corrections and
modify the SM prediction. Our study extends previous
works like Refs.~\cite{Deschamps:2009rh,Chowdhury:2017aav}.\footnote{
There is a vast literature on 2HDM studies, which we are not going to review here. Papers which have some minor overlap with our study are e.g.~Ref.~\cite{Arco:2020ucn}, which focuses on triple Higgs couplings in the 2HDM and uses many of the same constraints as we do here, with the addition of LHC searches for BSM Higgs searches, to set bounds on the allowed triple coupling values. Further studies of the 2HDM have also been carried out in e.g.~Refs.~\cite{Bhattacharyya:2015nca, Das:2015qva}.}
In 2009, Ref.~\cite{Deschamps:2009rh} studied bounds from quark flavour observables on the 2HDM-II within the framework of 
{\bf CKMfitter}. In particular, it was found that the 2HDM does not perform better in the fit than the SM and that the new charged scalar has to be heavier than 316~GeV. We will considerably extend this analysis by using updated measurements, theory predictions and 
by including additional flavour observables. In 2017, the authors of Ref.~\cite{Chowdhury:2017aav} used
{\bf HEPfit}
to constrain the 2HDM-II with Higgs data, electroweak precision observables and some flavour observables (the radiative decay $b \to s \gamma$ and the mass difference of neutral $B$ mesons $\Delta m_q$), as well as theoretical constraints like perturbativity. The main findings  are
lower mass bounds such as
$m_{H^+} >740$ GeV
and a lower limit on the mass splittings of the heavy Higgses of the order of 100 GeV.
Higgs signal strengths allow only small deviations from the alignment limit,  
$|\beta - \alpha - \pi/2| < 0.055$ - this bound is further reduced to 
$|\beta - \alpha - \pi/2| < 0.02$ within the overall fit (driven by theory constraints).
Finally, the authors of Ref.~\cite{Chowdhury:2017aav} found 
that the parameter $\tan \beta$ lies in the preferred range $[0.93; 5.0]$. 
We will extend this analysis by updating experimental values and considering a large number of flavour observables. Moreover we confirm the observation made in Ref.~\cite{Chowdhury:2017aav}, that the wrong-sign limit is excluded by data. 

Throughout this work, we use the Python program package \textbf{flavio} \cite{Straub:2018kue} for flavour physics for our predictions of observables in the 2HDM-II.
New physics models are not explicitly included in \textbf{flavio}. 
Instead, they impact observables through their contributions to the Wilson coefficients of dimension-6 operators in the relevant effective theory: the Weak Effective Theory (WET) below the electroweak scale, or the Standard Model Effective Field Theory (SMEFT) above.
%In Section~\ref{sec:signals}, the relevant 2HDM-II contributions are included in the coefficients of the SMEFT operators, and 
In Section~\ref{sec:flavour}, the relevant 2HDM-II contributions are included in the coefficients of the WET operators with five active flavours (there are only three active flavours in the WET-3 basis used for the anomalous magnetic moment of the muon $a_\mu$).
For a full list of operators in the \textbf{flavio} basis for each effective theory, see the documentation of the \textbf{WCxf} package \cite{Aebischer:2017ugx}.

Once 2HDM-II contributions to the relevant coefficients have been included, we use \textbf{flavio} to construct fits in 2HDM parameter spaces, where we show in colour the allowed regions of our parameter space within the confidence levels stated in the captions.
The package \textbf{flavio} expresses experimental measurements and input parameters as various one- and two-dimensional probability distribution functions. 
The errors of theoretical predictions are constructed by calculating the prediction for each of a number (we use $N=10^4$) of randomly-selected values of the input parameters within their probability distributions, and then computing the standard deviation of these values. 
We work with the ``fast likelihood" method in \textbf{flavio} for constructing likelihood functions of the form $\mathcal{L}=e^{-\chi^2(\vec{\xi})/2}$, where it is assumed that the set of fit parameters $\vec{\xi}$ contributing to the observables entering the likelihood function are taken at their central values.
This method uses the combined experimental and theoretical covariance matrix of the observables in the fit.
For further information on the ``fast likelihood" method and the package's construction, we refer to \textbf{flavio}'s documentation~\cite{Straub:2018kue} or the \textbf{flavio} webpage\footnote{ https://flav-io.github.io}.

The program package \textbf{flavio} is used in Section~\ref{sec:flavour} for the flavour observables, where we work in two-dimensional parameter spaces.
{ We make use of {\bf flavio} to construct likelihoods and combine measurements in Section~\ref{sec:signals} for the Higgs signal strengths, however for the calculations of the signal strengths themselves, we use analytical expressions written in Python natively.}
The constraints discussed in Section~\ref{sec:theory} and \ref{sec:STU} stemming from theoretical considerations  and the electroweak precision observables cannot be expressed in terms of coefficients of dimension-6 operators.
Therefore in Section \ref{sec:theory}, the results are taken from a native Python code generating Monte Carlo scans of the full parameter space.
The likelihood functions for the electroweak precision observables, taking into
account the experimental correlations were translated into native Python code to 
be properly combined in Section~\ref{sec:global} with all other observables and
theory constraints using some customisations to the {\bf flavio} package.

The paper is organised as follows.
Our notation for the 2HDM-II is set up in
Section~\ref{sec:2HDM}.
In Section~\ref{sec:theory}
theoretical constraints like perturbativity, vacuum
stability, and unitarity are studied, while Section~\ref{sec:STU} introduces
electroweak precision observables in the form of the
oblique parameters.
Constraints stemming from the SM Higgs production and decay at the LHC are studied in
Section~\ref{sec:signals}.
Most of our observables stem from the quark flavour sector, see
Section~\ref{sec:flavour}.
In Section~\ref{sec:leps}
we investigate leptonic and semi-leptonic tree-level decays, where the charged Higgs is predominantly responsible for the new contributions. 
In this section we further study a potential pollution of the determination of CKM elements from semi-leptonic and leptonic decays by the 2HDM-II. 
$B$-mixing is considered in Section~\ref{sec:mix} and loop-induced $b$ quark decays are studied in Section~\ref{sec:bsll}.
Section~\ref{sec:rad} contains a brief description of the $b \to s \gamma$ transition, known to give a lower bound on the mass of the charged Higgs boson. The leptonic decays
$B_q \to \mu^+ \mu^-$ get also sizable contributions from the new neutral Higgses, see
Section~\ref{sec:bsmumu}.
In Section~\ref{sec:b-to-s-ell-ell}
we study semi-leptonic 
$b \to s \ell^+ \ell^-$-transitions, where
the so-called flavour anomalies 
%(see e.g. \cite{Hurth:2021nsi,Alguero:2021anc,Cornella:2021sby,Altmannshofer:2021qrr,Geng:2021nhg,Ciuchini:2020gvn,MunirBhutta:2020ber,Biswas:2020uaq})
have been observed in recent
years. These anomalies can be best explained by vector-like new effects, while we consider
here the effects of new scalar couplings.
The results of our global fit, combining all constraints discussed so far are presented in 
Section~\ref{sec:global}.
In Section~\ref{sec:muon} we comment on the lepton flavour observable, $a_\mu$, the anomalous magnetic moment 
of the muon, where recent measurements at Fermilab~\cite{Abi:2021gix} have confirmed the older BNL
value~\cite{Bennett:2006fi}. We study two scenarios based on using the SM prediction from the theory initiative
\cite{Aoyama:2020ynm} or a recent lattice evaluation \cite{Borsanyi:2020mff}.
Using the program package {\bf BSMPT} \cite{Basler:2018cwe,Basler:2020nrq} we investigate in
Section~\ref{sec:Phase}  the question of whether our allowed parameter space can also lead to a 
first order phase transition in the early Universe.
Finally we conclude in Section~\ref{sec:Conclusion}, while all our inputs are collected in Appendix~\ref{sec:inputs}.

\section{The Two Higgs Doublet Model of Type II}
\label{sec:2HDM}
In the SM, a conjugate of the Higgs doublet is used to ensure both up-type and down-type quarks acquire mass. This can also be achieved by using two distinct complex scalar doublets 
\begin{equation}
    \Phi_i = \begin{pmatrix} \phi^+_i \\ (v_i + \phi^0_i + iG^0_i)/\sqrt{2}
    \end{pmatrix}, 
\end{equation}
with $i =1,2$. These acquire different vacuum expectation values (VEVs) $v_i$ after electroweak symmetry breaking (EWSB), that are related to the SM Higgs VEV $v$ by $v_1^2 + v_2^2 = v^2$ \cite{Lee:1973iz, Gunion:1989we, Branco:2011iw}. 
This construction has 8 degrees of freedom, 3 of which become 
the longitudinal polarisations of the massive $W^\pm$ and $Z$ bosons, leaving 5 physical Higgs bosons. Two of these are charged, $H^\pm$, two are neutral scalars, $H^0$ and $h^0$, and one is a neutral pseudoscalar, $A^0$. Typically $h^0$ is taken to be the lighter of the two neutral scalars. 
The potential for a general 2HDM with a softly broken $\mathbb{Z}_2$ symmetry is, in the lambda basis \cite{Branco:2011iw},
\begin{equation}
\begin{split}
    V(\Phi_1,\Phi_2)=m_{11}^2\Phi_1^\dagger\Phi_1+m_{22}^2\Phi_2^\dagger\Phi_2
    -m_{12}^2(\Phi_1^\dagger\Phi_2+\Phi_2^\dagger\Phi_1)
    + \frac{\lambda_1}{2}(\Phi_1^\dagger\Phi_1)^2+\frac{\lambda_2}{2}(\Phi_2^\dagger\Phi_2)^2 \\ +
    \lambda_3 (\Phi^ \dagger_1\Phi_1) (\Phi^\dagger_2\Phi_2) + \lambda_4 (\Phi^\dagger_1\Phi_2) (\Phi^\dagger_2 \Phi_1) +
    \frac{\lambda_5}{2} \left[(\Phi^\dagger_1\Phi_2)^2+(\Phi^\dagger_2\Phi_1)^2\right].
\end{split}
\end{equation}
The $\mathbb{Z}_2$ symmetry is required to forbid tree-level flavor changing neutral currents (FCNC). Constraints on the $\lambda_i$ from a theoretical perspective are considered in Section \ref{sec:theory}.
The parameters $m_{11}^2$ and $m_{22}^2$ are related to $v_1$ and $v_2$ as~\cite{Basler:2016obg}
\begin{equation}
    m_{11}^2=m_{12}^2 \, \frac{v_2}{v_1}-\lambda_1\frac{v_1^2}{2}-(\lambda_3+\lambda_4+\lambda_5)\frac{v_2^2}{2},
    \label{eq:m_11}
\end{equation}
\begin{equation}
    m_{22}^2=m_{12}^2 \, \frac{v_1}{v_2}-\lambda_2 \frac{v_2^2}{2}-(\lambda_3+\lambda_4+\lambda_5)\frac{v_1^2}{2}.
    \label{eq:m_22}
\end{equation}
In the remainder of this work we make use of the mass basis, through a transformation \cite{Arnan:2017lxi} 
\footnote{We differ with the expressions given in \cite{Branco:2011iw} for $m_{A^0}$ and $m_{H^\pm}$.}
\begin{align}
    \label{eq:mH0}
    \begin{split}
    m_{H^0}^2 &= \frac{m_{12}^2}{\sin\beta\cos\beta}\sin^2(\beta - \alpha) \\ 
    &\quad + v^2\left[\lambda_1\cos^2\alpha\cos^2\beta + \lambda_2 \sin^2 \alpha \sin^2 \beta + \frac{\lambda_3 + \lambda_4 + \lambda_5}{2} \sin2 \alpha \sin2 \beta \right], 
    \end{split} \\
    \label{eq:mh0}
    \begin{split}
    m_{h^0}^2 & = \frac{m_{12}^2}{\sin\beta\cos\beta}\cos^2(\beta - \alpha) \\  
    &\quad + v^2\left[ \lambda_1 \sin^2 \alpha \cos^2 \beta+ \lambda_2 \cos^2\alpha\sin^2\beta-\frac{\lambda_3 + \lambda_4 + \lambda_5}{2}\sin2\alpha\sin2\beta\right], \end{split} \\
    \label{eq:mA0} 
    m_{A^0}^2 & = \frac{m_{12}^2}{\sin\beta\cos\beta} - \lambda_5 v^2, \\
    \label{eq:mHpm}
    m_{H^\pm}^2 & = \frac{m_{12}^2}{\sin \beta \cos \beta} - 
    \frac{\lambda_4 + \lambda_5}{2} v^2.
\end{align}
The mass basis allows us to concentrate on the 8 physical parameters: 
the masses of the new particles, $m_{H^\pm}$, $m_{h^0}$, $m_{H^0}$, $m_{A^0}$, 
the mixing angles $\alpha$ and $\beta$, the softly $\mathbb{Z}_2$ breaking term $m_{12}$,
and the VEV~$v$. We identify $h^0$ as the observed SM-like boson 
with a mass of $125.1 \pm 0.14$ GeV \cite{Zyla:2020zbs} and take~$v=246$~GeV throughout this work.  
The angles $\alpha$ and $\beta$, describing the mixing of the neutral and charged Higgs fields, respectively, satisfy the following relations~\cite{Arnan:2017lxi}
\begin{align}
    \tan \beta & = \frac{v_2}{v_1}, 
    \label{eq:tan-beta} 
    \\
    \tan 2 \alpha &= \frac{2 \left(- m_{12}^2 + (\lambda_{3} + \lambda_{4} + \lambda_5) v_1 v_2 \right)}{m_{12}^2\left(v_2/v_1- v_1/v_2 \right)+\lambda_1v_1^2-\lambda_2v_2^2}.
    \label{eq:tan-alpha}
\end{align}
One may revert Eqs.~\eqref{eq:mH0} - \eqref{eq:tan-alpha} 
and express explicitly the $\lambda_i$ parameters in terms of the physical parameters, see e.g.~Ref.~\cite{Kling:2016opi}. We will show below that data strongly favours the small
$\cos(\beta-\alpha)$ limit, where the expressions for the $\lambda_i$ become particularly simple \cite{Han:2020zqg,Kling:2016opi}:
\begin{align}
    v^2\lambda_1 &= m_{h^0}^2 - \frac{\tan\beta(m_{12}^2-m_{H^0}^2\sin\beta\cos\beta)}{\cos^2\beta}, 
    \label{eq:lambda1} \\
    v^2\lambda_2 &= m_{h^0}^2 - \frac{m_{12}^2-m_{H^0}^2\sin\beta\cos\beta}{\tan\beta\sin^2\beta}, 
    \label{eq:lambda2} \\
    v^2\lambda_3 &= m_{h^0}^2 + 2m_{H^+}^2 - m_{H^0}^2 - \frac{m_{12}^2}{\sin\beta\cos\beta}, 
    \label{eq:lambda3} \\
    v^2\lambda_4 &= m_{A^0}^2 - 2m_{H^+}^2  + \frac{m_{12}^2}{\sin\beta\cos\beta}, 
    \label{eq:lambda4} \\
    v^2\lambda_5 &=  - m_{A^0}^2 + \frac{m_{12}^2}{\sin\beta\cos\beta}.
    \label{eq:lambda5}
\end{align}
The Lagrangian for the Higgs and Yukawa sectors in the 2HDM is given by  
\begin{equation}
    \mathcal{L}^{\rm 2HDM}_{\rm H+Y} =
    \sum_i|D_\mu\Phi_i|^2 - V(\Phi_1, \Phi_2) + \mathcal{L}_{\text{Y}},   
\end{equation}
where $\mathcal{L}_{\text{Y}}$ depends on the details of the quark-Higgs couplings in the 2HDM. 
There are four main types of 2HDM, based on which particles each of the doublets couple to.
In this work, the 2HDM of Type II (2HDM-II) is investigated as it gives quark mass terms in the Lagrangian that are similar to the form in the SM. This in turn yields a similar CKM matrix through which all flavour-changing interactions
can be described, with tree-level FCNCs forbidden by the imposition of the $\mathbb{Z}_2$ symmetry (which is only softly broken by $m_{12}^2$). The 2HDM-II model has the doublet $\Phi_1$ coupling to down-type quarks and leptons, while $\Phi_2$ is coupling to up-type quarks, so
\begin{equation}
    \mathcal{L}_{\text{Y}} = -  Y^{(u)}_{ij} \bar{Q^i_L} \tilde{\Phi}_2 u^j_R - Y^{(d)}_{ij}\bar{Q^i_L}\Phi_1 d^j_R  
    - Y^{(\ell)}_{ij}\bar{L_L^i}\Phi_1 \ell^j_R 
    + {\rm h.c.},
    \label{eq:Yakawa-2HDM-II}
\end{equation}
where $\tilde{\Phi}_{i} = i\sigma_2\Phi^*_i$, $Y_{ij}^{(u,d,\ell)}$ are the Yukawa couplings, $i,j = 1,2,3$ run over the generations of the fermions, 
and the left-handed $SU(2)$-doublets of quarks and leptons are, respectively,
\begin{equation*}
    Q_{L}^i = \begin{pmatrix} u_L^i \\ d_L^i \end{pmatrix}, \qquad
    L_{L}^i = \begin{pmatrix} \nu_L^i \\ \ell_L^i \end{pmatrix}.
\end{equation*}
After EWSB, rotating to the mass basis and focusing on the gauge and Yukawa couplings of~$h^0$ gives Lagrangian terms of a very similar form to the SM, with factors $\kappa_i$ relating the coupling constants to those in the SM \cite{Han:2020lta}:
\begin{equation}
    \mathcal{L}_{h^0} = \kappa_V\frac{m_Z^2}{v}h Z_\mu Z^\mu
    + \kappa_V\frac{2m_W^2}{v}h W^+_\mu W^{\mu -}
    - \sum_{f = u, d, \ell}\kappa_f\frac{m_f}{v}h\Bar{f}f,
\label{eq:Lh0}
\end{equation} 
where, in the 2HDM-II,
\begin{equation}
\begin{aligned}
    \kappa_{\small{V}} &= \sin{(\beta - \alpha)},\\     
    \kappa_{u} &= \sin{(\beta - \alpha)} + \cot{\beta}\cos{(\beta - \alpha)}, \\
    \kappa_{d,\ell} &= \sin{(\beta - \alpha)} - \tan{\beta}\cos{(\beta - \alpha)}.
\end{aligned}
\label{eq:kappas}
\end{equation}
\\
Naturally, in the SM all $\kappa_i = 1$, which is consistent with LHC data \cite{Aad:2019mbh, ATLAS:2020qdt, CMS:2020gsy}. Setting $\cos{(\beta - \alpha)} = 0$ recovers this and thus matches the phenomenology of $h^0$ in the 2HDM with the SM Higgs. As such, this is known as the {\bf alignment limit}. The dependence of these, and numerous other, couplings on $\cos{(\beta - \alpha)}$ and $\tan\beta$ leads us to choose these as parameters of the model, in place of $\alpha$ and $\beta$. 

The addition of new Higgs fields introduces new interactions, where the charged Higgs fields can replace $W^\pm$ fields in flavour-changing charged interactions, as well as direct couplings of the new bosons to the weak bosons themselves. 
The part of the Yukawa Lagrangian describing interactions of 
$H^\pm$ bosons with the fermions is given by \cite{Gunion:1989we, Branco:2011iw} \footnote{Here and hereafter, 
for notation simplicity, we omit indices $i, j$ indicating the generation of 
the up- and down-type quarks $u = u, c, t$ and $d = d, s, b$.}
\begin{equation}
    \mathcal{L}_{H^\pm} = \frac{\sqrt{2} \, V_{u d}}{v} H^+
    \bar u \left[ 
    m_{u} \cot \beta \, P_L + m_{d} \tan \beta \, P_R
    \right] d
    + \frac{\sqrt{2}}{v} m_{\ell} \tan \beta \, H^+ \left(\bar \nu_\ell  P_R \, \ell \right)
    + {\rm h.c.},
\label{eq:LHpm}
\end{equation}
where $P_{L,R} = (1 \mp \gamma_5)/2$, $m_u$ and $m_d$ are the masses of the 
up- and down-type quarks, $m_{\ell}$ ($\ell = e, \mu, \tau$) is the charged-lepton mass,
and $V_{ud}$ denotes the corresponding element of the quark-mixing 
Cabibbo-Kobayashi-Maskawa (CKM) matrix. 

The new Higgs fields, discussed above, will affect many measured 
particle physics quantities including the electroweak precision parameters, 
Higgs signal strengths and flavour physics observables.
These effects are investigated in Sections~\ref{sec:STU}, \ref{sec:signals}, 
and \ref{sec:flavour}
respectively and used to set bounds on the parameter space of the model.
However, we start with the discussion of purely theoretical constraints.

\section{Theoretical Constraints}
\label{sec:theory}
\subsection{Perturbativity}
\label{sec:theory_lambda}
Within the lambda basis, perturbativity  in the scalar sector can be simply expressed as  \cite{Chen:2018shg,Ginzburg:2005dt} 
\begin{equation}
    |\lambda_i|\leq 4\pi, \qquad 
    i = 1, \ldots 5.
    \label{perturbativity}
\end{equation}
Clearly this is not a strict experimental constraint like the bounds studied below; it is more related to our
ability to make meaningful predictions in the 2HDM - and the numerical value of the bound on the $\lambda_i$ contains a certain amount of arbitrariness. Thus we will also consider the less conservative bound of
$ |\lambda_i|\leq 4$ - inspired by the results in 
Refs.~\cite{Grinstein:2015rtl,Cacchio:2016qyh}.
None of our conclusions will actually be changed by modifying the perturbativity bound from 
$4 \pi $ to $4$; to be able to study some more extreme scenarios in Section~\ref{sec:Phase}
we will, however, use the more conservative bound $4\pi$.

Writing the expressions for $\lambda_i$ in terms of the additional Higgs masses we find that for low masses, i.e. $M \ll 1000$ GeV, no bounds (below the particle mass $M$) are set on the mass differences of the 
new Higgs particles, while for high masses a form of degeneracy has to hold.
In our analysis, we will consider heavy Higgs masses ranging between $10^{2.5} \approx 300 $ GeV and $10^5$ GeV. 
Masses of the charged Higgs boson as low as 300 GeV are clearly ruled out by flavour observables, in particular by $b \to s \gamma$, as discussed below. The  experimental lower bound on the charged 
Higgs mass from direct searches is actually only $m_{H^+}\gtrsim160\,$GeV \cite{Aaboud:2018gjj}.
The upper bound has been chosen to be equal to the centre-of-mass energy of a future 100 TeV collider.

Looking at the couplings in Eq.~(\ref{eq:LHpm}), one can derive two further perturbativity constraints from the Yukawa sector
\begin{equation}
\begin{aligned}
\frac{\sqrt{2} \, V_{tb} \, m_t \cot \beta}{2 v} \le { \sqrt{4 \pi}} & \quad \Rightarrow \quad \tan \beta > {0.14} \, ,
\\
\frac{\sqrt{2} \, V_{tb} \,  m_b \tan \beta}{2 v} \le{ \sqrt{4 \pi}} & \quad \Rightarrow \quad \tan \beta < { 300} \, ,
\label{eq:pert_Yukawa}
\end{aligned}
\end{equation}
where we again have chosen a very conservative range.
For the range of $\log (\tan \beta)$ we will consider the conservative lower bound $\tan \beta = 10^{-1.5} = 0.03$, and the upper bound $\tan \beta = 10^{+2.5} = 300$.

\subsection{Vacuum Stability and Unitarity}
\label{sec:theory_vacuum}
Next we apply the conditions for a stable vacuum as set out in Ref.~\cite{Deshpande:1977rw}:
\begin{equation}
\begin{aligned}
    \lambda_{1,2} & > 0, \\
    \lambda_3 & > - (\lambda_1\lambda_2)^{1/2}, \\
    \lambda_3+\lambda_4-|\lambda_5|
    & > -(\lambda_1\lambda_2)^{1/2}  .
\end{aligned}
\label{theoretical constraints}
\end{equation}
Demanding the vacuum to be the global minimum of the potential, we further find \cite{Barroso:2013awa}
\begin{equation}
    m_{12}^2\left(m_{11}^2-m_{22}^2\left(\frac{\lambda_1}{\lambda_2}\right)^\frac{1}{2}\right)\left(\tan\beta -\left(\frac{\lambda_1}{\lambda_2}\right)^\frac{1}{4}\right)>0.
\end{equation}
We also consider the conditions from tree-level unitarity, see Refs.~\cite{Ginzburg:2005dt,Arhrib:2000is}\footnote{{ For similar discussion in an alternative lambda basis see \cite{Horejsi:2005da}.} },
alongside NLO unitarity and the condition that NLO corrections to partial wave amplitudes are suppressed relative to LO contributions, see  Refs.~\cite{Cacchio:2016qyh,Grinstein:2015rtl}.

\begin{figure}[ht]
    \setlength{\belowcaptionskip}{-10pt}
    \centering
    \includegraphics[width=0.45\textwidth]{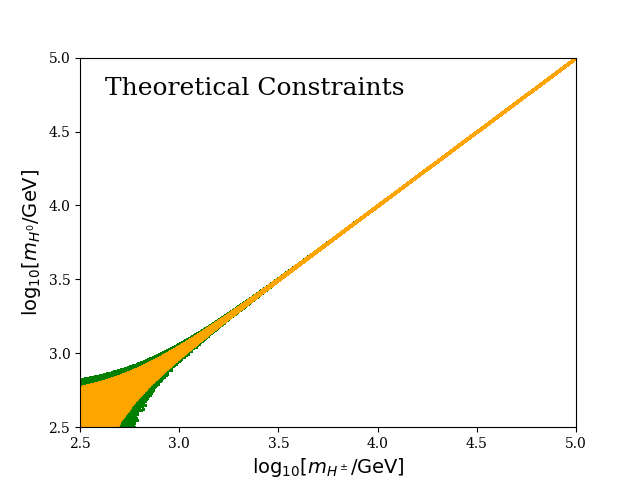} \quad
    \includegraphics[width=0.45\textwidth]{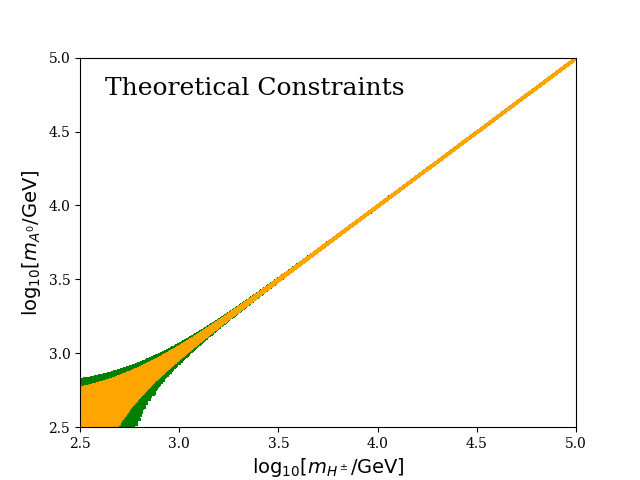}
    \\
    \includegraphics[width=0.45\textwidth]{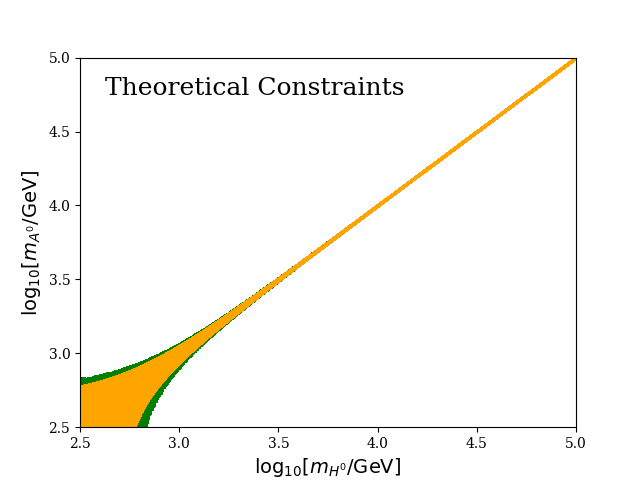} \quad
    \includegraphics[width=0.45\textwidth]{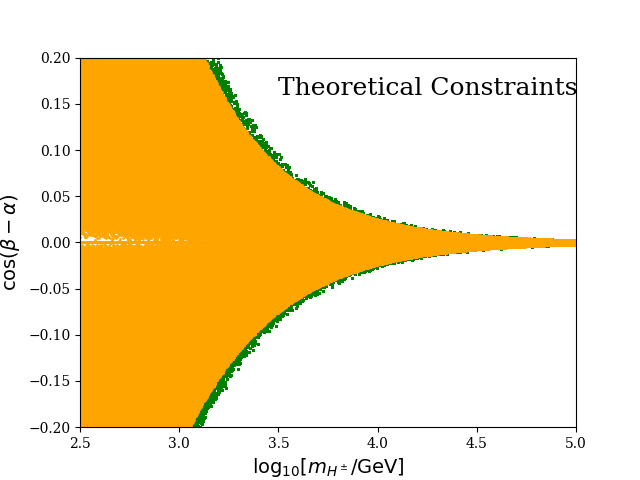}
    \\
    \includegraphics[width=0.45\textwidth]{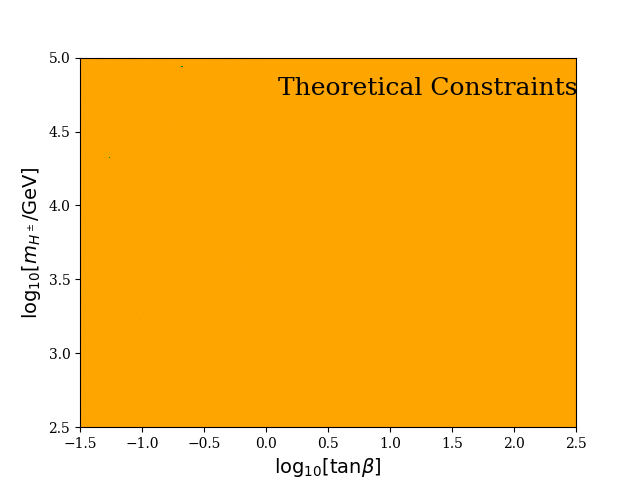} \quad
    \includegraphics[width=0.45\textwidth]{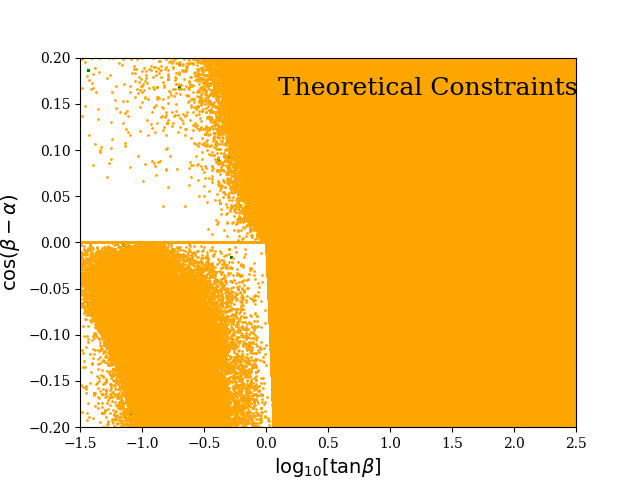}
    \caption{Bounds on the heavy Higgs masses $m_{H^0}$, $m_{A^0}$ and $m_{H^+}$, $\tan \beta$ and $\cos (\beta - \alpha)$ 
    stemming from the theory constraints (perturbativity, vacuum stability and unitarity conditions). For the plots $10^8$ points were generated and only values of $|\delta | < 0.5$ were considered. The starting values $|\lambda_i|<4\pi$ are given in green,  while $|\lambda_i|<4$ is shown in  amber.}
    \label{MontyCarloResults}
\end{figure}

To implement these constraints alongside the perturbativity conditions, we carry out Monte Carlo scans. 
Starting with the allowed sets of $\lambda_i$ (via Eq.~\eqref{perturbativity}) we then use Eqs.~\eqref{eq:mh0} and \eqref{eq:tan-alpha}
to get a relation among the parameters
$\lambda_1$,..., $\lambda_5$, $\beta $ and the small parameter $\delta:=\beta-\alpha-\pi/2$\footnote{Higgs data suggests $|\delta|<$0.05; see Section~\ref{sec:signals}.}, which is explicitly given in Appendix~\ref{delta_formula}.
Expanding the expressions up to second order in $\delta$, 
Eqs.~\eqref{eq:mH0}-\eqref{eq:mHpm} give the allowed ranges of Higgs masses
depending on the value of $\beta$ and $\delta$. We show the effect of the theory constraints on the Higgs masses $m_{H^+}$,
$m_{H^0}$, $m_{A^0}$, $\tan \beta$ and $\cos (\beta - \alpha)$
as two-dimensional projections in Fig.~\ref{MontyCarloResults}.
The starting values $|\lambda_i|<4\pi$ are depicted in green, while points from $|\lambda_i|<4$ are shown in amber -- 
the latter obviously leads to more constrained regions.
Here we note that due to the imposition of NLO unitarity constraints, for starting values $|\lambda_i|<4\pi$, we find most successful points for $|\lambda_i|\lesssim2\pi$.
Moreover, we find that for charged Higgs masses above 1 TeV - as implied by the flavour observables - 
the choice of $|\lambda_i|<4\pi$ or $|\lambda_i|<4$  has only a very minor impact.

The two upper plots and the left of the middle plots of Fig.~\ref{MontyCarloResults} show that
theory constraints require the new Higgses to be more or less degenerate in mass. The heavier the mass, the stronger this requirement
becomes, which can also be nicely read off from Table \ref{tab:theory-constraints}.
The right plot of the middle row in Fig. \ref{MontyCarloResults} shows that $|\delta |$ will be constrained to small values, if the 
new Higgs particles are heavier than about a TeV (such a bound  on the charged Higgs mass will follow from flavour constraints
discussed below). Again, the constraints on $|\delta |$ become stronger the heavier the new Higgs masses become. Quantitative
values for the bounds on $|\delta | $ are given in Table~\ref{tab:theory-constraints}.
In the lower row of Fig.~\ref{MontyCarloResults} we see that $\tan \beta$ is not
really constrained by the theory constraints, apart from the restrictions given in
Eq.~\eqref{eq:pert_Yukawa}.

In the alignment limit,
the value of $\alpha$ is fixed by the value of $\tan \beta$, 
and we find a similar requirement for a mass degeneracy of the new Higgs particles as in the general case. Moreover, we
do not find any bound on $\tan \beta$ within the regions of interest.

\begin{table}[htbp]
    \centering
    \begin{tabular}{|c|c| c| c| c|}
        \hline
        $m_{H^\pm}$/TeV$\in$ &$\Delta_{A^0H^0}$/GeV $\in$& $\Delta_{A^0H^\pm}$/GeV $\in$& $\Delta_{H^0H^\pm}$/GeV $\in$ &$\delta\in$\\ 
        \hline
        $[ 0.0 , 0.5 )$ & $( -536 , 521 )$ & $( -420 , 574 )$ & $( -351 , 532 )$ & $( -0.5 , 0.5 )$ \\ 
        \hline  
        $[ 0.5 , 1.0 )$ & $( -297 , 296 )$ & $( -328 , 268 )$ & $( -351 , 268 )$ & $( -0.5 , 0.5 )$ \\ 
        \hline  
        $[ 1.0 , 1.5 )$ & $( -163 , 158 )$ & $( -123 , 157 )$ & $( -118 , 154 )$ & $( -0.30 , 0.25 )$ \\ 
        \hline  
        $[ 1.5 , 2.0 )$ & $( -109 , 105 )$ & $( -74 , 110 )$ & $( -73 , 105 )$ & $( -0.19 , 0.16 )$ \\ 
        \hline  
        $[ 2.0 , 2.5 )$ & $( -82 , 84 )$ & $( -52 , 82 )$ & $( -46 , 79 )$ & $( -0.14 , 0.13 )$ \\
        \hline  
        $[ 2.5 , 3.0 )$ & $( -67 , 64 )$ & $( -42 , 67 )$ & $( -36 , 65 )$ & $( -0.11 , 0.09 )$ \\
        \hline  
        $[ 3.0 , 3.5 )$ & $( -53 , 55 )$ & $( -35 , 56 )$ & $( -34 , 53 )$ & $( -0.09 , 0.08 )$ \\
        \hline  
        $[ 3.5 , 4.0 )$ & $( -47 , 47 )$ & $( -30 , 47 )$ & $( -27 , 48 )$ & $( -0.07 , 0.07 )$ \\
        \hline  
        $[ 4.0 , 4.5 )$ & $( -42 , 40 )$ & $( -30 , 41 )$ & $( -22 , 43 )$ & $( -0.07 , 0.06 )$ \\
        \hline  
        $[ 4.5 , 5.0 )$ & $( -35 , 35 )$ & $( -23 , 36 )$ & $( -18 , 37 )$ & $( -0.06 , 0.06 )$ \\
        \hline  
    \end{tabular}
    \caption{A numerical summary of the effects of imposing the theory constraints with starting values $|\lambda_i|<4\pi$. These intervals are found by carrying out a $10^8$ points Monte Carlo scan with $|\delta|<0.5$, $m_{H^\pm},m_{A^0},m_{H^0}<5000\,\mathrm{GeV}$ and $-1.5<\log_{10}(\tan\beta)<2.5$. We have defined $\Delta_{ij}:=m_i-m_j$. }
    \label{tab:theory-constraints}
\end{table}

\section{Electroweak Precision Observables}
\label{sec:STU}
The oblique parameters first proposed in Ref.~\cite{Peskin:1990zt} are defined as
\begin{align}
    \label{sdef}
    S &:=\frac{4e^2}{\alpha}\frac{d}{dq^2}\big[\Pi_{33}(q^2)-\Pi_{3Q}(q^2)\big]\big|_{q^2=0}, \\
    \label{tdef}
    T &:=\frac{e^2}{\alpha\sin^2(\theta_W)m^2_W}\big[\Pi_{11}(0)-\Pi_{33}(0)\big], \\
    \label{udef}
   U &:=\frac{4e^2}{\alpha}\frac{d}{dq^2}\big[\Pi_{11}(q^2)-\Pi_{33}(q^2)\big]\big|_{q^2=0}.
\end{align}
$\Pi_{XY}(q^2)$ represents the self energies of the electroweak gauge bosons $X$ and $Y$ with 4-momentum~$q^\mu$, 
the subscript $i$ represents the $W_i$ field and 
the subscript $Q$ represents the $B$ field. 
Here we stress that in defining these oblique parameters a subtraction of the SM contributions is made; $S = S_{\rm NP} - S_{\rm SM}$ and similarly for $T$ and $U$, such that, by construction, in the SM $S = T = U = 0$.

While most of the constraints we consider below depend on the particular type of the 2HDM, the oblique parameters 
are, at one loop order, unaffected by the Yukawa couplings and therefore universal for all 2HDMs.
The expressions for $S$, $T$, and $U$ in the 2HDM (derived from Ref.~\cite{Grimus:2008nb})
have been collected in Appendix~\ref{sec:Oblique_formula}. 

New physics generally has only a small effect on $U$; we therefore 
follow the approach outlined in Ref.~\cite{Zyla:2020zbs} and set $U$ to zero. 
This effectively reduces the error on the experimental result for $T$, due to the correlation between $T$ and $U$.
The oblique parameters will be
included in our global fit.

\section{Higgs Signal Strengths}
\label{sec:signals}

Signal strengths are a key measurement in determining if the observed Higgs boson 
is indeed that of the SM, or if new physics is involved. For a single production mode 
$i$ with a cross section $\sigma_i$, and decay channel $f$ with a branching 
fraction ${\cal B}_f$, the signal strength $\mu^f_i$ is defined as
\begin{equation}
    \mu^f_i = \frac{(\sigma_i \cdotp {\cal B}_f)_\text{Exp.}}{(\sigma_i \cdotp {\cal B}_f)_\text{SM}},
\end{equation}
as the cross section and branching fraction cannot be separately measured without further assumptions. Evidently if the SM is accurate, all these signal strengths should take a value of 1. 

In practice, the small branching fractions and cross sections render some channels
impossible to measure currently, but there are recent measurements available for 31
channels from the ATLAS and CMS collaborations at $\sqrt{s} = 13\;\text{TeV}$ with an integrated luminosity of up to 139 fb$^{-1}$ and 137~fb$^{-1}$ respectively \cite{Sirunyan:2018koj, Sirunyan:2019qia, Aad:2019mbh, Aad:2020plj, Aad:2020xfq, CMS:2020gsy, ATLAS:2020qdt}. 
These include various combinations of the four main $h^0$ production channels at the LHC: gluon-gluon fusion (ggF), 
vector boson fusion (VBF), in association with a weak vector boson (Vh) or
a top-antitop pair (tth), and the decay channels $\gamma\gamma$, $ZZ^*$, $WW^*$, $Z\gamma$, $\tau^+\tau^-$, $b\Bar{b}$ and~$c\bar{c}$. 
The recent observations of $h^0 \to \mu^+\mu^-$ \cite{Aad:2020xfq, Sirunyan:2020two} are also included
here. In cases such as $h^0 \to \mu^+\mu^-$ where a combined signal strength across all channels is given,
it is interpreted as being purely from gluon-gluon fusion as this mechanism dominates the Higgs production
cross section at the LHC. In instances where a correlation matrix between measurements is given this is
also included in the analysis. For full details on which channels have been used and their experimental
values, see Table~\ref{tab:List-of-Higgs-signal-strengths} in Appendix~\ref{sec:inputs}. { Where multiple measurements are available for a single channel we take an average, which significantly improves our sensitivity (see {\bf flavio}'s documentation for details \cite{Straub:2018kue}).}
 
The couplings of $h^0$ to the massive gauge bosons and fermions are multiplied by factors $\kappa_i$
compared to the SM Higgs, as set out in Section~\ref{sec:2HDM}, with the relations for $\kappa_i$ 
given in Eq.~\eqref{eq:kappas}. As the couplings are modified, the signal strength data can be used 
to place constraints on the two parameters on which the $\kappa_i$ depend: $\cos{(\beta - \alpha)}$ 
and $\tan\beta$.
A scenario in which \begin{align}
    \label{eq:WSL_def}
    \kappa_{\small{V}} = 1, \qquad 
    \kappa_{u} = 1, \qquad 
    \kappa_{d, \ell} = -1,
\end{align} has been introduced as one of the ways a fourth sequential chiral generation of fermions may still exist\footnote{Simply adding a fourth sequential, chiral, perturbative generation of fermions, without any additional new degrees of freedom is definitely ruled out, see e.g. Refs.~\cite{Eberhardt:2012gv,Djouadi:2012ae}.} while remaining hidden from current LHC data~\cite{Das:2017mnu}. By enforcing $\kappa_u = -\kappa_d = 1$ one finds  
\begin{equation}
    \label{eq:WSL_cond}
    \cos{(\beta - \alpha)} = \sin{2\beta} = \frac{2 \tan \beta}{1 + \tan^2 \beta} \, .
\end{equation}
The full criteria of the wrong sign limit are satisfied exactly only when this equation is followed at very large $\tan\beta$, giving the set of couplings in Eq.~\eqref{eq:WSL_def}. The contributions of the additional Higgs bosons to the loop-level processes have been neglected as they are negligible for the allowed masses found in Section \ref{sec:flavour}. 

\begin{figure}[bt]
    \centering
    \includegraphics[width=0.6\textwidth]{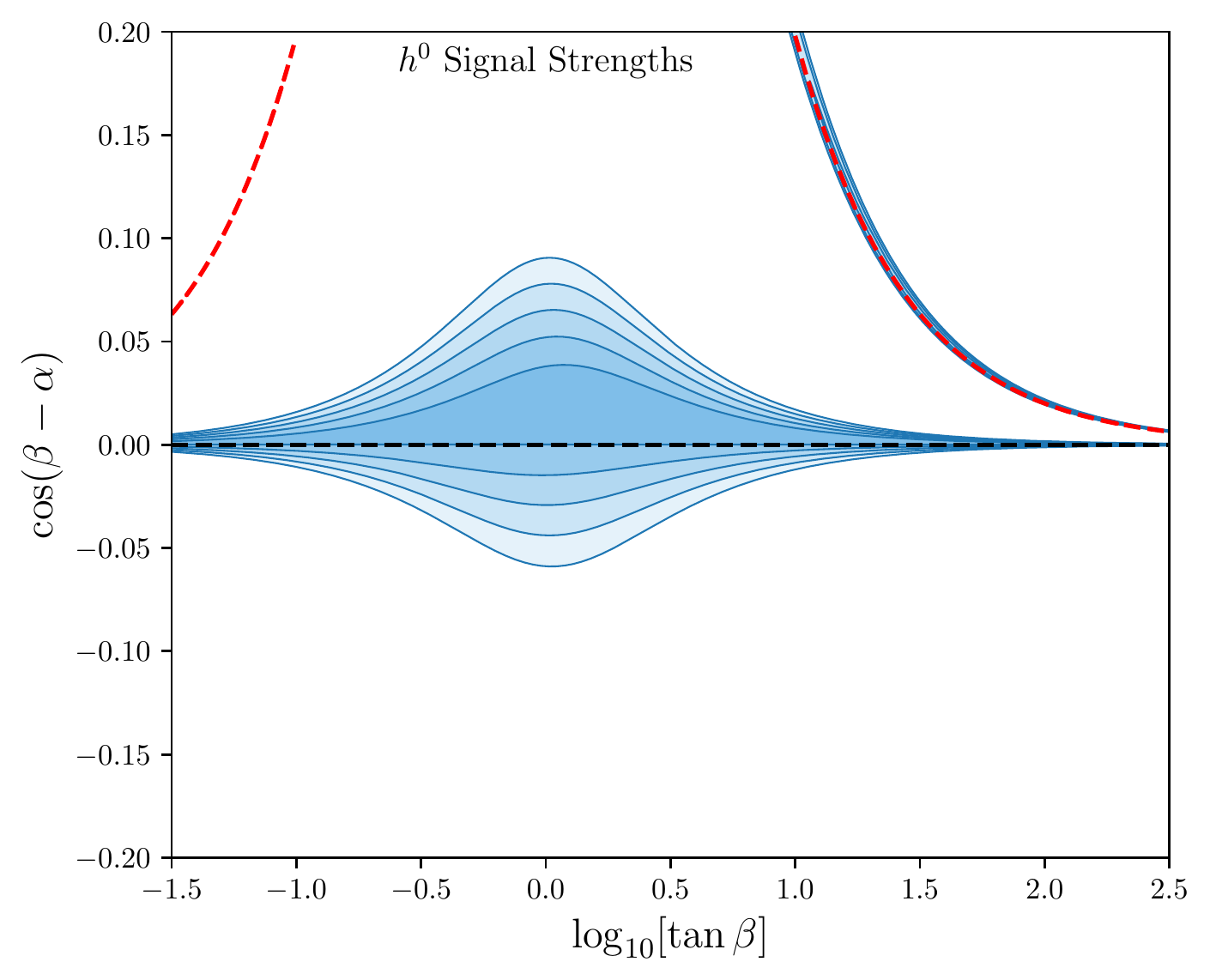}
    \caption{\label{fig:combSS} 
    Contour plot of the allowed 2HDM-II parameter space in the $(\tan\beta-\cos(\beta-\alpha))$ plane from the Higgs signal strengths, with the dashed black line indicating the alignment limit ($\cos(\beta-\alpha)=0$) and the dashed red line the wrong sign limit of Eq.~(\ref{eq:WSL_cond}). The contours indicate the allowed space at 1, 2, 3, 4, 5$\sigma$ from dark to lighter.}
\end{figure}

The results of the analysis including all the Higgs signal strengths 
listed in Table~\ref{tab:List-of-Higgs-signal-strengths} 
are shown in Fig.~\ref{fig:combSS} (see also Table~\ref{tab:Comb_fit_res});
we will later include the 
Higgs signal strengths in our global fit. 
% This fit was performed with \textbf{flavio}~\cite{Straub:2018kue} (see Section~\ref{sec:introduction} for details) using the Warsaw SMEFT operator basis~\cite{Grzadkowski:2010es}. The \textbf{flavio} package calculates the signal strengths using numeric results, apart from in the key channels of gluon fusion production and diphoton decay. In these channels one loop effects are accounted for analytically, including scope for modified Higgs couplings, exactly as we require for this work. The details of the \textbf{flavio} procedure for Higgs signals calculations are explained fully in Ref.~\cite{Falkowski:2019hvp}. We calculate the relevant Wilson coefficients that undergo modifications due to the altered couplings of $h^0$ in the 2HDM. 
This fit is performed using the analytic expressions for the Higgs production and decay channels in the 2HDM found in Ref.~\cite{Gunion:1989we}.
As outlined above, the coupling modifications, the $\kappa_i$, are functions of $\tan\beta$ and $\cos(\beta-\alpha)$, allowing us to constrain this plane. 
%{\blue Additionally, in order to verify our results, we have performed a full analytical fit  outside the \textbf{flavio} framework using the expressions in Ref.~\cite{Gunion:1989we}, and find good agreement between the two fits.}
For some recent studies that include a similar analysis in the 2HDM, see, for instance,
Refs.~\cite{Chowdhury:2017aav, Dawson:2020oco, Han:2020zqg, Han:2020lta, Ellis:2018gqa}. 
We observe that the alignment limit must be closely followed, which is unsurprising given that almost 
all of the experimental data included here matches the SM within~2$\sigma$, with the majority agreeing
at~1$\sigma$. As a result, the 2HDM must closely replicate the SM behaviour for $h^0$. At 2$\sigma$ we find
\begin{equation}
    \label{eq:MaxCos}
    \big|\cos{(\beta - \alpha)}\big| \leq 0.050,
\end{equation}
with the maximum allowed value of $\cos{(\beta - \alpha)}$ occurring at $\tan\beta$ $\approx$ 1. Due to the dependence of $\kappa_{d,\ell}$ on $\tan\beta$, this maximum value falls significantly once away from $\tan\beta$ $\approx$ 1, to the point where, at large values of $\tan\beta$, the alignment limit must be strictly followed and it is valid to set $\cos{(\beta - \alpha)} = 0$. At 3$\sigma$ and above we find the wrong sign limit to be allowed, which substantially increases this upper bound, provided Eq.~\eqref{eq:WSL_cond} is closely followed.

\begin{figure}[bt]
    \centering
    \includegraphics[width=0.6\textwidth]{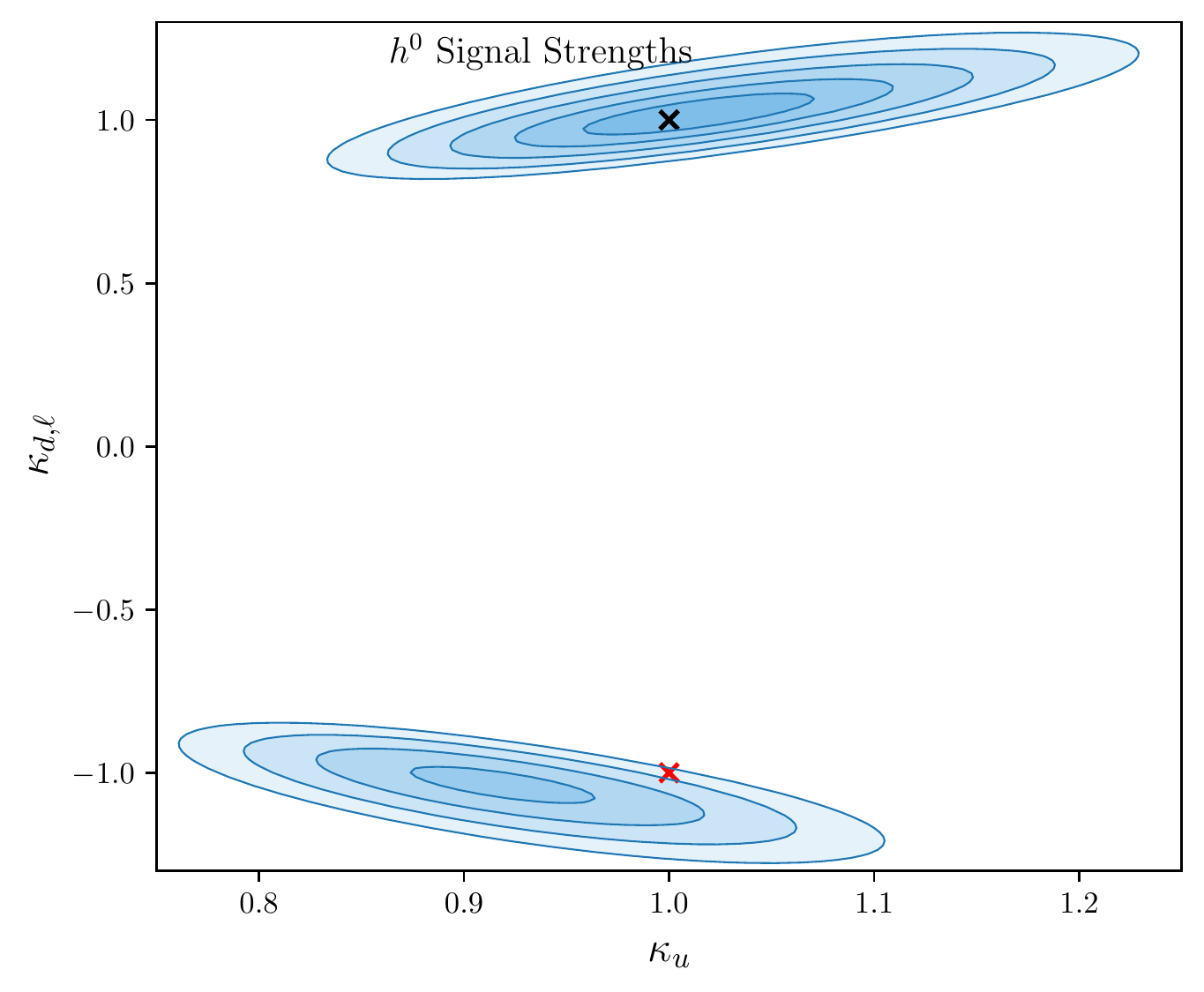}
    \caption{\label{fig:kaps_ud} Contour plot of the coupling constant modifiers, in which we have fixed to its best fit value $\kappa_V =1.036$ and marked the alignment (wrong-sign) limit in black (red). The contours indicate the allowed space at 1, 2, 3, 4, 5$\sigma$ from dark to lighter.}
\end{figure}

The wrong sign condition from enforcing $\kappa_u = -\kappa_d = 1$, Eq.~\eqref{eq:WSL_cond}, which is shown in Fig.~\ref{fig:combSS} as a dashed red line, is here found to be excluded at 2.7$\sigma$
by the signal strength data, an improvement over previous works such as
Refs.~\cite{Han:2020zqg, Aad:2019mbh, Bertrand:2020lyb}, due in part to our inclusion of a larger data set than in such works, which improves our sensitivity. 

To further probe the wrong-sign limit, we examine the possible values of the coupling modification factors, independently from Eq.~\eqref{eq:kappas}. We first perform a fit in terms of the three $\kappa_i$ to find the best fit point, and then fix one of the $\kappa_i$ to its best fit value, allowing a 2D fit in terms of the remaining $\kappa_{j,k\neq i}$ to be performed. We show the results from such a fit in the $(\kappa_u-\kappa_{d,\ell})$ plane in Fig.~\ref{fig:kaps_ud}, where we have fixed to the best fit value $\kappa_V = 1.036$, and stress again that these results do not make use of Eq.~\eqref{eq:kappas}. 
This fit does not exclude $\kappa_d = -1$, in line with the current lack of sensitivity to the sign of $\kappa_d$ at the LHC. We then look to find solutions in terms of $\tan\beta$ and $\cos(\beta-\alpha)$ that give values of $\kappa_u$ and $\kappa_d$ that lie within the 2$\sigma$ allowed region close to $\kappa_d =-1$ (bottom part of Fig.~\ref{fig:kaps_ud}) by using Eq.~\eqref{eq:kappas}. However, the only solutions we find require that $|\cos(\beta-\alpha)| \gtrsim 0.96$. Such extreme values of $\cos(\beta-\alpha)$ are not consistent, in the 2HDM-II, with the fixed value of $\kappa_V = 1.036$ from the best fit point that we use in the scan. Therefore this $2\sigma$ allowed region in the $(\kappa_u-\kappa_{d,\ell})$ plane cannot be attained in the 2HDM-II. 
Thus, while we do not in general exclude $\kappa_d = -1$ we can, in the 2HDM-II that we examine here, rule out the arrangement $\kappa_u = -\kappa_d = 1$ (given by Eq.~\eqref{eq:WSL_cond}) at 2.7$\sigma$ from Fig.~\ref{fig:combSS}. 

More generally, we observe that when $\kappa_{d,\ell} = -1$, small deviations away from $\kappa_u = 1$ have a significant impact, causing $\cos(\beta-\alpha)$ to be large, which is incompatible with the signal strength data. The debate in the literature over the possibility of the wrong-sign limit will hopefully be resolved by future collider data, with suggestions on fruitful search channels made in Refs.~\cite{Modak:2016cdm} and~\cite{Raju:2020hpe}, with the latter also including the possibility of a hidden sequential fourth generation of fermions.

\section{Flavour Observables}
\label{sec:flavour}
In this section, we present our fits of the 2HDM-II for selected flavour observables. 
We show fits for groups of observables individually, and then 
in Section~\ref{sec:global}, we combine these with all other observables in the global fit. 
For each fit, we use the standard $(\tan\beta-m_{H^+})$ parameter space in a log-log scale. 
All fits have been performed using the \textbf{flavio} package as described in Section~\ref{sec:introduction}.

\subsection{Leptonic and Semi-leptonic Tree-Level Decays}
\label{sec:leps}

\begin{figure}[H]     
    \centering 
    \includegraphics[width=0.40\textwidth]{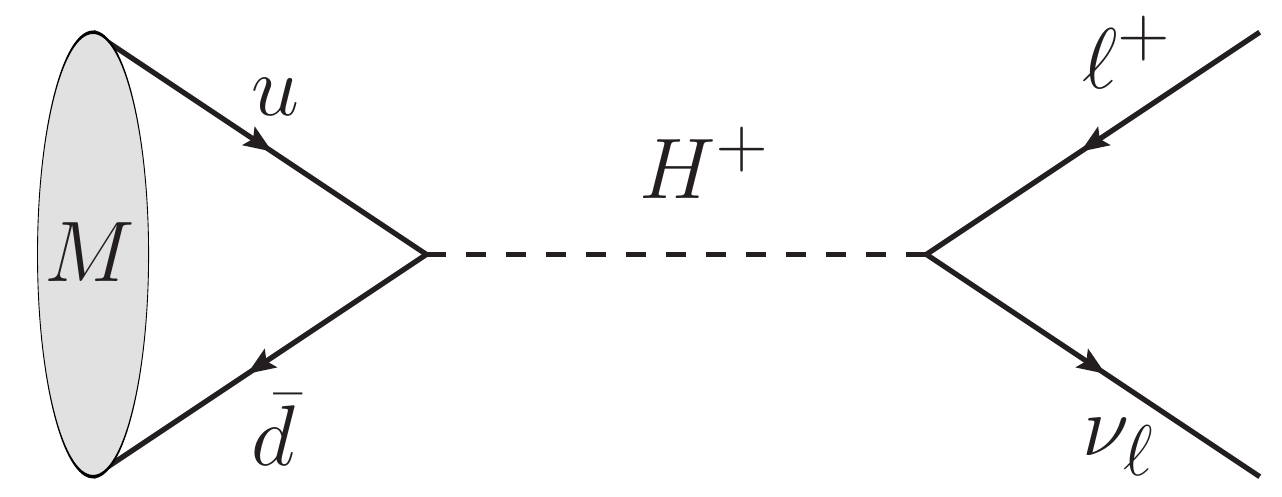}
    \qquad \qquad 
    \includegraphics[width=0.40\textwidth]{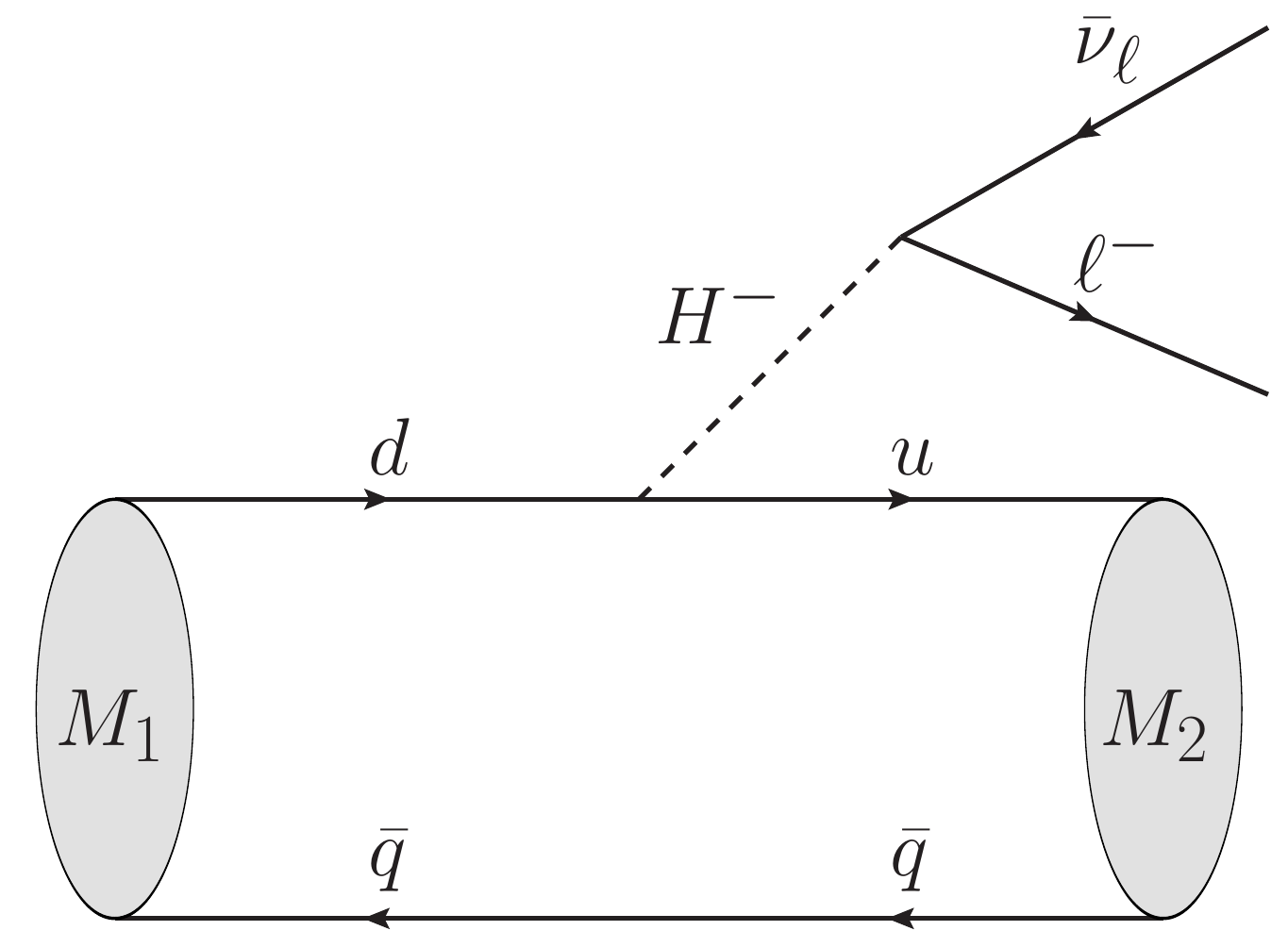}
    \caption{Diagrams contributing to leptonic (left)  
    and semi-leptonic (right) decays in the 2HDM.}        
     \label{fig:l-and-SL-decays-2HDM}
\end{figure}

\begin{figure}[ht]
    \centering
    \includegraphics[width=0.49\textwidth]{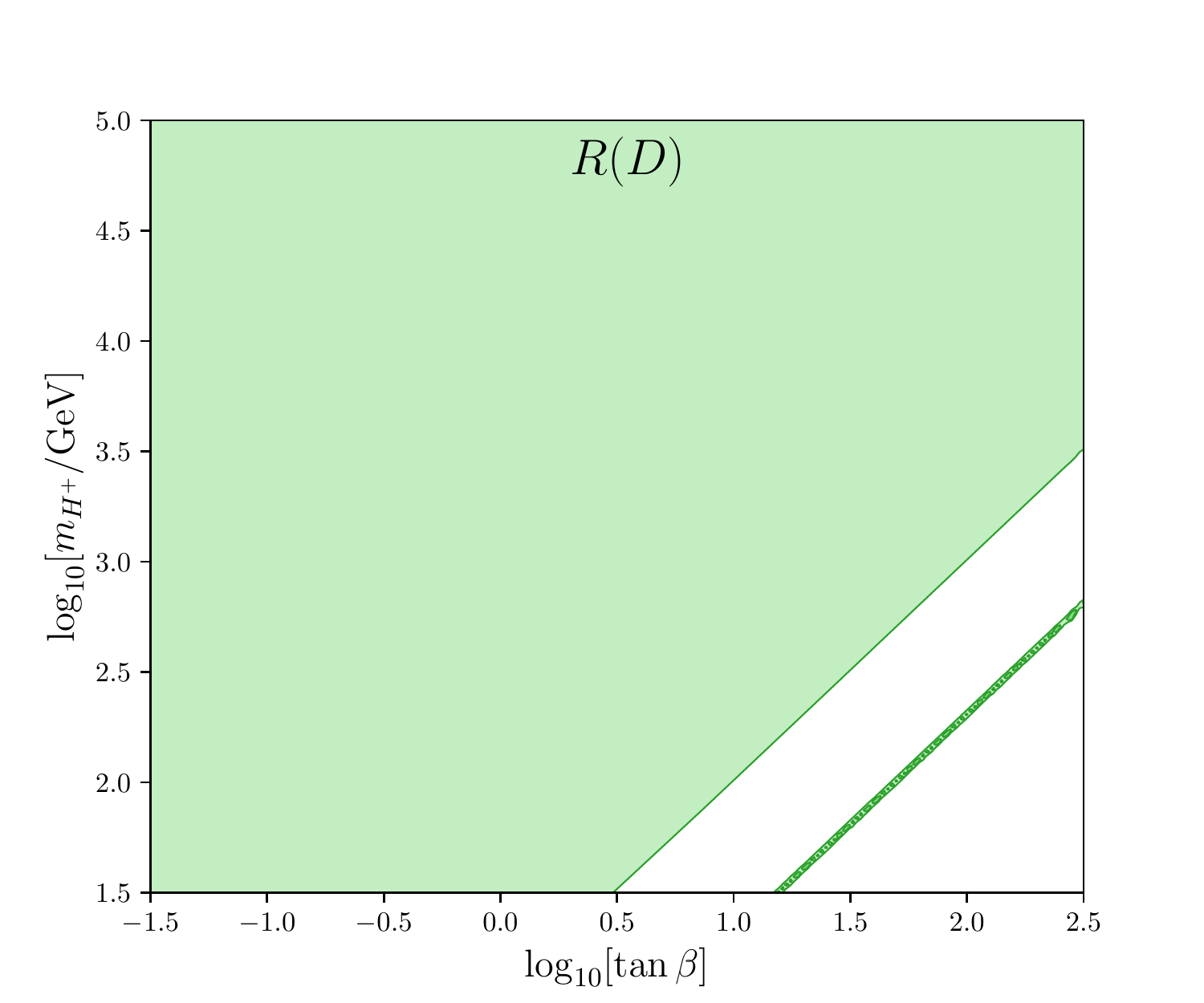}
    \includegraphics[width=0.49\textwidth]{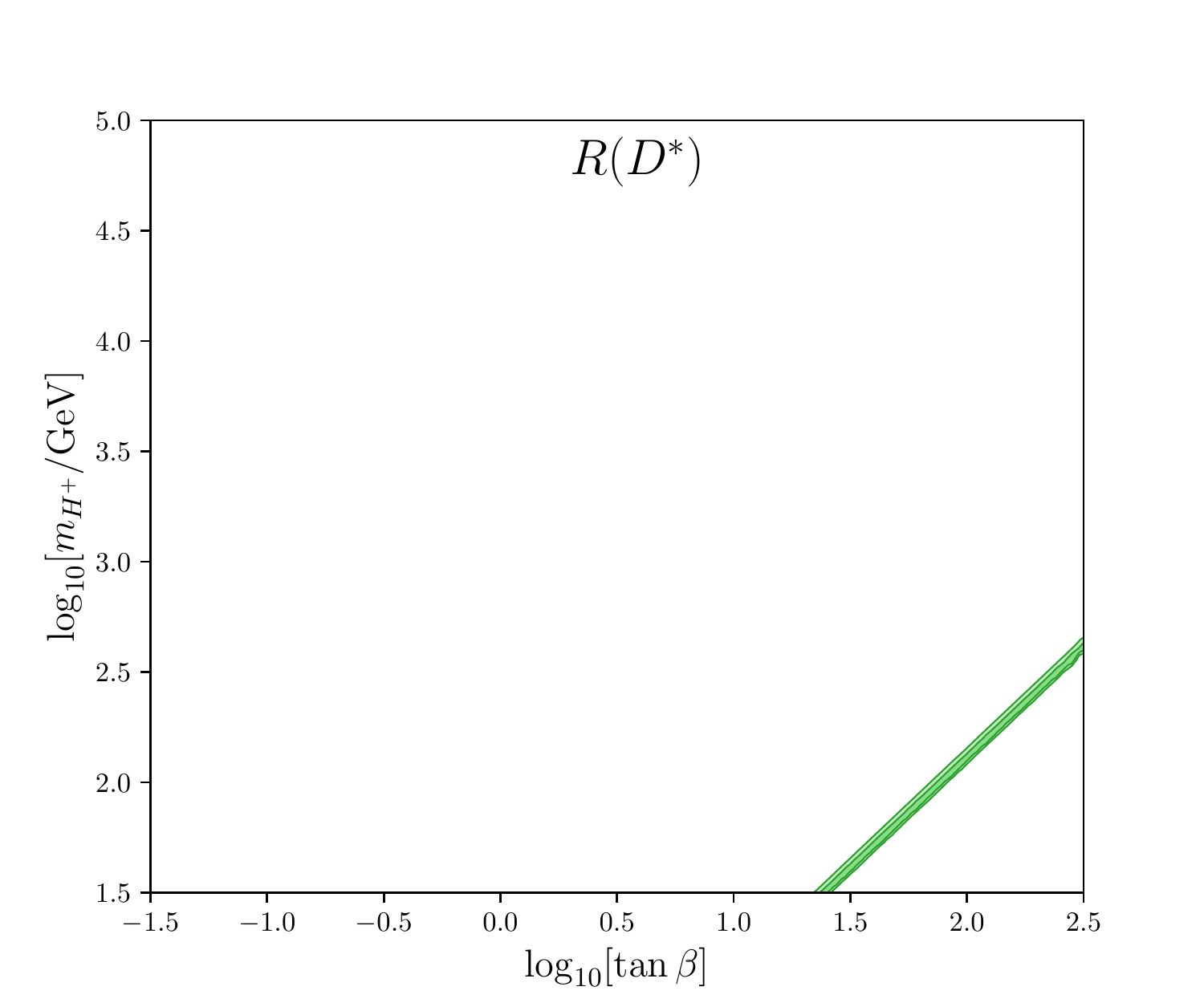}
    \caption{Contour plot of the allowed 2HDM-II parameter space in the $(\tan \beta - m_{H^+})$ plane, originating from $R(D)$ (left) and $R(D^*)$ (right). 
    The lighter contour indicates the allowed parameter space at $2\sigma$ confidence level while the darker contour corresponds to $1\sigma$.}
    \label{fig:RD} 
\end{figure}

The Yukawa term in the 2HDM-II Lagrangian 
(\ref{eq:Yakawa-2HDM-II}) leads to additional contributions
to the tree-level flavour-changing charged transitions.
Integrating out the heavy charged Higgs boson $H^\pm$,
one obtains the effective Hamiltonian describing the $d \to u \, \ell^- \bar \nu_\ell$ 
or $u \to \, d \ell^+ \nu_\ell$ transition in the 2HDM: 
\begin{equation}
{\cal H}^{\rm eff}_{H^+}
= - \frac{4 G_F}{\sqrt 2} V_{u d} 
(C_{S-P} \, {\cal O}_{S-P} + C_{S+P} \, {\cal O}_{S+P}) + {\rm h. c.,}
\label{eq:Leff-L-and-SL}
\end{equation}
where the new effective operators are
\begin{align}
    \label{eq:O-SP}
    \mathcal{O}_{S - P} = (\bar{u} P_L \, d)  (\bar{\ell} P_L \nu_{\ell}), 
    \qquad
    \mathcal{O}_{S + P}   = (\bar{u} P_R \, d) (\bar{\ell} P_L \nu_{\ell}).
\end{align}
The corresponding Wilson coefficients are related to the 2HDM-II parameters as follows:
\begin{align}
    \label{eq:C-SP}
    C_{S - P}  = \frac{m_{u} \, m_\ell}{m_{H^+}^2} \, , 
    \qquad
    C_{S + P} & = \frac{m_{d} \, m_\ell \tan^2 \beta}{m_{H^+}^2} \, .
\end{align}
For use in \textbf{flavio}, 
we convert the latter coefficients to $C_{SL}^{du\ell\nu_\ell},C_{SR}^{du\ell\nu_\ell}$, respectively,
in the \textbf{flavio} WET basis, finding
\begin{align}
    C_{SL}^{du\ell\nu_\ell} = - C_{S - P}  \, ,
    \qquad
    C_{SR}^{du\ell\nu_\ell} &= - C_{S + P} \, ,
\label{eq:Coef-to-Flavio-SL}
\end{align}
where the negative sign in front of these coefficients 
is due to the convention for the effective operators 
adopted in \textbf{flavio}'s basis.
Feynman diagrams describing the leptonic and semi-leptonic transitions in the 2HDM 
are shown in Fig.~\ref{fig:l-and-SL-decays-2HDM}.
\noindent
The full list of leptonic and semi-leptonic modes included in our fit is given 
in Table~\ref{tab:List-of-flavour-observables},
where the corresponding SM predictions (produced in {\bf flavio}) are based on
\cite{Bernlochner:2017jka,Caprini:1997mu,Sakaki:2013bfa,Bharucha:2015bzk,Gubernari:2018wyi,Detmold:2016pkz,Bernard:2006gy,Bernard:2009zm,Antonelli:2010yf}. 
In addition to previously measured semi-leptonic decay channels,
we consider for the first time the $B_s \to D_s^{(*)} \mu \bar \nu_\mu$ modes
measured recently by the LHCb collaboration~\cite{Aaij:2020hsi}, 
and the corresponding $B_s \to D_s^{(*)}$ form factors are determined using 
Lattice QCD \cite{McLean:2019qcx}.
Moreover, we consider the Lepton-Flavour Universality (LFU) observables 
$R (D^{(*)}) \equiv {\cal B} (B \to D^{(*)} \tau \bar \nu_\tau)/{\cal B} (B \to D^{(*)} \ell \bar \nu_\ell)$, 
where $\ell = e$ or $\mu$.
The experimental measurements of the latter were found to be 
in tension with the corresponding SM predictions, giving hints of possible LFU violation. 
In the 2HDM, this violation is caused by different couplings
of the charged Higgs boson to the lepton pair, which 
are proportional to the lepton mass~$m_\ell$.
Using the HFLAV averages for $R(D^{(*)})$ \cite{Amhis:2019ckw},
we present in Fig.~\ref{fig:RD} the allowed $(\tan \beta-m_{H^+})$ regions 
at 1 and 2$\sigma$ levels, where the left (right) plot corresponds 
to $R(D)$ ($R (D^*)$).
Note that the semi-leptonic $b \to c \ell \bar \nu_\ell$ transitions
receive negative corrections compared to the SM contributions, therefore 
$R (D)$ and $R (D^*)$ move further away 
from the measurements (apart from a narrow region of the 
$(\tan \beta - m_{H^+})$ parameter space (see Fig.~\ref{fig:RD}), 
where the 2HDM-II correction becomes more than the twice the size of the SM contribution proportional to the scalar form factor). 
Since $R(D)$ is in fact only $1.2\sigma$ 
away from the SM prediction (see Table~\ref{tab:List-of-flavour-observables}), 
one finds a wide allowed range of $\tan \beta$ and $m_{H^+}$ at the $2\sigma$ level 
(the left plot in Fig.~\ref{fig:RD}). 
On the other side, $R(D^*)$ is in greater tension ($2.8 \sigma$) with the SM, therefore  one obtains just a 
narrow region with large $\tan \beta$ and small $m_{H^+}$ (the right plot in Fig.~\ref{fig:RD}). 
The~experimental combination of both $R(D)$ and $R(D^*)$  \cite{Amhis:2019ckw}
is more than three standard deviations away from the SM prediction, 
and within 2$\sigma$ one finds in the 2HDM-II only a very narrow region at
very low masses of the charged Higgs, $m_{H^+} \sim 1 \, {\rm GeV}$, which is far from the physical domain.
We find that the 2HDM-II is not able to accommodate
the experimental data on both $R(D)$ and $R(D^*)$ within 
$3.5 \, \sigma$; the corresponding tension in the SM (using {\bf flavio}) is $3.2 \, \sigma$.

\begin{figure}[ht]
    \centering
    \includegraphics[width=0.49\textwidth]{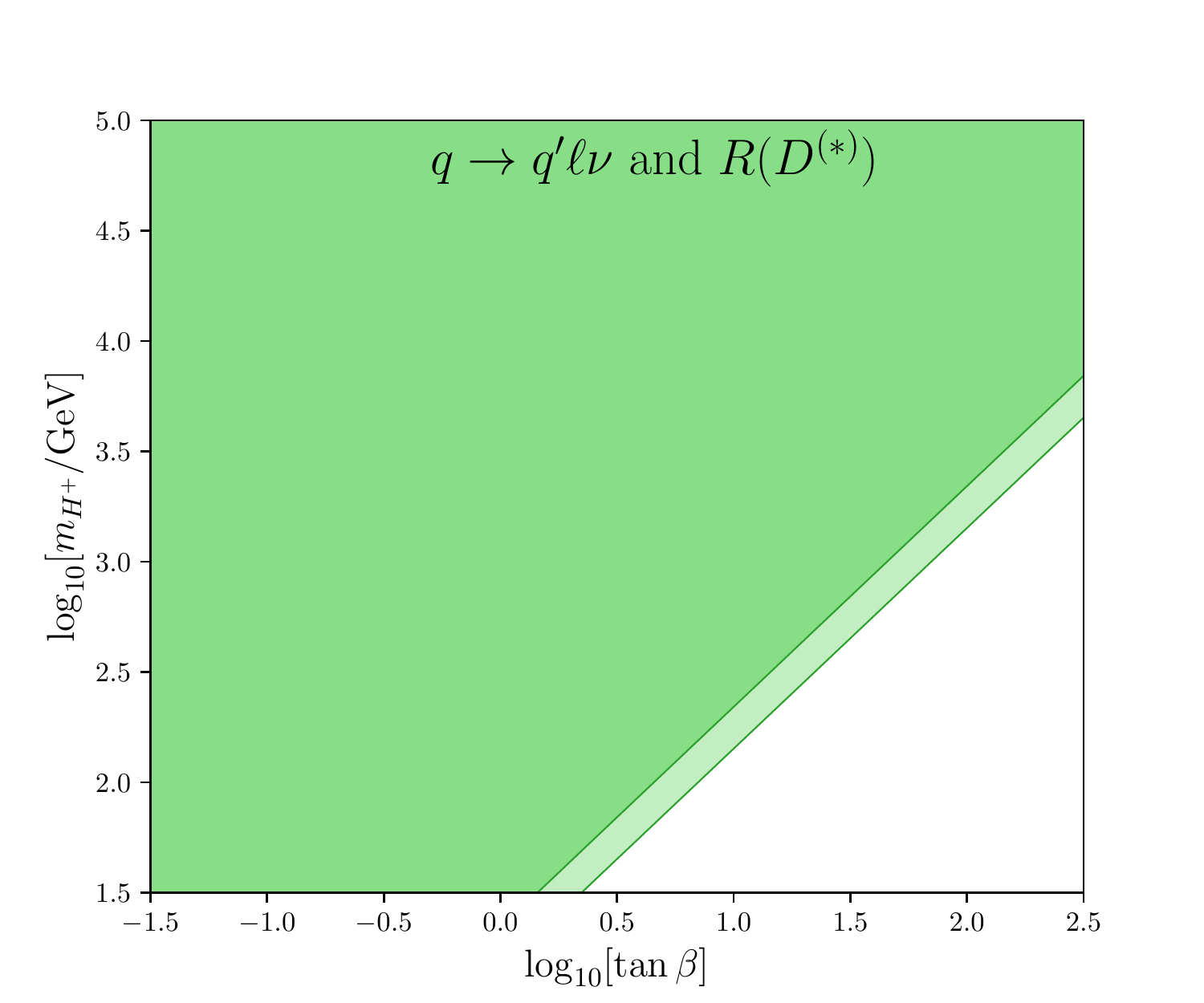}
    \setlength{\belowcaptionskip}{-12pt}
    \caption{Contour plot of the allowed 2HDM-II parameter space in the $(\tan \beta - m_{H^+})$ plane, originating from the combination 
    of tree-level leptonic and semi-leptonic decays of $B, B_s, D, D_s, K$, and $\pi$ mesons 
    and the hadronic decays of $\tau$ leptons to $K$ and $\pi$ mesons with a tau neutrino as well as $R(D)$ and $R(D^*)$, see Table~\ref{tab:List-of-flavour-observables}. 
    The lighter contour indicates the allowed parameter space at $2\sigma$ confidence level while the darker contour corresponds to $1\sigma$.}
    \label{fig:L-and-SL} 
\end{figure}

Combining all the leptonic and semi-leptonic tree-level decay channels indicated in Table~\ref{tab:List-of-flavour-observables} yields the allowed range of $\tan \beta$  and $m_{H^+}$ at $1$ and $2\sigma$ levels as shown in Fig.~\ref{fig:L-and-SL}.

In addition, we would like to make an important observation regarding the extraction of the CKM matrix
elements from experimental data. Consider e.g. the branching fraction of a leptonic meson decay in the
2HDM, 
\begin{equation}
    \label{eq:Br-Lept-2HDM}
    \mathcal{B}_{\rm 2HDM} = 
    \mathcal{B}_{\rm SM} \times (1+\delta_{\rm 2HDM})^2, 
\end{equation}
where $\delta_{\rm 2HDM}$ is the 2HDM correction factor. Conventionally, the corresponding CKM element is determined from the experimental measurement, assuming $\mathcal{B}_{\rm exp} = \mathcal{B}_{\rm SM}$. But if the 2HDM is realistic
then $\mathcal{B}_{\rm exp} = \mathcal{B}_{\rm 2HDM}$ and the CKM element extracted from measurement is actually proportional to $(1 + \delta_{\rm 2HDM})$.
Therefore, the fraction $1/(1 + \delta_{\rm 2HDM})$ is then interpreted as the modification factor that 
the measured CKM element must receive in the 2HDM to be the true CKM element.
A similar argument holds for the semi-leptonic meson decays.  

Using Eq.~\eqref{eq:Br-Lept-2HDM}, and a similar relation for the semi-leptonic decays, the size of the modification factor $1/(1 + \delta_{\rm 2HDM})$ can be scanned across the 2HDM-II parameter space $(\tan\beta-m_{H^+})$ to test the significance of the factor for each of the CKM matrix elements, where the accepted values are commonly taken from leptonic and semi-leptonic tree-level decays.
We perform tests of this significance for $V_{us}$, $V_{ub}$ and $V_{cb}$ using the tree-level parameterisation to consider the propagated effects to all CKM elements, and we also take into account unitarity of the CKM matrix.
We find negligible effects to all CKM elements for most of the parameter space considered here, and the small areas where more noticeable effects arise are excluded from theoretical constraints, direct searches, 
and by the global fit in Section~\ref{sec:global}.

Were the 2HDM-II proven to be physical, it could be the case that there is some small modification to CKM elements that, although is not enough to impact results here, would need to be considered. 
This would be dependent on the specific values found for $m_{H^+}$ and $\tan\beta$ from experiment.

\subsection{Neutral $B$-Meson Mixing}
\label{sec:mix}
\begin{figure}[ht]  
    \centering                                      
    \includegraphics[scale=0.5]{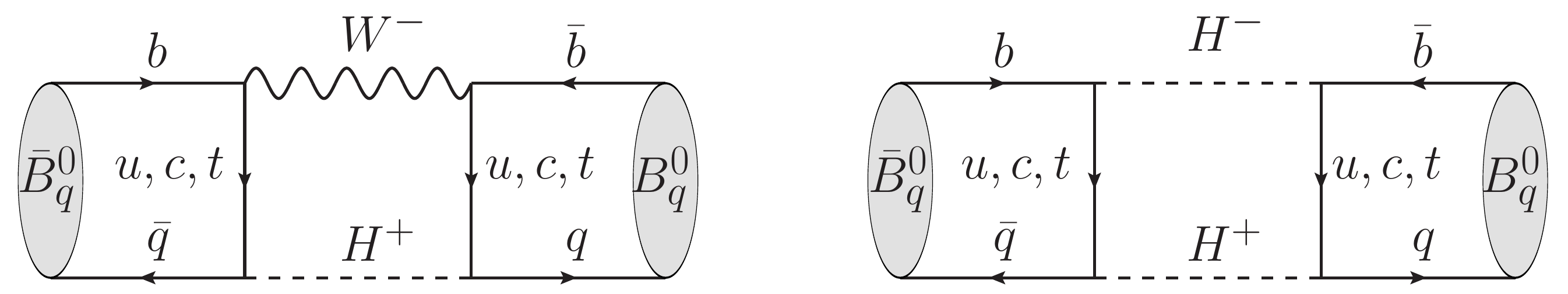}
    \caption{Examples of box diagrams describing $B_d$- and $B_s$-meson mixing in the 2HDM.} 
    \label{fig:mix} 
\end{figure}

\begin{figure}[t]
    \centering
    \includegraphics[width=0.6\textwidth]{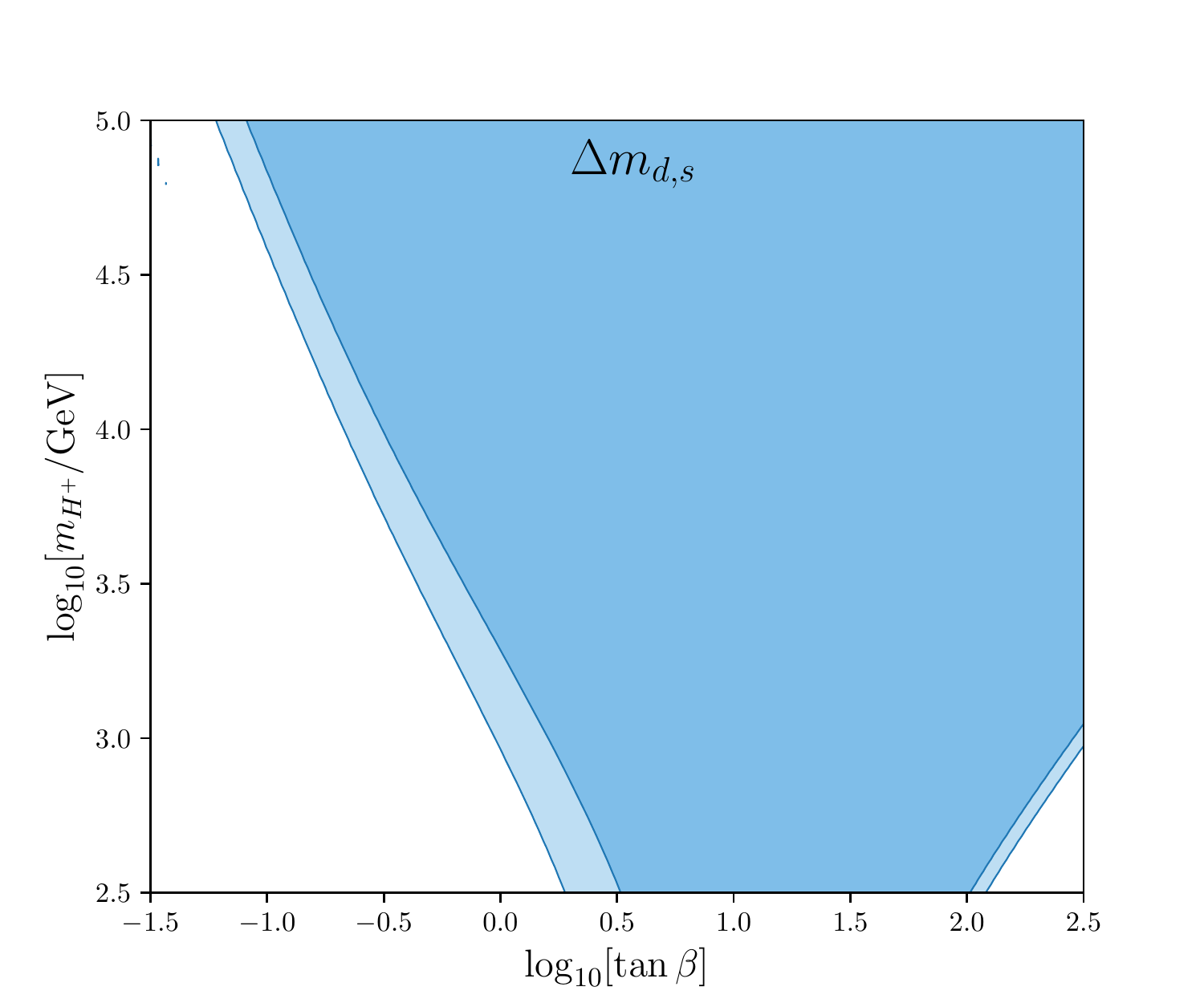}
    \caption{Contour plot of the allowed 2HDM-II parameter space in the $(\tan \beta - m_{H^+})$ plane for the mass difference 
    in the mixing of $B_d$ and $B_s$ mesons.
    The lighter contour indicates the allowed parameter space at $2\sigma$ confidence level while the darker contour corresponds to $1\sigma$.}
    \label{fig:mixplot}
\end{figure}

Neutral $B$-meson mixing plays an important role
in flavour physics. For a recent discussion 
on $B$-mixing in the SM and within New Physics
models see, for example, Ref.~\cite{DiLuzio:2019jyq}.
In the 2HDM, $B$-mixing gains additional
contributions from new box diagrams where one or
both of the virtual $W^\pm$ bosons are replaced
by charged Higgses $H^\pm$, see Fig.~\ref{fig:mix}.
Here we consider the mass difference $\Delta m_{d,s}$
between the heavy and light eigenstates in both $B_d$
and $B_s$ meson mixing.
\noindent
The effective 2HDM Hamiltonian for $\Delta B=2$ processes can be expressed as \cite{Crivellin:2019dun}:
\begin{equation}
    \mathcal{H}_{\rm eff}^{\Delta B=2}
    = \sum_{k=1}^5 C_k \mathcal{O}_k + \sum_{k=1}^3 C_k^\prime \mathcal{O}_k^\prime,
    \label{eq:Heff-mixing}
\end{equation}
where the operators are defined (for $q=d,s$) as
\begin{equation}
\label{eq:mixing-operators}
\begin{aligned}
    \mathcal{O}_1^{(')} 
    & = (\bar{q}^\alpha \gamma^\mu P_{L(R)}b^\alpha) (\bar{q}^\beta\gamma_\mu P_{L(R)}b^\beta), 
    & \qquad
    & \\
    \mathcal{O}_2^{(')} 
    & = (\bar{q}^\alpha P_{L(R)}b^\alpha)(\bar{q}^\beta P_{L(R)} b^\beta), 
    & \mathcal{O}_4 
    & = (\bar{q}^\alpha P_L b^\alpha)(\bar{q}^\beta P_R b^\beta),
    \\
    \mathcal{O}_3^{(')} 
    & = (\bar{q}^\alpha P_{L(R)} b^\beta)(\bar{q}^\beta P_{L(R)} b^\alpha), 
    & \mathcal{O}_5 
    & = (\bar{q}^\alpha P_L b^\beta)(\bar{q}^\beta P_R b^\alpha),
\end{aligned}
\end{equation}
with $\alpha$ and $\beta$ denoting the colour indices.
The contributions of these operators to $B$-meson mixing in the 2HDM 
are well-known throughout literature, see e.g. Refs.~\cite{Geng:1988bq, Urban:1997gw, Crivellin:2019dun}.
In our work, for the Wilson coefficients $C^{(\prime)}_k$ in
Eq.~\eqref{eq:Heff-mixing}
we use expressions calculated in Ref.~\cite{Crivellin:2019dun}.
We convert these to the \textbf{flavio} WET basis \cite{Aebischer:2017ugx}, where we find
\begin{equation}
\begin{aligned}
    C_{VLL(RR)}^{bqbq} &= C_1^{(')}, & \qquad C_{SLR}^{bqbq} &= C_4, \\
    C_{SLL(RR)}^{bqbq} &= C_2^{(')}, & C_{VLR}^{bqbq} &= -\frac12\, C_5. 
\end{aligned}
\end{equation}
The operators $\mathcal{O}_3^{(')}$ do not give any contributions (at LO in QCD) to mixing 
in the 2HDM \cite{Crivellin:2019dun}, so we do not need to consider conversion for $C_3^{(\prime)}$ here.
The largest theory uncertainties by far stem from the
non-perturbative values of the matrix elements of the
$\Delta B = 2 $ operators given in
Eq.~(\ref{eq:mixing-operators}). In our analysis we will use the averages presented in Ref.~\cite{DiLuzio:2019jyq}, which are based on HQET Sum Rule evaluations 
\cite{King:2019lal,Kirk:2017juj,Grozin:2016uqy}
and lattice simulations
\cite{Dowdall:2019bea,Boyle:2018knm,Bazavov:2016nty}.
The perturbative SM corrections are known and implemented to NLO-QCD accuracy \cite{Buras:1990fn}.
In the numerical analysis we further use the most recent experimental averages for $\Delta m_d$ and $\Delta m_s$ \cite{Amhis:2019ckw}\footnote{Not yet including the recent, most precise value of $\Delta m_s$ from LHCb \cite{Aaij:2021jky}.} (see Table~\ref{tab:List-of-flavour-observables}).
The results of the fit are shown in Fig.~\ref{fig:mixplot}, indicating that the mass differences constrain
the allowed $\tan \beta$ region from below.

\subsection{Loop-Level $b\to s,d$ Transitions}
\label{sec:bsll}
\begin{figure}[ht]
    \centering
    \includegraphics[scale=0.5]{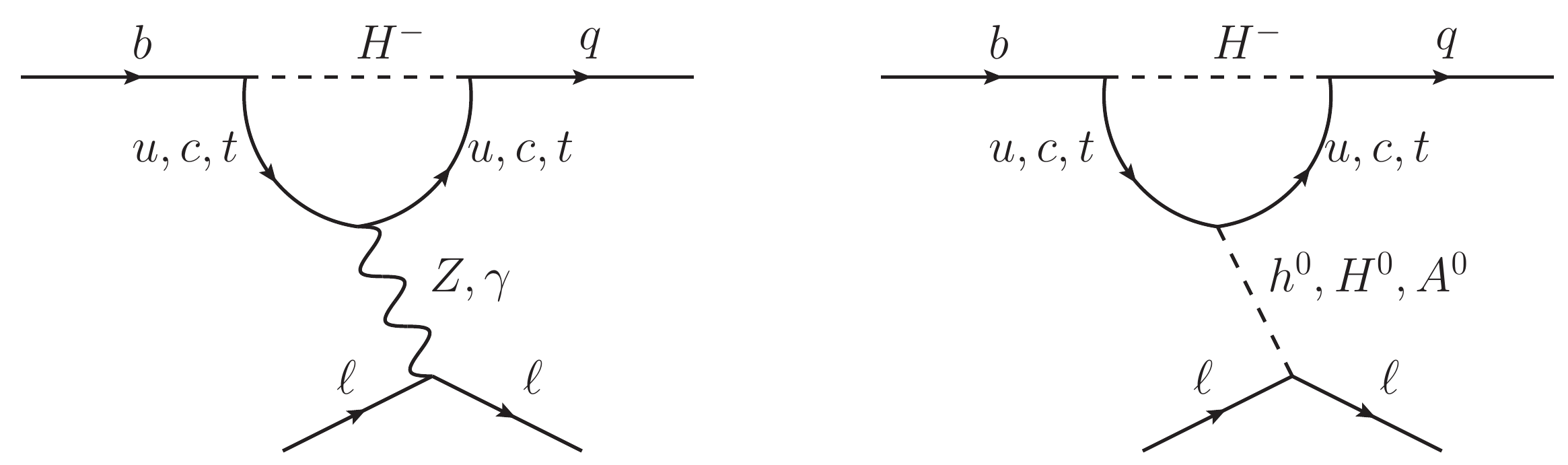} \\[5mm]
    \includegraphics[scale=0.5]{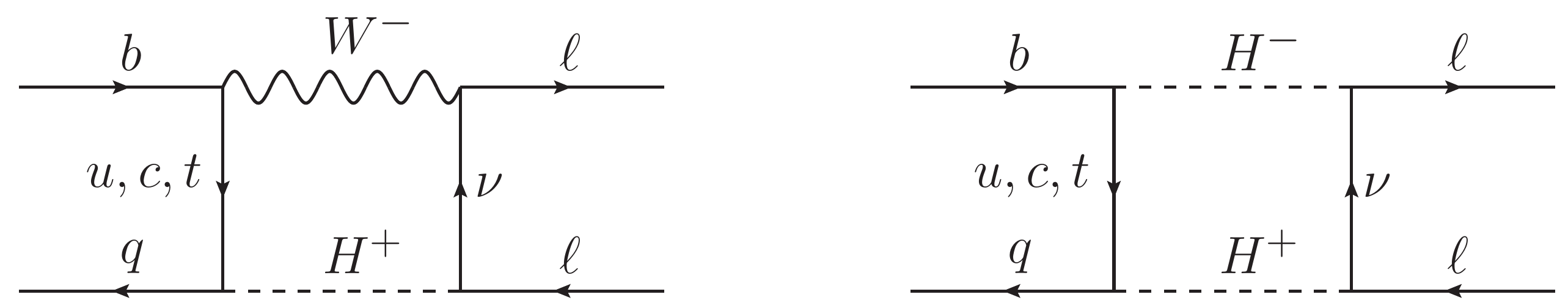}
    \caption{Example of diagrams (penguins and boxes) describing  the $b \to q \ell^+ \ell^-$ transitions in the 2HDM.}
    \label{fig:b-to-q-ell-ell-2HDM}
\end{figure}

The effective Hamiltonian for flavour-changing neutral current (FCNC) $b\to q \ell^+\ell^+$ and $b \to s\gamma$  (for $q = d ,s $) 
processes is defined as \cite{Crivellin:2019dun} 
\begin{equation}
    \mathcal{H}_{\rm eff}^{b \to q \ell \ell} = -\frac{4G_F}{\sqrt{2}}V_{tb}V_{tq}^*
    \left(\sum_{k =7,8} C_k^{(')} \mathcal{O}_k^{(')} + 
    \sum_{k = 9, 10, S, P} \!\!\!
    C_k^{(')} \mathcal{O}_k^{(')}\right) + {\rm h.c.}.
    \label{eq:Heff-b-to-s-ell-ell}
\end{equation}
The operators in this Hamiltonian are
\begin{equation}
\label{eq:bsll-operators}
\begin{aligned}
    \mathcal{O}_7^{(')} 
    &  = \frac{e \, m_b}{16\pi^2}\left(\bar{q}\sigma^{\mu\nu}P_{R(L)} b\right) \! F_{\mu\nu}, 
    & \mathcal{O}_8^{(')} 
    & = \frac{g_s \, m_b}{16\pi^2}\left(\bar{q}\sigma^{\mu\nu} P_{R(L)} T^a b \right) \! G_{\mu\nu}^a, \\
    \mathcal{O}_9^{(')} &= \frac{e^2}{16\pi^2}(\bar{q}\gamma_\mu P_{L(R)}b)(\bar{\ell}\gamma^\mu\ell), & \mathcal{O}_{10}^{(')} &= \frac{e^2}{16\pi^2}(\bar{q}\gamma_\mu P_{L(R)}b)(\bar{\ell}\gamma^\mu\gamma_5\ell), \\
    \mathcal{O}_S^{(')} &= \frac{e^2}{16\pi^2}(\bar{q}P_{L(R)}b)(\bar{\ell}\ell), 
    & \mathcal{O}_P^{(')} &= \frac{e^2}{16\pi^2}(\bar{q}P_{L(R)}b)(\bar{\ell}\gamma_5\ell).
\end{aligned}
\end{equation}
The expressions for 2HDM-II contributions to each operator's Wilson coefficient are given e.g.~in Section~3 of \cite{Crivellin:2019dun}.
Since all the operators in Eq.~\eqref{eq:Heff-b-to-s-ell-ell} are defined in the same way
as in the \textbf{flavio} WET basis (for either $\ell=e,\mu$), no Wilson coefficient conversions are required here.
Some example diagrams describing the $b \to q \ell^+ \ell^-$ transitions in the 2HDM are shown in Fig.~\ref{fig:b-to-q-ell-ell-2HDM}.

\subsubsection{Radiative Decay $\bar{B}\to X_s\gamma$}
\label{sec:rad}
\begin{figure}[ht]
    \centering
    \includegraphics[width=0.45\textwidth]{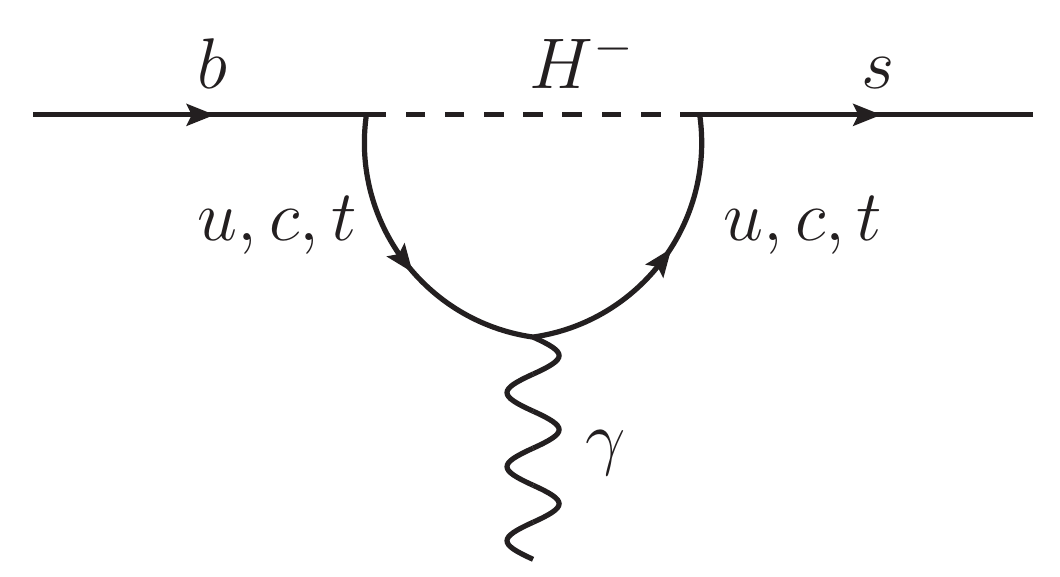}
    \caption{One-loop penguin contributions to $b\to s\gamma$ processes in the 2HDM.}
\end{figure}
\begin{figure}[ht]
    \centering
    \includegraphics[scale=0.6]{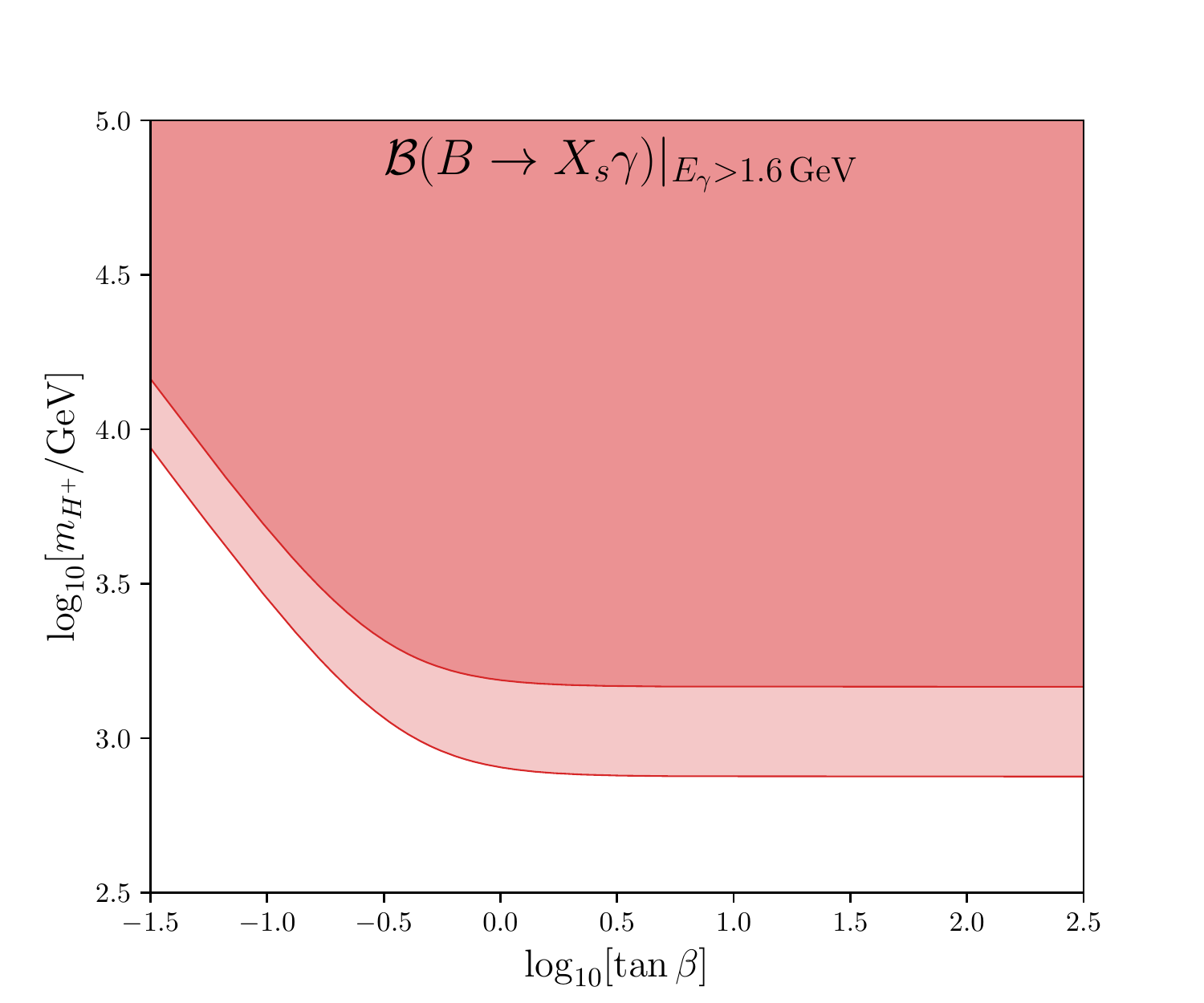}
    \caption{Contour plot of the allowed 2HDM-II parameter space in the $(\tan \beta - m_{H^+})$ plane for the radiative decay $B\to X_s\gamma$.
    The lighter contour indicates the allowed parameter space at $2\sigma$ confidence level while the darker contour corresponds to $1\sigma$.}
    \label{fig:gamma} 
\end{figure}

\noindent
The radiative decay $\bar{B}\to X_s\gamma$ is an important observable in the 2HDM-II. Historically, it has played a significant role in giving a lower bound on the charged Higgs mass.  
A recent update of the SM value with NNLO accuracy in QCD~\cite{Misiak:2020vlo}
(based on e.g. Refs.~\cite{Misiak:2006ab,Misiak:2015xwa}),
\begin{equation}
    \mathcal{B}^{\rm SM} (\bar{B}\to X_s\gamma)\big|_{E_\gamma > 1.6 \, {\rm GeV}} 
    = (3.40 \pm 0.17) \times 10^{-4},
\end{equation}
leads to a stronger constraint on the charged Higgs mass: 
$m_{H^+} \geq 800\,$GeV (95\% CL) \cite{Misiak:2020vlo}.
The experimental average from \cite{Amhis:2019ckw}
is based on \cite{Chen:2001fja,Lees:2012ym,Belle:2016ufb}.
The operators which contribute to the radiative decays $\bar{B}\to X_s\gamma$ are $\mathcal{O}_7,\mathcal{O}_8$, given by Eq.~\eqref{eq:bsll-operators}.
As expected, our fit for this process provides an important lower bound for the charged Higgs mass, predicted here as (see also Fig.~\ref{fig:gamma})
\begin{equation}
    m_{H^+} \gtrsim 790\, (1510) \,\text{GeV} \; \text{ at } 2\sigma\,(1\sigma).
\end{equation}
We slightly differ from Ref.~\cite{Misiak:2020vlo} in that our 2HDM-II contributions are taken at NLO \cite{Borzumati:1998nx}, while they use NNLO results \cite{Hermann:2012fc} and we also use a different statistical treatment.
In addition, as can be seen in Fig.~\ref{fig:gamma}, the bound for $m_{H^\pm}$ becomes even stronger for lower values of $\tan \beta$.

\subsubsection{Leptonic $B_{d,s}\to\mu^+\mu^-$ Decays}
\label{sec:bsmumu}
The FCNC leptonic meson decays $B_{d,s}\to\mu^+\mu^-$ are particularly sensitive to scalar operator contributions 
in NP models, and therefore can be excellent probes of the effects of the 2HDM.
These processes are particularly worthy of consideration now, due to the recently-produced experimental combinations which we use in our fit \cite{Altmannshofer:2021qrr}, and recent ATLAS, CMS, and LHCb results
\cite{Aaij:2017vad,Aaboud:2018mst,Sirunyan:2019xdu,LHCb-CONF-2020-002,LHCb:2021awg,LHCb:2021vsc}.
The SM prediction is based on a perturbative element \cite{Buchalla:1993bv,Bobeth:2013uxa,Beneke:2019slt} and a
non-perturbative determination of the decay constants, see e.g. Ref.~\cite{Bussone:2016iua,Bazavov:2017lyh,Hughes:2017spc}.
It has been common in the past to study these decays in the large $\tan\beta$ 
limit ($\tan\beta\gg\sqrt{m_t/m_b}$, see e.g. Ref.~\cite{Logan:2000iv}) where the Yukawa coupling to $b$ quarks is large and there can be simplifications to the 2HDM contributions of the operators which affect this process: $\mathcal{O}_{10}^{(')}, \mathcal{O}_S^{(')}, \mathcal{O}_P^{(')}$. 
However, a big part of the allowed parameter space of the 2HDM-II lies not within the
large  $\tan\beta$, thus we use only the general expressions given in 
Ref.~\cite{Crivellin:2019dun}.
\begin{figure}[ht]
    \centering
    \includegraphics[width=0.6\textwidth]{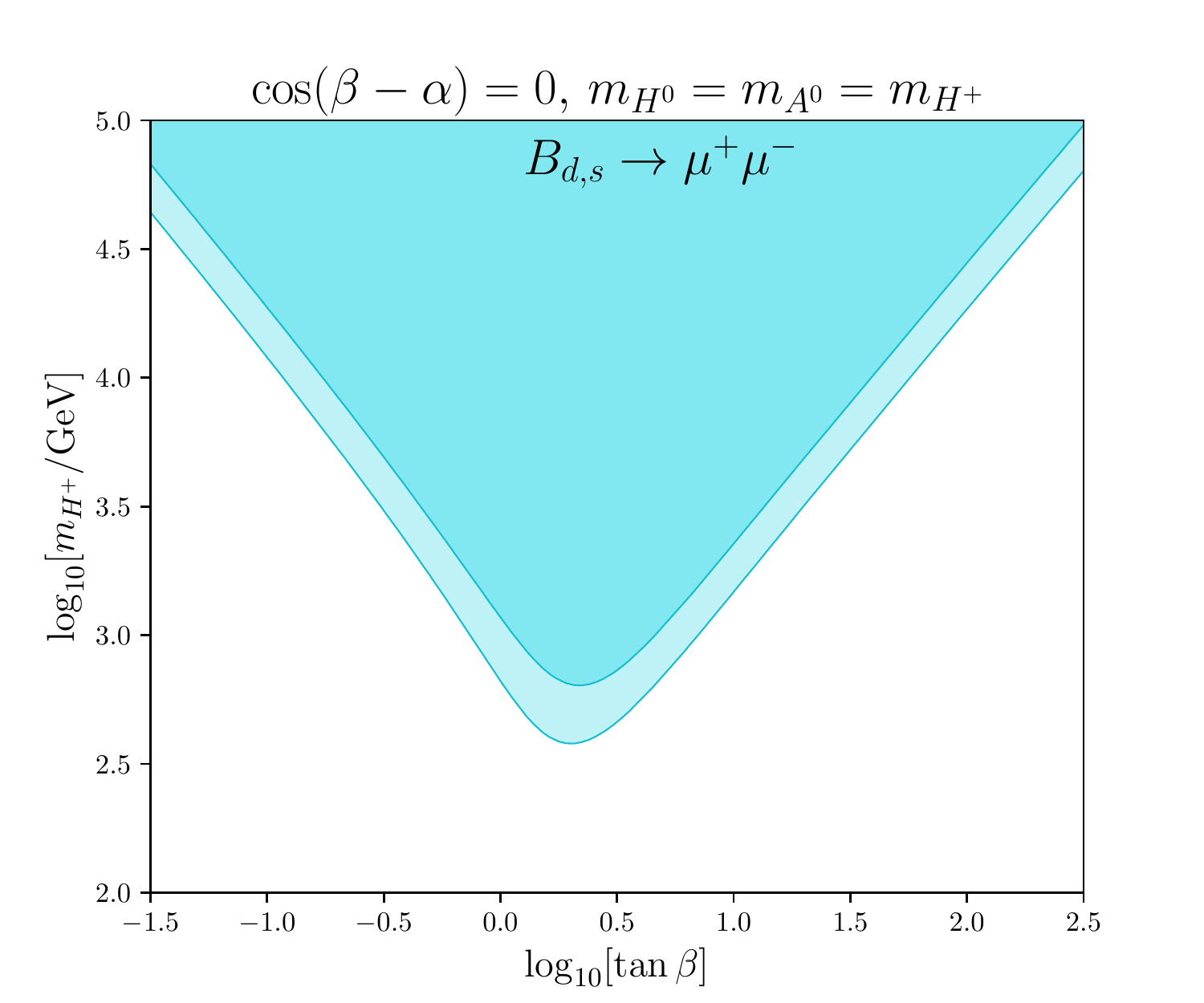}
    \setlength{\belowcaptionskip}{-12pt}
    \caption{ 
    Contour plot of the allowed 2HDM-II parameter space in the $(\tan \beta - m_{H^+})$ plane for the FCNC leptonic $B_{d,s}\to\mu^+\mu^-$ decays, fixing the additional parameters as
    $\cos (\beta -\alpha) = 0$, $m_{H^0} = m_{A^0} = m_{H^+}$.
    The lighter contour indicates the allowed parameter space at $2\sigma$ confidence level while the darker contour corresponds to $1\sigma$.}
    \label{fig:bmumu} 
\end{figure}
These generalised expressions bring three further parameters of the 2HDM into play: $\alpha$, $m_{H^0}$, and $m_{A^0}$. 
In Section~\ref{sec:global}, we will combine all observables in a fit of the five independent parameters: $\tan\beta$, $m_{H^+}$, $m_{H^0}$, $m_{A^0}$ and $\cos(\beta-\alpha)$.
Here, in order to present a 2D fit in the $(\tan\beta-m_{H^+})$ parameter space, we fix the other three parameters 
as $\cos (\beta -\alpha) = 0$, as the alignment limit is preferred 
(see  Section~\ref{sec:global}) by the global fit, and $m_{H^0} = m_{A^0} = m_{H^+}$, which is justified from theoretical constraints (see Section~\ref{sec:theory}). 
The~corresponding contour plot is shown in Fig.~\ref{fig:bmumu}.
From there one can see that the mass of charged Higgs is also constrained from below, roughly:
\begin{equation}
    m_{H^+} \gtrsim 300\, {\rm GeV},
\end{equation}
however this bound is not as strong as that from the 
$B\to X_s\gamma$ decay.
Furthermore, $B_{d,s} \to \mu^+ \mu^-$ yields a strong correlation between
possible values of $\tan\beta$ and $m_{H^+}$.

\subsubsection{Semi-leptonic $b \to s \ell^+\ell^-$ Transitions}
\label{sec:b-to-s-ell-ell}
For the semi-leptonic $b\to s\ell^+\ell^-$ processes we are considering, all of the operators from Eq.~\eqref{eq:bsll-operators} contribute.
Therefore all the 2HDM parameters, $m_{H^+}$, $m_{H^0}$, $m_{A^0}$, $\tan\beta$ and $\cos(\beta-\alpha)$, will affect these processes.

The first group of observables are the LFU ratios $R_K$ and $R_{K^{*}}$ 
\cite{Bobeth:2007dw}, defined as
\begin{equation}
    R_{K^{(*)}} \equiv \frac{\mathcal{B} [B\to K^{(*)} \mu^+\mu^-]}{\mathcal{B} [B \to K^{(*)} e^+e^-]},
\label{eq:RK}
\end{equation}
where the branching fractions are integrated over bins of squared dilepton invariant mass~$q^2$. These quantities are
theoretically very clean, since almost all hadronic effects drop out in the ratio.
In the SM, the values of $R_{K^{(*)}}$ are very close to 1 with tiny uncertainties. The electromagnetic corrections for
$R_{K^{(*)}}$ worked out in Ref.~\cite{Bordone:2016gaq} were found to be very small, $\approx 1-2 \% $. 
In our analysis, we consider the ten binned LFU ratios listed in Table~\ref{tab:RKtab}. The recent measurement by LHCb shows a deviation 
from the SM of $3.1\sigma$ for $R_{K^+}$~\cite{Aaij:2021vac}.
\begin{figure}[ht]
    \centering
    \includegraphics[width=0.6\textwidth]{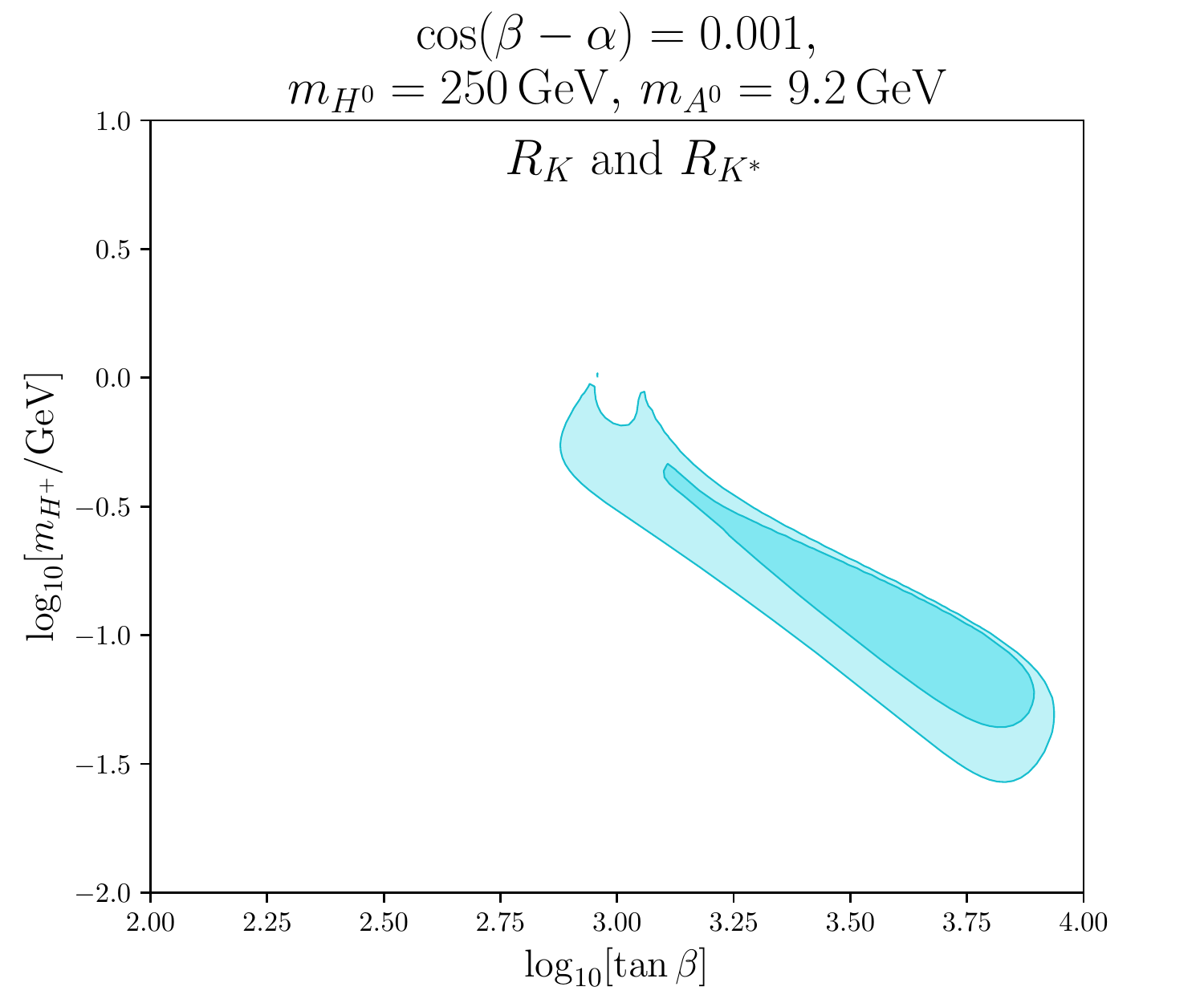}
    \setlength{\belowcaptionskip}{-15pt}
    \caption{
    Contour plot of the allowed 2HDM-II parameter space in the $(\tan \beta - m_{H^+})$ plane for the $R_{K^{(*)}}$ observables and fixing the additional parameters at their best fit point in Table~\ref{tab:Comb_fit_res}.
    The lighter contour indicates the allowed parameter space at $2\sigma$ confidence level while the darker contour corresponds to $1\sigma$.}
    \label{fig:rks}
\end{figure}
\noindent In Fig.~\ref{fig:rks} we present the resulting $(\tan \beta-m_{H^+})$ contour plot
for $R_{K^{(*)}}$ observables,
where we fixed the three other 2HDM parameters at their best fit values for the 10 $R_{K^{(*)}}$ bins:
$\cos (\beta -\alpha) = 0.001$, $m_{H^0} = 250\,{\rm GeV}, m_{A^0} = 9.2\,$ GeV (see Table~\ref{tab:Comb_fit_res}).
As one can see, $R_{K^{(*)}}$ 
could only be accommodated within the 2HDM-II only at 
very low $m_{H^+} \lesssim 1\, {\rm GeV}$ and very high $\tan\beta \gtrsim 300$,
which are beyond the physical domain 
(see Section~\ref{sec:theory}).
Therefore, the 2HDM-II (as well as the SM) is unable to accommodate the current experimental values of $R_{K^{(*)}}$. 
The contours in Fig.~\ref{fig:rks} are not only found in regions which are in disagreement with the allowed
contours in Figs.~\ref{fig:L-and-SL}, \ref{fig:mixplot}, \ref{fig:gamma}, \ref{fig:bmumu}, but also with the constraints from direct searches
\cite{Aaboud:2018gjj}.
Moreover, in deriving the formulae for the relevant Wilson coefficients \cite{Crivellin:2019dun}, it was assumed
that $m_{H^+}$ is at least of the order of the electroweak scale, meaning that the validity of contours much lower than $\log_{10}[m_{H^+}/\text{GeV}] \sim 2$ must be questioned. 

The combined fit of the 10 $R_{K^{(*)}}$ observables is in disagreement with our combined fits of all observables to $4.2 \, \sigma$ confidence.
Due to this disagreement between $R_{K^{(*)}}$ and other flavour observables, we work with two approaches for our fits: 
excluding and including~$R_{K^{(*)}}$.

\begin{figure}[ht]
    \centering
    \includegraphics[width=0.6\textwidth]{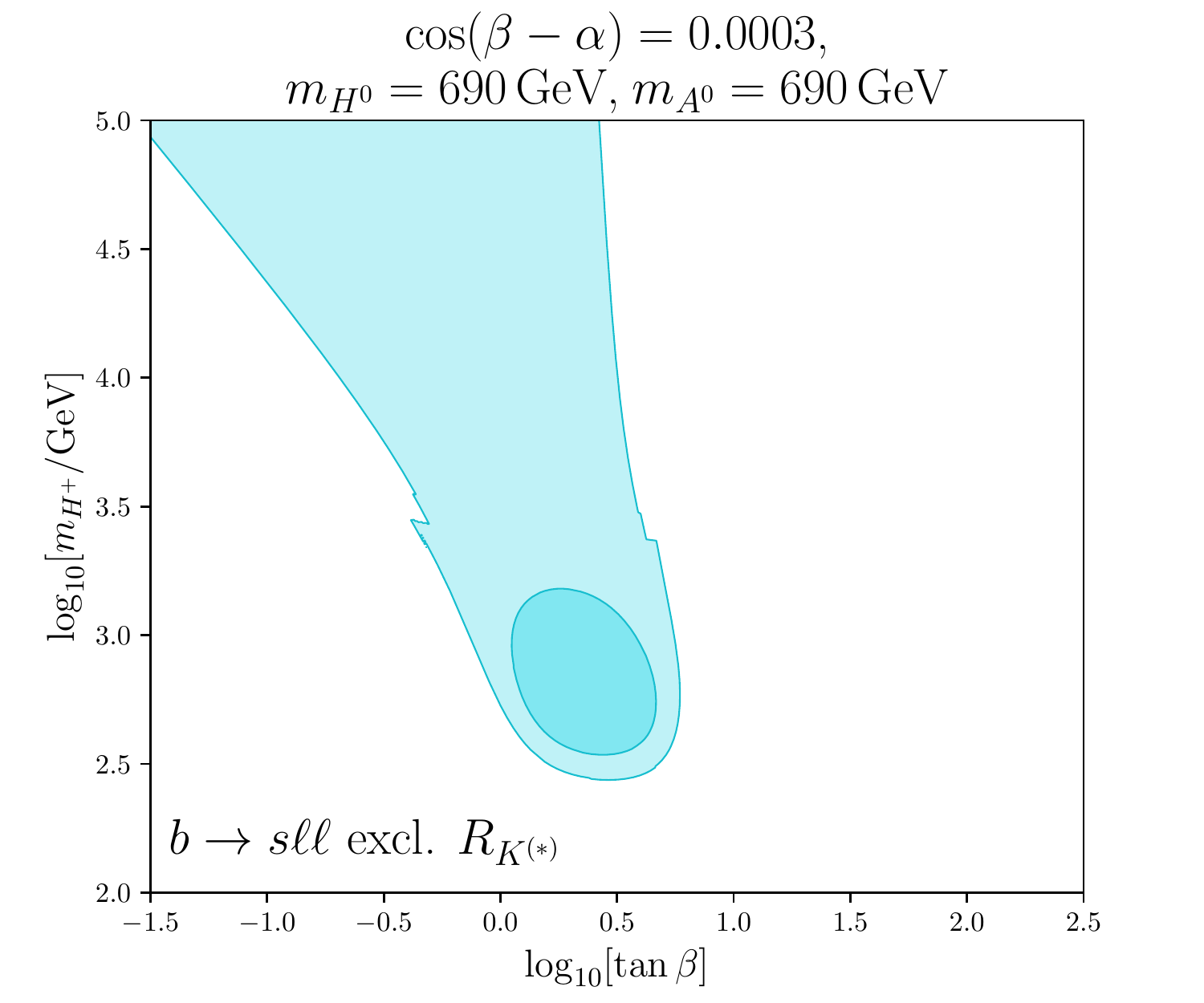}
    \setlength{\belowcaptionskip}{-10pt}
    \caption{Contour plot of the allowed 2HDM-II parameter space in the $(\tan \beta - m_{H^+})$ plane for the leptonic $B_{d,s} \to  \mu^+ \mu^-$ decays and the semi-leptonic $b\to s\ell^+\ell^-$ observables listed in Table \ref{tab:bsll} excluding the $R_{K^{(*)}}$ observables and fixing the additional parameters to the best fit point from the $b\to s\ell^+\ell^-$ fit in Table~\ref{tab:Comb_fit_res}. The lighter contour indicates the allowed parameter space at $2\sigma$ confidence level while the darker contour corresponds to $1\sigma$.}
    \label{fig:bsll}
\end{figure}

Next we include further $b\to s\ell^+\ell^-$ observables 
(binned branching ratios, angular distributions, asymmetries, etc.) -- some of these also deviate from the SM predictions,
see Table~\ref{tab:bsll}.
For more detailed analyses of these processes and their NP implications, see e.g. Refs.~\cite{Khodjamirian:2010vf, Khodjamirian:2012rm, Bharucha:2015bzk, Khodjamirian:2017fxg, Alguero:2019ptt, Hurth:2020ehu, Alok:2019ufo, Ciuchini:2019usw, Hurth:2020rzx, Hurth:2021nsi,Alguero:2021anc,Cornella:2021sby,Altmannshofer:2021qrr,Geng:2021nhg,Ciuchini:2020gvn,MunirBhutta:2020ber,Biswas:2020uaq}.
For all $b\to s \ell^+ \ell^-$ processes listed in Table \ref{tab:bsll} and the leptonic decays $B_q \to \mu^+ \mu^-$ (see
Table \ref{tab:List-of-flavour-observables}), but excluding
the $R_{K^{(*)}}$ observables, one gets the allowed regions for $\tan\beta$ and $m_{H^+}$ shown in Fig.~\ref{fig:bsll}.
Deviations of experiment from SM predictions now introduce an upper bound on the charged Higgs mass at $1\sigma$. 
A further inclusion of the $R_{K^{(*)}}$ observables does not change the contours of allowed regions in Fig.~\ref{fig:bsll} 
sizeably, but does worsen the corresponding fit and the $p$-value, see Table~\ref{tab:Comb_fit_res}.

\section{Global Fit}
\label{sec:global}
\subsection{Results}
Combining all the observables collected in 
Table~\ref{tab:List-of-Higgs-signal-strengths} (Higgs signal strengths),  
Table~\ref{tab:List-of-oblique-parameters} ($S,T,U$), 
Table~\ref{tab:List-of-flavour-observables} (flavour observables without $b \to s \ell^+ \ell^-$),
Table~\ref{tab:bsll}  ($b \to s \ell^+ \ell^-$ observables)
and Table~\ref{tab:RKtab} ($R_{K^{(*)}}$) 
one obtains the results shown in Table~\ref{tab:Comb_fit_res}.
We emphasize that in searching for the best-fit point of the corresponding likelihood function we have applied the theory constraints 
for the Higgs mass differences (see Table~\ref{tab:theory-constraints}). 
As one can see from Table~\ref{tab:Comb_fit_res},
including all the observables yields a relatively poor fit with a low 
$p \approx 1.5 \%$-value. This is due to the fact that 
the $R_{K^{(*)}}$ and $R(D^{(*)})$
quantities cannot be accommodated within the 2HDM-II, 
as was already discussed in Sections~\ref{sec:leps}
and~\ref{sec:b-to-s-ell-ell}. 
Nevertheless the 2HDM-II still achieves a higher level of agreement than the SM and we find a corresponding pull of $2.3 \sigma$.
Not considering the observables $R_{K^{(*)}}$  yields a better fit for the 2HDM-II with a significantly higher $p \approx 7 \%$-value,
again marking a better performance than the SM, with a corresponding pull~of~$1.8 \sigma$.

\begin{figure}[th]
    \centering
    \includegraphics[scale=0.6]{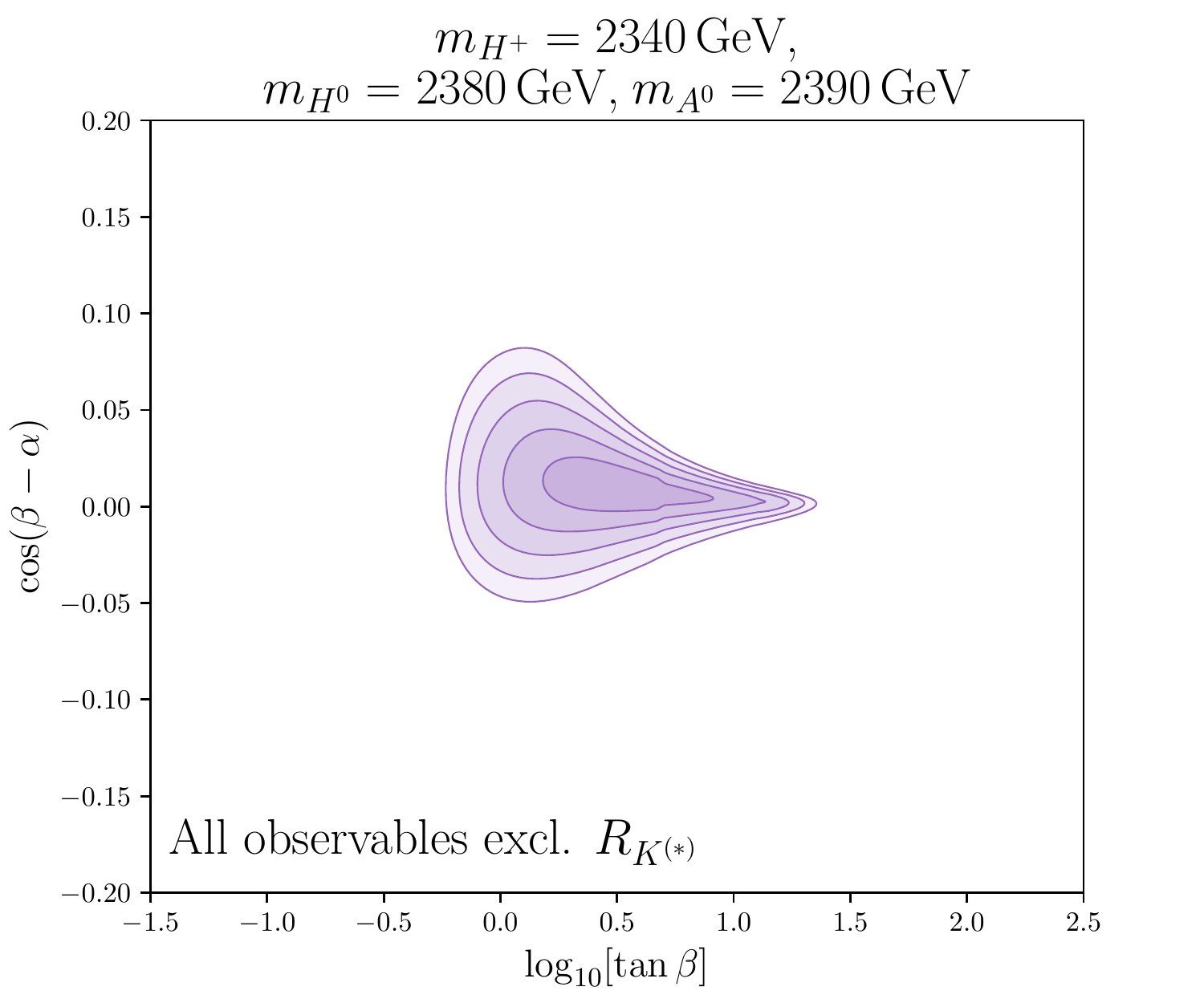}
    \caption{
    Contour plot of the allowed 2HDM-II parameter space in the $(\tan \beta - \cos(\beta-\alpha))$ plane obtained by combining all observables
    excluding $R_{K^{(*)}}$, fixing the additional parameters to the best fit point from Table~\ref{tab:Comb_fit_res}. The contours indicate the allowed parameter space at $1,2,3,4,5\sigma$ confidence level going from darker to lighter.}
    \label{fig:Global-fit-cba-vs-tanbeta}
\end{figure}

Next we find from  Table~\ref{tab:Comb_fit_res}
and Fig.~\ref{fig:Global-fit-cba-vs-tanbeta} 
%($\cos (\beta - \alpha)$ vs $\tan \beta$, with the heavy Higgs masses set to the values of the best fit)
that the global fit prefers the alignment limit, 
$\cos (\beta - \alpha) = 0$, allowing only small deviations from zero:
\begin{equation}
\max \{ |\cos (\beta - \alpha)|\} 
\leq 0.02 \, (0.04), \qquad {\rm at} \, \,  1\sigma \, (2 \sigma).
\end{equation}
This bound is considerably stronger than the one we obtained in Eq.~(\ref{eq:MaxCos}) using only data on Higgs signal strengths.
Regarding the wrong sign limit solution, we found using only the Higgs signal strengths this to be allowed at $2.7\sigma$ or above; using the global fit of all observables (excluding $R_{K^{(*)}}$), 
this is now excluded within $6\sigma$ of our best fit point with $p$-value $\approx 7\%$.
We also perform fits to all observables requiring the wrong sign limit as defined in Eq.~\eqref{eq:WSL_cond}, where one can see the quality of these fits is significantly worsened compared to the general global fit with no input constraint on $\cos(\beta-\alpha)$, yielding a $p$-value $\approx 0.1\%$.

\begin{figure}[ht]
    \centering
    \includegraphics[scale=0.6]{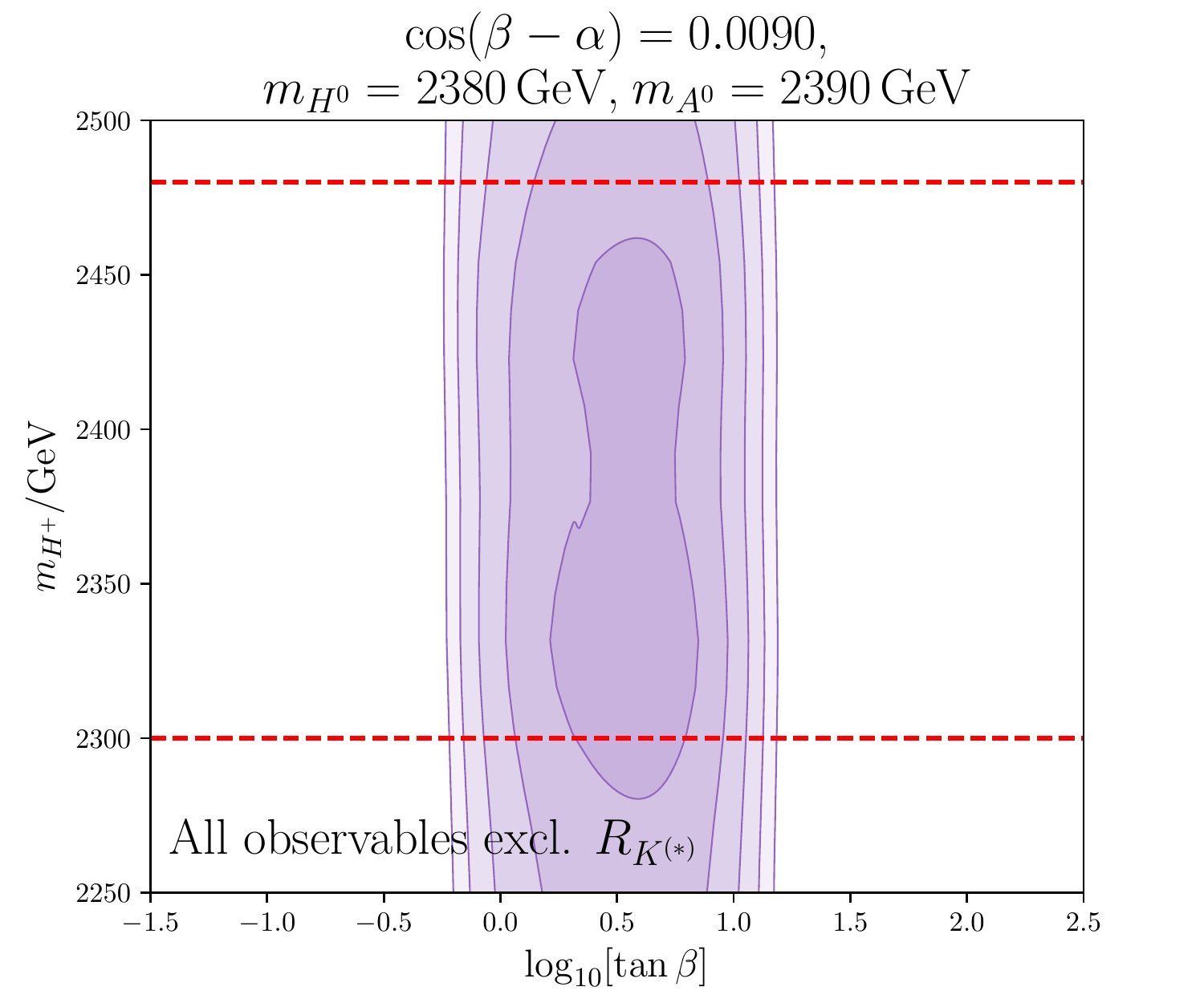}
    \caption{
    Contour plot of the allowed 2HDM-II parameter space in the $(\tan \beta - m_{H^+})$ plane, obtained by combining all observables
    excluding $R_{K^{(*)}}$, fixing the additional parameters to the best fit point from Table~\ref{tab:Comb_fit_res} and including the restrictions from the theory constraints in Table~\ref{tab:theory-constraints} (shown by the red dashed lines). The contours indicate the allowed parameter space at $1,2,3,4,5\sigma$ confidence level going from darker to lighter.}
    \label{fig:Global-fit-mHp-vs-tanbeta-and-th-constr}
\end{figure}

\begin{figure}[th]
    \centering
    \includegraphics[scale=0.6]{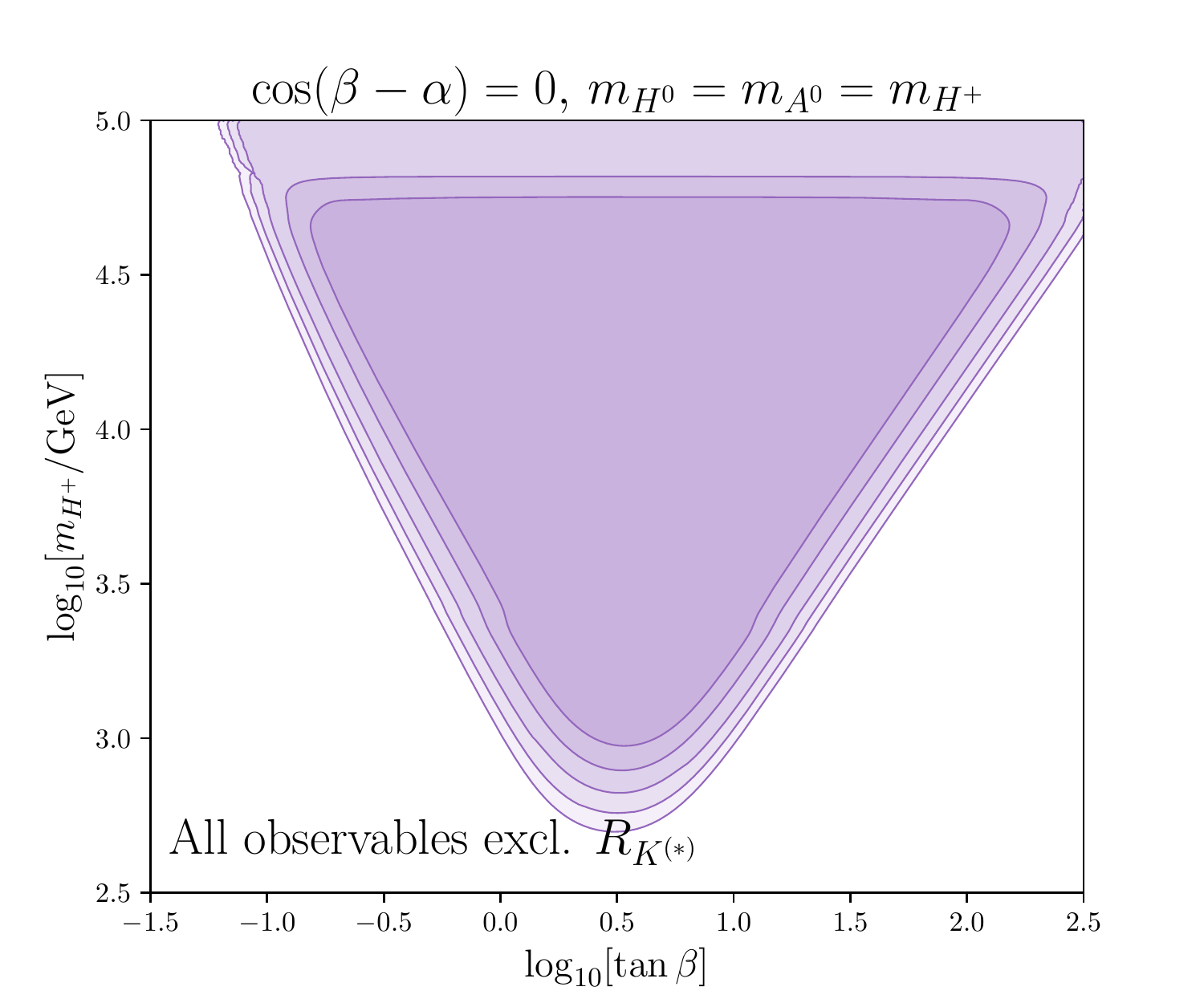}
    \caption{
    Contour plot of the allowed 2HDM-II parameter space in the $(\tan \beta - m_{H^+})$ plane, obtained by combining all observables
    excluding $R_{K^{(*)}}$, fixing the additional parameters as
    $\cos (\beta - \alpha) = 0$, $m_{H^+} = m_{H^0} = m_{A^0}$. 
    The contours indicate the allowed parameter space at $1,2,3,4,5\sigma$ confidence level going from darker to lighter.}
    \label{fig:Global-fit-mHp-vs-tanbeta-Sc-B}
\end{figure}

The best fit further favours values of $m_{H^+} \approx m_{H^0}  \approx  m_{A^0} \approx 2.3$~TeV and $\tan \beta \approx 4$.
Performing a scan around the best-fit point (excluding $R_{K^{(*)}}$) and keeping in mind the theory constraints in Table~\ref{tab:Comb_fit_res}, we find that the charged Higgs mass is constrained to be in the region
\begin{eqnarray}
76.3 \, {\rm TeV}  \geq m_{H^+} \geq 1.26 \, {\rm TeV} 
&& \mbox{at } 1\sigma,
\nonumber\\
m_{H^+} \geq  0.86 \, {\rm TeV} 
&& \mbox{at } 2\sigma,
\nonumber\\
m_{H^+} \geq  0.68 \, {\rm TeV} 
&& \mbox{at } 3\sigma,
\nonumber\\
m_{H^+} \geq  0.57\, {\rm TeV} 
&& \mbox{at } 4 \sigma,
\nonumber\\
m_{H^+} \geq  0.49\, {\rm TeV} 
&& \mbox{at } 5\sigma. \label{eq:H+-limits}
\end{eqnarray}
From the  $2 \sigma$-level onwards we do not see an upper bound within the area considered 
($m_{H^+} \le 100 \, {\rm TeV} $).
The parameter $\tan \beta$  has a strong correlation
with $m_{H^+}$. For the best fit values of  $\cos (\beta - \alpha)$, $m_{H^0}$ and $m_{A^0}$, we find the allowed regions in 
the ($\tan \beta - m_{H^+}$) plane given in 
Fig.~\ref{fig:Global-fit-mHp-vs-tanbeta}  
and Fig.~\ref{fig:Global-fit-mHp-vs-tanbeta-and-th-constr}  
(the latter zoomed into the best-fit point and shown in combination with the theory constraints). 
Assuming both the alignment limit, $\cos (\beta - \alpha) = 0$, 
and the degeneracy for the Higgs masses, $m_{H^+} = m_{H^0} = m_{A^0}$,
the contour plot of Fig.~\ref{fig:Global-fit-mHp-vs-tanbeta-Sc-B} shows that $\tan \beta$ is strongly constrained 
for low $m_{H^+} \sim 1 \, {\rm TeV}$ while the bounds on $\tan \beta$ become weaker with an 
increase of $m_{H^+}$ up to $ 100 \, {\rm TeV} $.
We stress again that these constraints above are obtained from 
the fit by {\it excluding} $R_{K^{(*)}}$. The  inclusion of $R_{K^{(*)}}$ would not significantly change the 
allowed parameter regions, while it would considerably worsen the quality of the fit.

Furthermore, we investigate several constrained scenarios in the fit, 
as indicated in Table~\ref{tab:Comb_fit_res}.  
First, we checked that the wrong-sign limit is absolutely
disfavored by the global fit. Second, in the exact alignment
limit the fit yields similar results as in the general case. 
Finally, assuming both the alignment limit, 
$\cos (\beta - \alpha) = 0$, 
and the degeneracy for the Higgs masses, 
$m_{H^+} = m_{H^0} = m_{A^0}$, again leads to similar results
due to the preference of both of these assumptions by the global fit 
and theory constraints.

\renewcommand{\arraystretch}{1.3}
\begin{table}[ht] %th
    \centering
    \begin{tabular}{|l|c|c|c|c|}
    \hline
    Scenario &  $\#$ & Best-fit point & 
    $\chi_{min}^2$ & $p$-value \\
    & Observables & $\{\tan\beta,m_{H^+},m_{H^0},m_{A^0},\cos(\beta-\alpha)\}$ & & \\
    \hline
    {All incl. $R_{K^{(*)}}$} 
    & {$275$}  
     & {$\{ 4.2, 2140\,{\rm GeV}, 2180\,{\rm GeV}, 2210\,{\rm GeV}, 0.0096 \}$} 
    & {$323$} 
    & {$\phantom{0}1.5\%$} \\
    {All excl. $R_{K^{(*)}}$} 
    & {$265$} 
     & {{$\{ 4.3, 2340\,{\rm GeV}, 2380\,{\rm GeV}, 2390\,{\rm GeV}, 0.0090 \}$}} 
    & {$295$} 
    & {$\phantom{0}6.6\%$} \\
    {Flavour incl. $R_{K^{(*)}}$}
    & {$241$} 
     & {{$\{ 4.2, 2310\,{\rm GeV}, 2300\,{\rm GeV}, 2290\,{\rm GeV}, 0.001 \}$}} 
    & {$297$} 
    & {$\phantom{0}0.4$\%} \\
    {Flavour excl. $R_{K^{(*)}}$} 
    & {$231$} 
    & {{$\{ 4.3, 2420\,{\rm GeV}, 2390\,{\rm GeV}, 2360\,{\rm GeV}, 0.0001 \}$}}
    & {$269$} 
    & {$\phantom{0}2.5$\%} \\
    {$b\to s\ell\ell$ incl. $R_{K^{(*)}}$} 
    & {$202$}
    & $\{ 4.0 , 820\,{\rm GeV}, 690\,{\rm GeV}, 690\,{\rm GeV}, 0.0003 \}$
    & $265$
    & $\phantom{0}0.1$\% \\
    {$b\to s\ell\ell$ excl. $R_{K^{(*)}}$} 
    & {$192$}
    & $\{ 4.0, 820\,{\rm GeV}, 690\,{\rm GeV}, 690\,{\rm GeV}, 0.0003 \}$
    & $238$
    & $\phantom{0}0.7$\% \\
    {Only $R_{K^{(*)}}$} 
    & {$10$} 
     & {{$\{ 2370, 0.14\,{\rm GeV}, 250\,{\rm GeV}, 9.2\,{\rm GeV}, 0.001 \}$}} 
    & {$8.5$} 
    & {$13.1$\%} \\
    {Higgs Signals} 
    & {$31$}
     & {{$\{\sim100 \pm 20, -, -, -, 0.0003 \}$}}
    & {$24$}
    & {$72.9$\%} \\
    \hline\hline
    
    \multicolumn{5}{|c|}{Wrong Sign Limit, $\cos(\beta-\alpha)=\sin2\beta$} \\
    \hline
    {All incl. $R_{K^{(*)}}$}
    & {$275$} 
    & {{$\{ 12.1, 2320\,{\rm GeV}, 2270\,{\rm GeV}, 2260\,{\rm GeV}\}$}}
    & {$380$} 
    & {$0.001\%$} \\
    {All excl. $R_{K^{(*)}}$}
    & {$265$} 
    & {{$\{ 12.2, 2320\,{\rm GeV}, 2270\,{\rm GeV}, 2260\,{\rm GeV}\}$}}
    & {$340$} 
    & {$0.1\%$} \\
    \hline\hline
    
    \multicolumn{5}{|c|}{Alignment Limit, $\cos(\beta-\alpha)=0$} \\
    \hline
    {All incl. $R_{K^{(*)}}$}
    & {$275$} 
    & {{$\{ 4.1, 2290\,{\rm GeV}, 2330\,{\rm GeV}, 2340\,{\rm GeV} \}$}}
    & {$326$} 
    & {$1.3$\%} \\
    {All excl. $R_{K^{(*)}}$}
    & {$265$} 
     & {{$\{ 4.2, 2260\,{\rm GeV}, 2300\,{\rm GeV}, 2310\,{\rm GeV} \}$}}
    & {$297$} 
    & {$6.4$\%} \\
    \hline
    \hline
    \multicolumn{5}{|c|}{$\cos(\beta-\alpha)=0$, $m_{H^+} = m_{H^0} = m_{A^0}$} \\
    \hline
    {All incl. $R_{K^{(*)}}$}
    & {$275$}
    & {{$\{ 4.0 , 2070\, {\rm GeV}  \}$}}
    & {$328$} 
    & {$1.3$\%} \\
    {All excl. $R_{K^{(*)}}$}
    & {$265$}
    & {{$\{ 4.2 , 2320\, {\rm GeV} \}$}}
    & {$300$} 
    & {$5.7$\%} \\
    \hline
    \end{tabular}
    \setlength{\belowcaptionskip}{-18pt}
    \caption{Best fit points of 2HDM-II parameter fits for various groups of observables using the constraints from theory to inform the physical parameter values.}
    \label{tab:Comb_fit_res}
\end{table}

\begin{figure}[ht]
    \centering
    \includegraphics[scale=0.6]{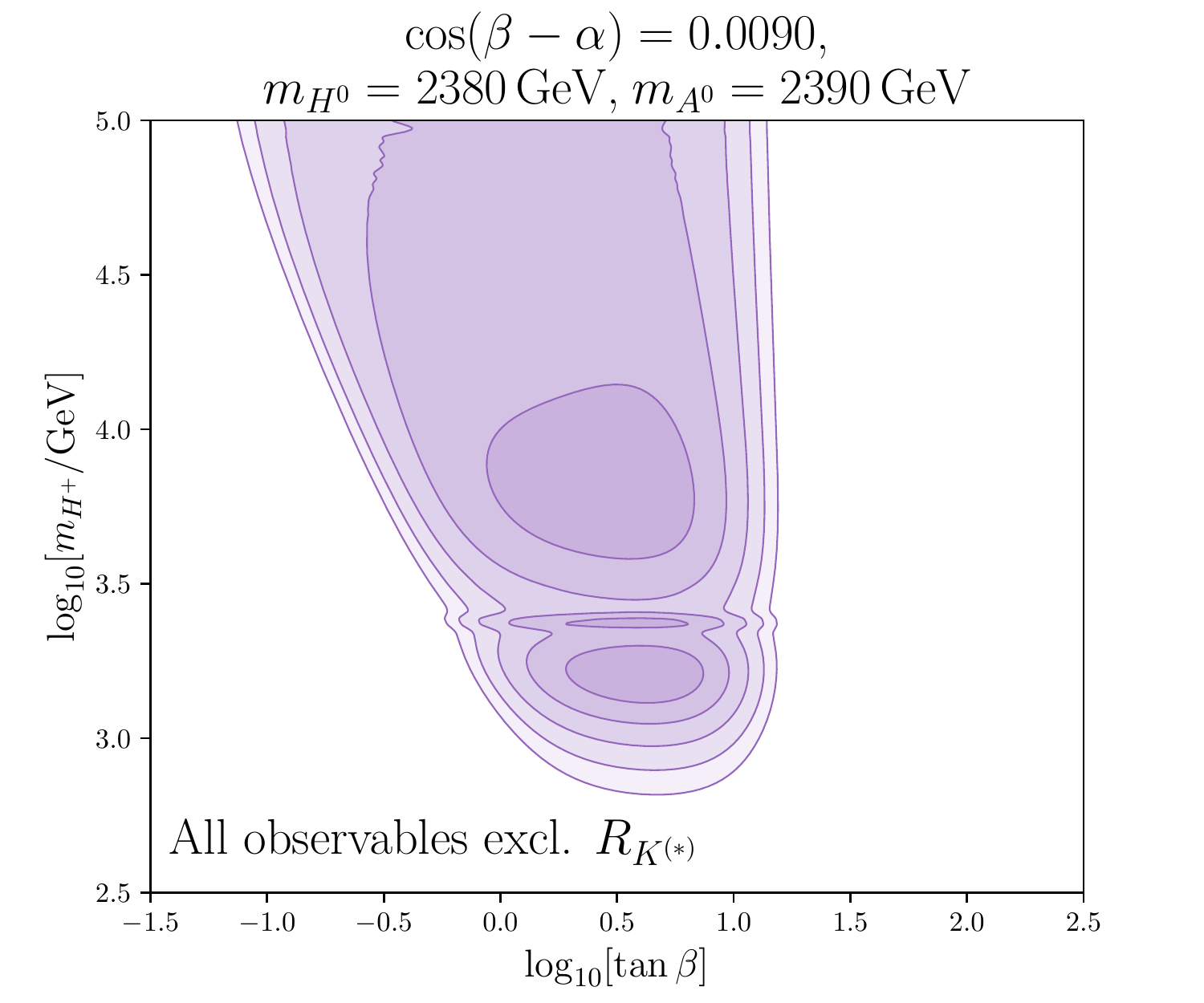}
    \setlength{\belowcaptionskip}{-15pt}
    \caption{
    Contour plot of the allowed 2HDM-II parameter space in the $(\tan \beta - m_{H^+})$ plane, obtained by combining all observables
    excluding $R_{K^{(*)}}$, fixing the additional parameters to the best fit point from Table~\ref{tab:Comb_fit_res}. The contours indicate the allowed parameter space at $1,2,3,4,5\sigma$ confidence level going from darker to lighter.}
    \label{fig:Global-fit-mHp-vs-tanbeta}
\end{figure}

\subsection{Comment on the Anomalous Magnetic Moment of the Muon}
\label{sec:muon}
\begin{figure}[ht]
    \centering
    \includegraphics[scale=0.4]{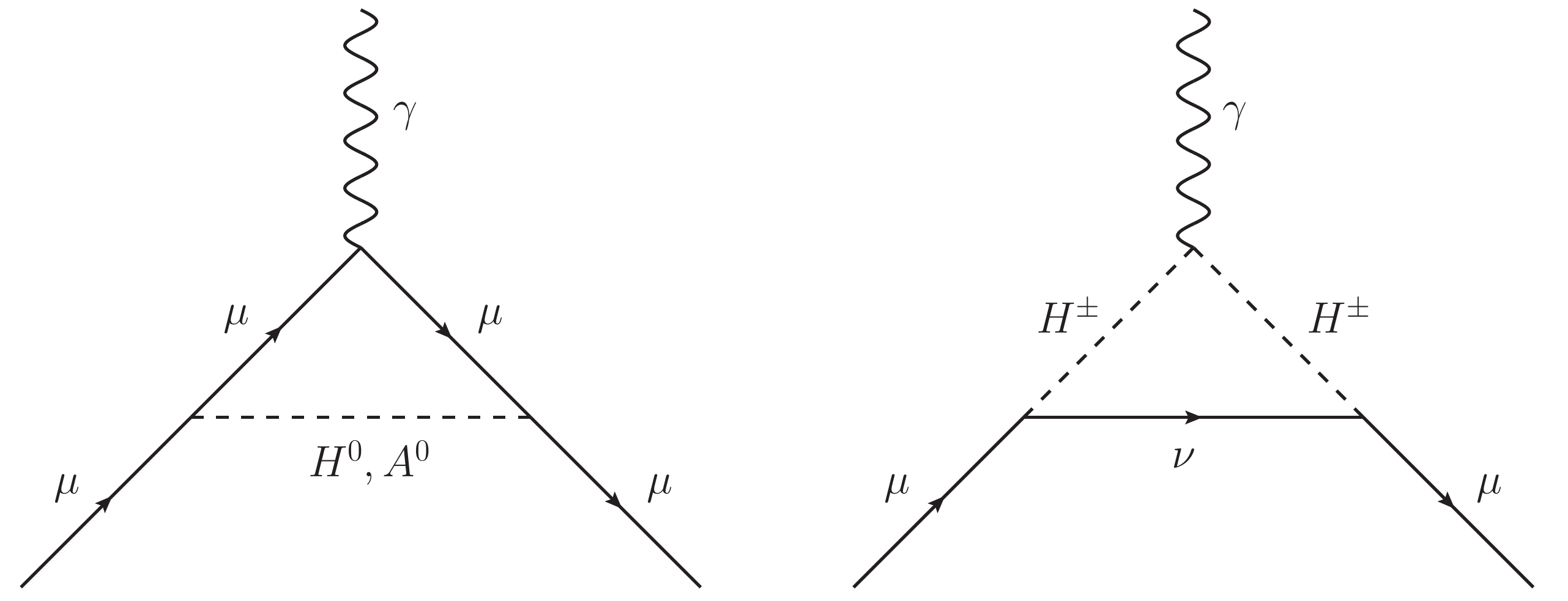} \\[5mm]
    \includegraphics[scale=0.4]{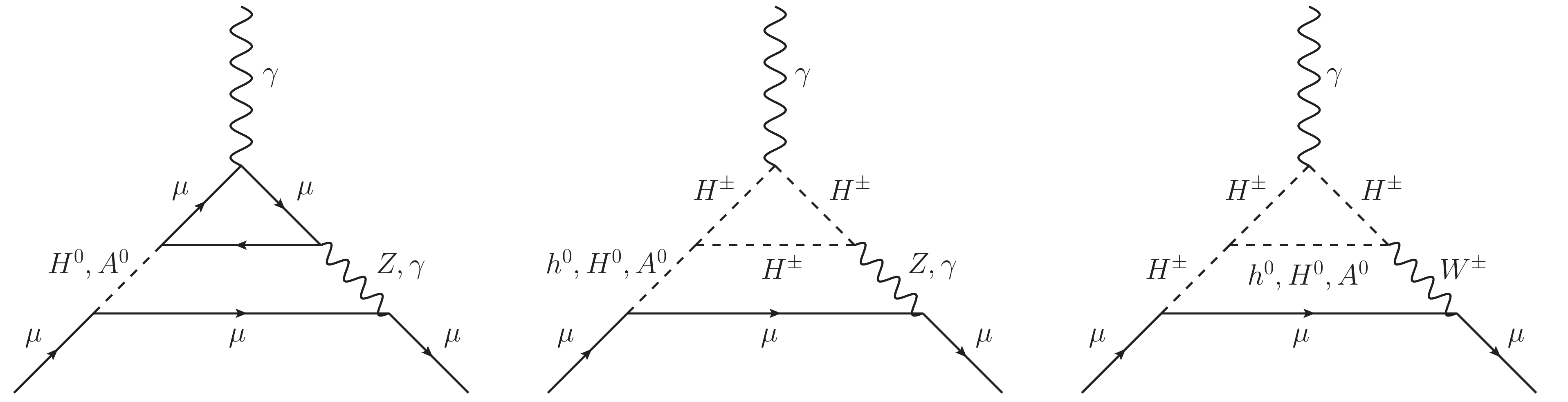}
    \caption{Examples of one- and two-loop 2HDM contributions to $\mu\to\mu\gamma$.}
    \label{fig:amu_feyn} 
\end{figure}
The anomalous magnetic moment of the muon,
\begin{equation}
    a_\mu = \frac{g_\mu-2}{2},
\end{equation}
measures the deviation of the muon gyromagnetic moment $g_\mu$ from 2 due to quantum loop effects. 
This can be accurately measured in experiment, with recent measurements at Fermilab \cite{Abi:2021gix} confirming the older BNL value \cite{Bennett:2006fi}.
It can also be accurately predicted in the Standard Model and the Theory Initiative (WP \cite{Aoyama:2020ynm}) yields a $4.2\sigma$ discrepancy with the updated combined experimental value. We will also consider a recent Lattice QCD evaluation (BMW Collaboration \cite{Borsanyi:2020mff}) that finds the $g_\mu - 2$ discrepancy to be only $1.6\sigma$.

This deviation in $a_\mu$ can be a strong motivation for potential new physics signals, such as supersymmetric particle loops or other BSM contributions, see e.g. Ref.~\cite{Athron:2021iuf}.
2HDM contributions to  $a_\mu$ form a subset of SUSY contributions to the anomalous
magnetic moment and they have been recently studied e.g. in Refs.~\cite{Botella:2020xzf,Jana:2020pxx,Ghosh:2020tfq}.

Matching \textbf{flavio}'s WET-3 basis \cite{Aebischer:2017ugx}, we write the effective Hamiltonian for $\ell\to\ell\gamma$ processes as
\begin{equation}
    \mathcal{H}_{\rm eff}^{\ell\to\ell\gamma} = -\frac{4 G_F}{\sqrt{2}}\frac{e}{16\pi^2} m_\mu C_7 (\bar{\ell}\sigma_{\rho\omega} P_R\ell) F^{\rho\omega} + {\rm h.c.}
\end{equation}
The 2HDM contribution to $a_\mu$ is then defined as
\begin{equation}
    a_\mu^{\rm 2HDM} = \frac{G_Fm_\mu^2}{\sqrt{2} \, \pi^2}C_7,
\end{equation}
where we take the contributions at one- and two-loop level 
given in \cite{Ilisie:2015tra}.
The two-loop contributions come from the Barr-Zee diagrams (\cite{Barr:1990vd}, see Fig. \ref{fig:amu_feyn}) with a fermion or boson
loop which can give rise to significant contributions to $a_\mu^{\rm 2HDM}$. All Higgs bosons are present in these loops, therefore $a_\mu^{\rm 2HDM}$ depends not only on $\tan\beta$ and $m_{H^+}$, but also on $m_{A^0}$ and $m_{H^0}$, with a further dependence on the mixing angle $\alpha$. 
\begin{figure}[ht]
    \centering
    \includegraphics[width=0.49\textwidth]{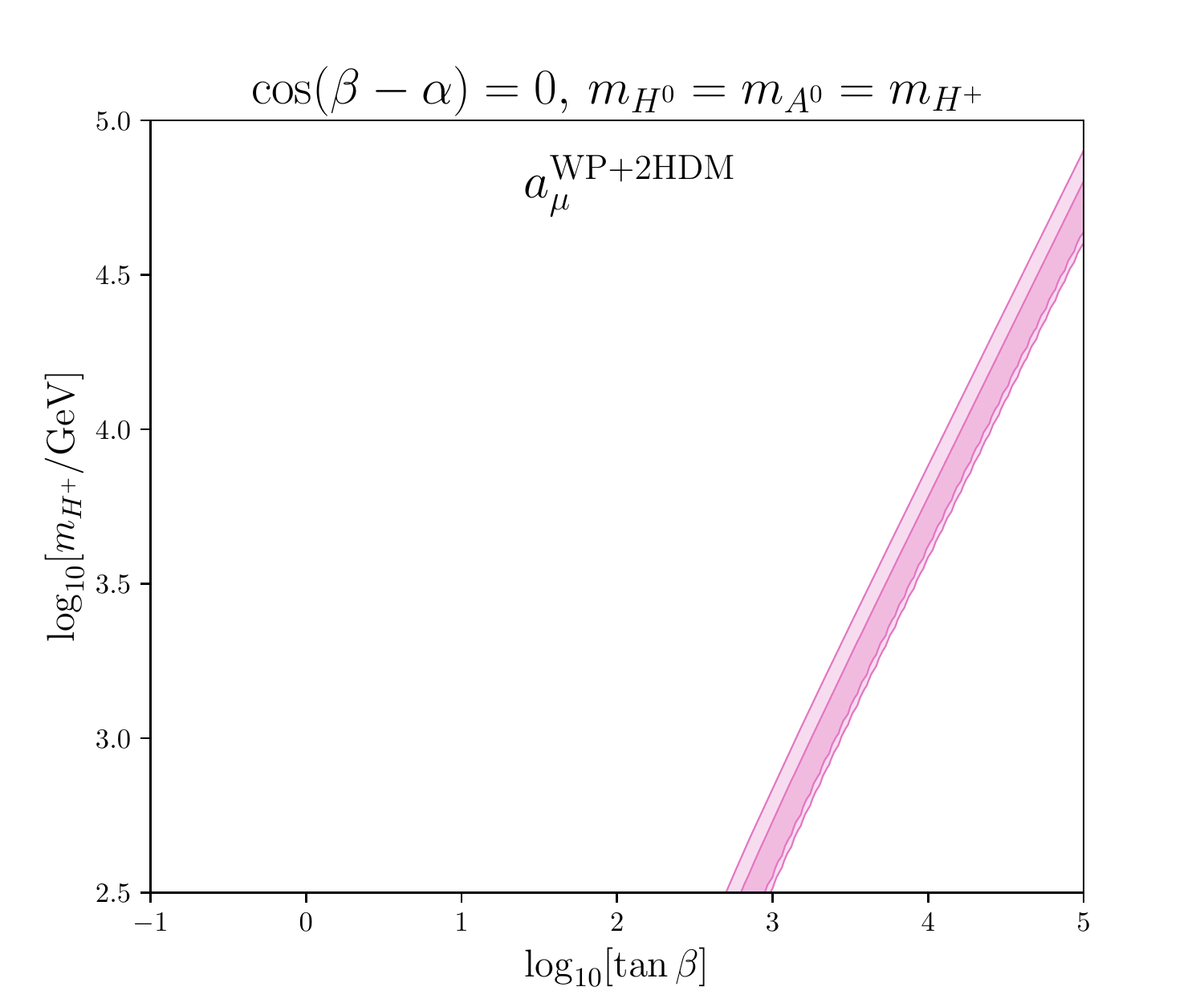}
    \includegraphics[width=0.49\textwidth]{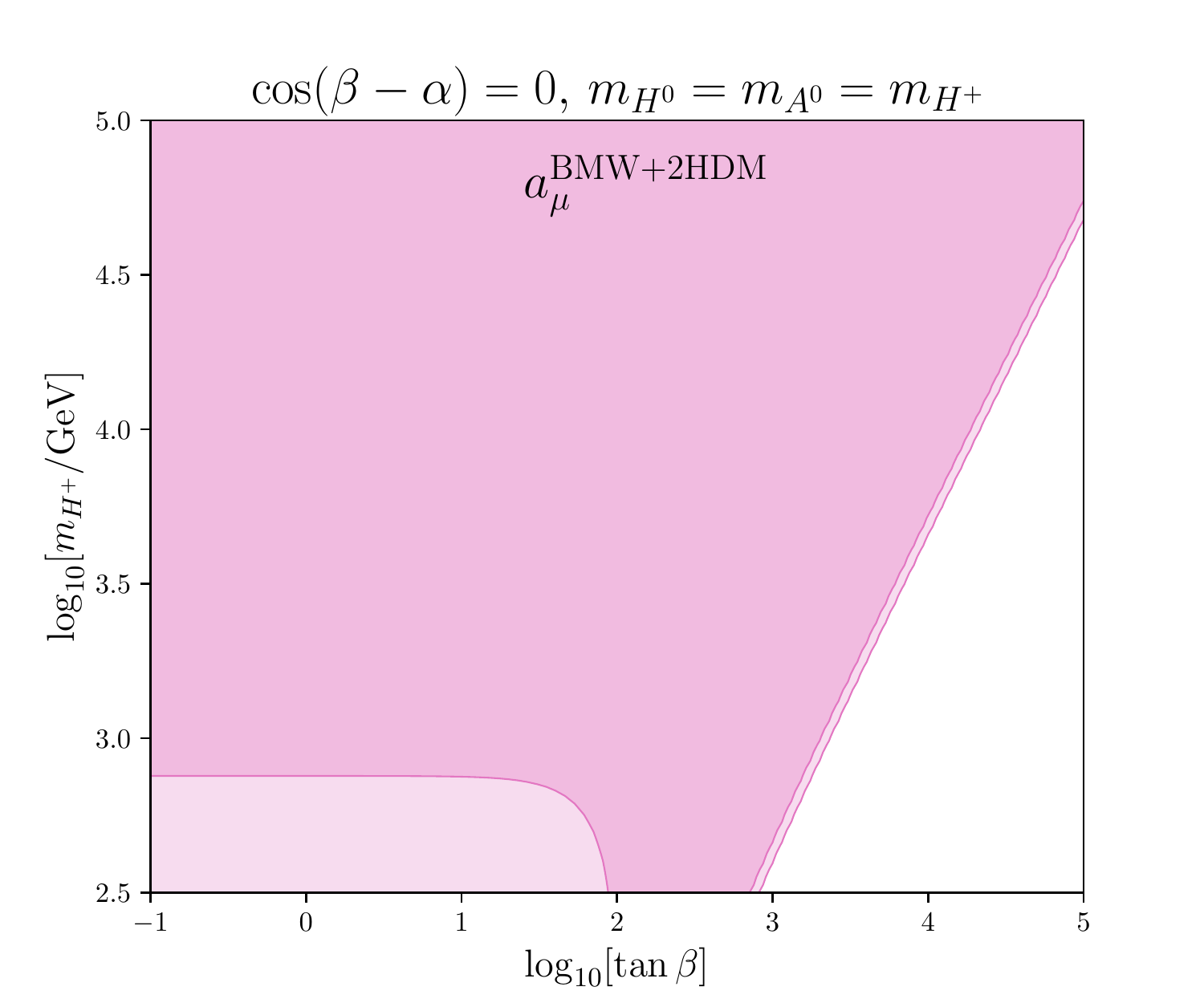}
    \setlength{\belowcaptionskip}{-12pt}
    \caption{Contour plots of the allowed 2HDM-II parameter space in the $(\tan \beta - m_{H^+})$ plane for~$a_\mu$, using the Standard Model prediction 
    by WP20 \cite{Aoyama:2020ynm} (BMW \cite{Borsanyi:2020mff}) (left (right) plot), 
    fixing the additional parameters as
    $\cos(\beta-\alpha)=0, \, m_{H^0} = m_{A^0} = m_{H^+}$. 
    The lighter contour indicates the allowed parameter space at $2\sigma$ confidence level while the darker contour corresponds to $1\sigma$.}
    \label{fig:g-2}
\end{figure}
\noindent
From Fig.~\ref{fig:g-2} we find that the 2HDM-II cannot explain the discrepancy between the experimental number for the anomalous magnetic moment of the muon and the white paper prediction as the required values of $\tan \beta$ lie in the 
non-perturbative region. Obviously the BMW prediction yields a consistency of experiment and theory within 2$\sigma$ and thus a large portion
of the 2HDM-II parameter space is still allowed.

\subsection{Electroweak Phase Transition}
\label{sec:Phase}
The electroweak phase transition (EWPT) is a mechanism which can generate the baryon asymmetry we observe in the Universe today through the process of EW baryogenesis, with the requirement that the EWPT is of strong first order (SFO). 
In the SM this could be achieved for
$m_{h^0}\lesssim70\,$~GeV~\cite{Kajantie:1996mn}, however the measurement of the SM
Higgs boson mass as $m_{h^0} = 125.1 \pm 0.14 \,$~GeV \cite{Zyla:2020zbs} means we
now require BSM physics to achieve a SFOEWPT; a 2HDM is in principle capable of
generating this.
For recent work testing the 2HDM for a SFOEWPT across large regions in its parameter
space, see e.g. Refs.~\cite{Basler:2016obg,Su:2020pjw,Dorsch:2016nrg}, in which the
authors find regions where the 2HDM-II could support a SFOEWPT, however only for
Higgs masses less than $1\,$TeV. 
In \cite{Wang:2021ayg},  the strength of the EWPT in the 2HDM-II is tested for Higgs
masses $\gtrsim 600\,$GeV; a FOEWPT can be found for these masses but it is not
strong enough to support EW baryogenesis.
Furthermore, a SFOEWPT appears to favour a mass splitting ($\sim200\,$GeV \cite{Su:2020pjw}) between $H^+$ and $H^0$, and $m_{A^0}\sim m_{H^+}$
\cite{Su:2020pjw,Dorsch:2016nrg}. 
This could still be allowed by our results in Section~\ref{sec:theory}, however only
for low charged Higgs masses, in tension with the global fit of Fig.~\ref{fig:Global-fit-mHp-vs-tanbeta}. 

In this work, we use the \textbf{BSMPT} package \cite{Basler:2018cwe,Basler:2020nrq}
to calculate the strength of the EWPT for selected parameter points across our fit.
The \textbf{BSMPT} package calculates the strength of the EWPT for the 2HDM in the
lambda parameter basis ($\lambda_{i}$, $m_{12}^2$, $\tan\beta$).
We have so far primarily worked in the mass basis
($m_{A^0},m_{H^0},m_{h^0},m_{H^+},\tan\beta,\cos(\beta-\alpha)$), and so must convert
to the lambda basis.
In the alignment limit, a sufficient but not necessary condition for vacuum stability is \cite{Kling:2016opi}
\begin{equation}
    \label{eq:m12}
     m_{12}^2 = m_{H^0}^2\sin\beta\cos\beta \,\cap\,m_{h^0}^2+m_{H^\pm}^2-m_{H^0}^2>0\,\cap\, m_{h^0}^2+m_{A^0}^2-m_{H^0}^2>0.
\end{equation}
We then use this, along with Eqs.~\eqref{eq:lambda1}-\eqref{eq:lambda5}, to perform the conversion, while making the assumption that $\cos(\beta-\alpha)$ is small, which is of course in line with our constraints. 
This is only appropriate when the conditions of Eq.~\eqref{eq:m12} are fulfilled, i.e. primarily for $m_{H^0} \lesssim m_{A^0}, m_{H^+}$. 

The \textbf{BSMPT} package parameterises the strength of the EWPT in the 2HDM by calculating
\begin{equation}
    \xi_c = \frac{\omega_c}{T_c},
\end{equation}
where $\omega_c$ is the high-temperature VEV ($\omega_c^2=\omega_1^2(T_c) +\omega_2^2(T_c)$ and $\omega_i (T=0) = v_i$) 
at the critical temperature $T_c$.
The \textbf{BSMPT} package only passes a result if $\xi_c>0$, with the desired SFOEWPT indicated by $\xi_c>1$.
\begin{table}[th] %th
\centering
    \begin{tabular}{|c|ccc|cccc||c|c|c|}
        \hline\hline
        \multirow{2}{*}{$\tan\beta$} & \multicolumn{3}{c|}{Mass Basis (GeV)} & \multicolumn{3}{c}{Lambda Basis} & $m_{12}^2$ & $\omega_c$ & $T_c$ & \multirow{2}{*}{$\xi_c$} \\
        & $m_{H^+}$ & $m_{H^0}$ & $m_{A^0}$ & $\lambda_3$ & $\lambda_4$ & $\lambda_5$ & (GeV$^2$) & (GeV) & (GeV) & \\
        \hline\hline 
        $4.2$ & $50000$ & $50000$ & $50000$ 
            & $0.26$ & $0$ & $0$ & $5.6\times10^{8}$ 
            & $0.66$ & $162$ & $0.004$ \\
            \hline
        $4.2$ & $2320$ & $2320$ & $2320$ 
            & $0.26$ & $0$ & $0$ & $1.2\times10^{6}$ 
            & $23$ & $162$ & $0.14$  \\
        $4.2$ & $2320$ & $2250$ & $2280$ 
            & $10.8$ & $-8.3$ & $-2.2$ & $1.1\times10^{6}$ 
            & $31$ & $186$ & $0.17$  \\
        $1.0$ & $2320$ & $2320$ & $2320$ 
            & $0.26$ & $0$ & $0$ & $2.7\times10^{6}$ 
            & $23$ & $162$ & $0.14$ \\
        $10$ & $2320$ & $2320$ & $2320$ 
            & $0.26$ & $0$ & $0$ & $5.3\times10^{5}$ 
            & $24$ & $162$ & $0.15$ \\
            \hline
        $4.2$ & $860$ & $710$ & $860$ 
            & $8.03$ & $-3.9$ & $-3.9$ & $1.1\times10^{5}$ 
            & $142$ & $175$ & $0.81$  \\
        $4.2$ & $860$ & $690$ & $860$ 
            & $8.95$ & $-4.3$ & $-4.3$ & $1.1\times10^{5}$ 
            & $177$ & $176$ & $1.00$ \\
        $4.2$ & $680$ & $470$ & $680$ 
            & $8.22$ & $-4.0$ & $-4.0$ & $5.0\times10^{4}$ 
            & $211$ & $149$ & $1.42$  \\
        $4.2$ & $570$ & $320$ & $570$ 
            & $7.60$ & $-3.7$ & $-3.7$ & $2.3\times10^{4}$ 
            & $226$ & $126$ & $1.79$  \\
        $4.2$ & $490$ & $250$ & $490$ 
            & $6.12$ & $-2.9$ & $-2.9$ & $1.4\times10^{4}$ 
            & $207$ & $126$ & $1.64$  \\
        $4.2$ & $490$ & $490$ & $490$ 
            & $0.26$ & $0$ & $0$ & $5.4\times10^{4}$ 
            & $24$ & $161$ & $0.15$  \\
        \hline\hline
    \end{tabular}
    \setlength{\belowcaptionskip}{-8pt}
    \caption{
    Table of results for the EWPT in the 2HDM.
    We consider the exact alignment limit, $\cos(\beta-\alpha)=0$, 
    and in addition we employ the condition by Eq.~\eqref{eq:m12}.
    In this limit, as follows from Eqs.~\eqref{eq:lambda1}-\eqref{eq:lambda5},
    one gets: $\lambda_1 = \lambda_2 = m_{h^0}^2/v^2 = 0.26$, i.e. 
    $\lambda_{1,2}$ are fixed and independent of the Higgs masses and $\tan\beta$ (and therefore not again shown in the table). 
    In addition, one obtains that $\lambda_{3,4,5}$ are independent of $\tan \beta$, and $\lambda_3 + \lambda_4 + \lambda_5 = m_{h^0}^2/v^2 = 0.26$. 
    { For similar discussions of the lambda basis, see e.g. \cite{BhupalDev:2014bir}.}}
    \label{tab:bsmpt} 
\end{table}

We choose $M=50\,$TeV as a benchmark for more extreme masses, the best-fit point of our degenerate mass fit, and lower limits in $m_{H^+}$ at $\sim 2,3,4,5\sigma$ where we select more favourable mass splittings as points to test for a SFOEWPT.
We also show some variations in $\tan\beta$ at the best-fitting degenerate masses.
In Table \ref{tab:bsmpt}, we present our selected parameter points in both bases, and then also the results produced by the \textbf{BSMPT} package. 
As is generally favoured by our fits, for higher masses we mostly test parameter points in the limit of degenerate masses, since this will satisfy the conditions of Eq.~\eqref{eq:m12} and the small mass splittings allowed at higher mass scales have only small effects, as is shown in one benchmark point. 
However for lower masses where these splittings are more significant, we consider further benchmark points allowing for mass splitting more favourable for a SFOEWPT \cite{Su:2020pjw}, and find that demanding degeneracy for these lower masses is a poorer choice.

We find that for most of our allowed parameter space, a SFOEWPT is not possible due to the large mass scales involved. 
From the benchmark points at higher masses, we find a maximum of $\xi_c=0.17$; much smaller than required. 
For our benchmark points at lower masses however, the outlook improves, where we first find a $\xi_c\geq1$ for a parameter point at $\sim2\sigma$ from the global best fit point. 
On the other hand, we note that all points in Table \ref{tab:bsmpt} 
with a ratio $\xi_c \geq 1$ require some of the couplings $|\lambda_i |> 4$, which could be in conflict with perturbativity, depending on what constraint we apply; 
see the beginning of Section~\ref{sec:theory_lambda}.
As we have only chosen a few benchmark points, we cannot rule out the possibility of a $\xi_c\geq1$ at some point closer to our best fit, however this seems unlikely given the trends shown here and in other works. 
Therefore to accommodate a SFOEWPT in the 2HDM-II, we would require much lower Higgs masses than are favoured by the global fit. But for such regions of the parameter space the 2HDM-II will no longer perform better in the overall fit compared to the SM.

\section{Conclusion}
\label{sec:Conclusion}
In this work we have investigated theoretical bounds on the 2HDM-II stemming from perturbativity,
unitarity and vacuum stability
as well as experimental constraints stemming from electroweak precision, Higgs signal strengths, flavour observables and the anomalous magnetic moment of the muon.
All in all, we find that the 2HDM-II can accommodate the data better than the SM.
The best fit point for the 2HDM-II lies around
\begin{equation}
\begin{aligned}
m_{H^+} \approx m_{H^0}  &\approx  m_{A^0}   \approx 2 \, \mbox{TeV} \, ,
\\
\tan \beta \approx 4, \quad & \quad
\cos (\beta - \alpha )  \approx 0.01  \, .
\end{aligned}
\end{equation}
For the charged Higgs mass we find a lower limit of 680 GeV at $3\sigma$ and the remaining Higgses have to be largely degenerate -- this requirement becomes stronger the heavier the Higgs masses become. 
For low values of the charged Higgs mass, $\tan \beta$ is severely constrained to lie around a value of 4, while for high values
of the charged Higgs mass (around 50 TeV) $\tan \beta$ should be in the region between 0.1 and 300.
We further find that the wrong-sign limit is disfavoured by Higgs signal strengths and excluded by the global fit by more than $5\sigma$ and that only small deviations from the alignment limit are possible.

Experiments currently observe some deviations from SM predictions, in particular in $R(D^{(*)})$,  $R_{K^{(*)}}$ and the anomalous  magnetic moment of the muon.
The tensions in  $R(D^{(*)})$ will either be enhanced in the 2HDM-II 
or  to be resolved, they would require parameter regions in the $(\tan \beta - m_{H^+})$ plane which are excluded by other flavour observables.
Eliminating the tension in $R_{K^{(*)}}$ would require charged Higgs
masses below 1 GeV and non-perturbative values of $\tan \beta$.
The 2HDM-II can also not resolve the discrepancy between the measurement
of the magnetic moment of the muon and the white paper prediction.

Finally we found that the allowed parameter space of the 2HDM-II strongly disfavours the existence of a  
sufficiently strong first order phase transition for EW baryogenesis. This requires mass splittings discouraged by theoretical constraints, and low masses of the charged Higgs -- at least $2\sigma$ away from our best fit point, which makes this model theoretically much less attractive.

As a next step we plan to study more types of 2HDMs and also investigate implications of our findings for future collider searches.

\acknowledgments{The work of M.B.  is supported by
Deutsche Forschungsgemeinschaft (DFG, German Research Foundation) through  TRR 257 “Particle Physics Phenomenology after the Higgs Discovery”.
O.A. is supported by a UK Science and Technology Facilities Council (STFC) studentship under grant ST/V506692/1. 
We would like to thank Otto Eberhardt, Christoph Englert, Martin Jung, Rusa Mandal,
Margarete Mühlleitner, Ulrich Nierste, Danny van Dyk for helpful discussions, 
Dipankar Das for very useful comments on the arXiv version of the paper
and Matthew Kirk, Peter Stangl and David Straub for guidance with {\bf flavio}. 
The fits were run on the Siegen OMNI cluster. 
}

\newpage
\appendix

\section{Inputs and Observables}
\label{sec:inputs}
\renewcommand{\arraystretch}{1.25}
    { 
    \begin{longtable}{|C{3cm}|c|C{2cm}|c|}
     \hline Observable & Experimental value & Source & SM prediction  \\
        \hline \hline
        \multicolumn{4}{|c|}{\bf Higgs signal strengths}\\
        \hline 
        \multirow{3}{*}{$\mu^{\gamma\gamma}_{\text{ggF}}$} 
        & $1.03 \pm 0.11$ & \cite{ATLAS:2020qdt} & \multirow{3}{*}{1} \\
        & $1.09^{+0.15}_{-0.14}$ & \cite{CMS:2020gsy} & \\ 
        & $1.07^{+0.12}_{-0.11}$ & \cite{Sirunyan:2021ybb} & \\ 
        \hline
        \multirow{2}{*}{$\mu_{\text{ggF}}^{ZZ}$} 
        & $0.94^{+0.11}_{-0.10}$ & \cite{ATLAS:2020qdt} & \multirow{2}{*}{1} \\ 
        & $0.98^{+0.12}_{-0.11}$ & \cite{CMS:2020gsy} &  \\
        \hline
        \multirow{3}{*}{$\mu^{WW}_{\text{ggF}}$} 
        & $1.08^{+0.19}_{-0.18}$ & \cite{ATLAS:2020qdt} & \multirow{3}{*}{1} \\
        & $1.28^{+0.20}_{-0.19}$ & \cite{CMS:2020gsy} &  \\
        & $1.20^{+0.16}_{-0.15}$ & \cite{ATLAS:2021upe} &  \\
        \hline
        $\mu^{Z\gamma}_{\text{ggF}}$ 
        & $2.0^{+1.0}_{-0.9}$ & \cite{Aad:2020plj} & 1 \\[1mm]
        \hline
        \multirow{2}{*}{$\mu^{\tau\tau}_{\text{ggF}}$} 
        & $1.02^{+0.60}_{-0.55}$ & \cite{ATLAS:2020qdt} & \multirow{2}{*}{1} \\
        & $0.39^{+0.38}_{-0.39}$ & \cite{CMS:2020gsy} & \\
        \hline
        \multirow{3}{*}{$\mu^{\mu\mu}_{\text{ggF}}$} 
        & $1.2 \pm 0.6$ & \cite{Aad:2020xfq} & \multirow{3}{*}{1}\\
        & $0.31^{+1.82}_{-1.81}$ & \cite{CMS:2020gsy} &  \\
        & $1.19^{+0.42}_{-0.41}$ & \cite{Sirunyan:2020two} & \\
        \hline
         \multirow{2}{*}{$\mu^{bb}_{\text{ggF}}$} 
        & $2.51^{+2.43}_{-2.01}$ & \cite{ATLAS:2020qdt} & \multirow{2}{*}{1}\\
        & $2.45^{+2.53}_{-2.35}$ & \cite{CMS:2020gsy} & \\
        \hline
        \multirow{3}{*}{$\mu^{\gamma\gamma}_{\text{VBF}}$} 
        & $1.31^{+0.26}_{-0.23}$ & \cite{ATLAS:2020qdt} & \multirow{3}{*}{1} \\
        & $0.77^{+0.37}_{-0.29}$ & \cite{CMS:2020gsy} & \\
        & $1.04^{+0.34}_{-0.31}$ & \cite{Sirunyan:2021ybb} & \\
        \hline
        \multirow{2}{*}{$\mu^{ZZ}_{\text{VBF}}$} 
        & $1.25^{+0.50}_{-0.41}$ & \cite{ATLAS:2020qdt} & \multirow{2}{*}{1} \\
        & $0.57^{+0.46}_{-0.36}$ & \cite{CMS:2020gsy} & \\
        \hline
        \multirow{3}{*}{$\mu^{WW}_{\text{VBF}}$} 
        & $0.60^{+0.36}_{-0.34}$ & \cite{ATLAS:2020qdt} & \multirow{3}{*}{1} \\
        & $0.63^{+0.65}_{-0.61}$ & \cite{CMS:2020gsy} & \\
        & $0.99^{+0.24}_{-0.20}$ & \cite{ATLAS:2021upe} & \\
        \hline
        \multirow{2}{*}{$\mu^{\tau\tau}_{\text{VBF}}$}  
        & $1.15^{+0.57}_{-0.53}$ & \cite{ATLAS:2020qdt} & \multirow{2}{*}{1} \\
        & $1.05^{+0.30}_{-0.29}$ & \cite{CMS:2020gsy} & \\
        \hline
        $\mu^{\mu\mu}_{\text{VBF}}$  
        & $3.18^{+8.22}_{-7.93}$ & \cite{CMS:2020gsy} & 1 \\
        \hline
        \multirow{2}{*}{$\mu^{bb}_{\text{VBF}}$}  
        & $3.03^{+1.67}_{-1.62}$ & \cite{ATLAS:2020qdt} & \multirow{2}{*}{1} \\ 
        & $0.95^{+0.38}_{-0.36}$ & \cite{ATLAS:2020bhl} &
        \\
        \hline
        $\mu^{\gamma\gamma}_{\text{Wh}}$ 
        & $3.76^{+1.48}_{-1.35}$ & \cite{Sirunyan:2018koj} & 1 \\
        \hline
        $\mu^{WW}_{\text{Wh}}$ 
        & $2.85^{+2.11}_{-1.87}$ & \cite{CMS:2020gsy} & 1 \\
        \hline
        $\mu^{\tau\tau}_{\text{Wh}}$ 
        & $3.01^{+1.65}_{-1.51}$ & \cite{CMS:2020gsy} & 1 \\
        \hline
        \multirow{2}{*}{$\mu^{bb}_{\text{Wh}}$} 
        & $1.27^{+0.42}_{-0.40}$ & \cite{CMS:2020gsy} & \multirow{2}{*}{1} \\
        & $0.95^{+0.27}_{-0.25}$ & \cite{Aad:2020jym} & \\
        \hline
        $\mu^{\gamma\gamma}_{\text{Zh}}$ 
        & $0.00 \pm 1.14$ & \cite{Sirunyan:2018koj} & 1 \\
        \hline
        $\mu^{WW}_{\text{Zh}}$ 
        & $0.90^{+1.77}_{-1.43}$ & \cite{CMS:2020gsy} & 1 \\
        \hline
        $\mu^{\tau\tau}_{\text{Zh}}$ 
        & $1.53^{+1.60}_{-1.37}$ & \cite{CMS:2020gsy} & 1 \\
        \hline
        \multirow{2}{*}{$\mu^{bb}_{\text{Zh}}$} 
        & $0.93^{+0.33}_{-0.31}$ & \cite{CMS:2020gsy} & \multirow{2}{*}{1} \\
        & $1.08^{+0.25}_{-0.23}$ & \cite{Aad:2020jym} & \\
        \hline
        $\mu^{cc}_{\text{Zh}}$ 
        & $37^{+20}_{-19}$ & \cite{Sirunyan:2019qia} & 1 \\
        \hline
        \multirow{2}{*}{$\mu^{\gamma\gamma}_{\text{Vh}}$} 
        & $1.32^{+0.33}_{-0.30}$ & \cite{ATLAS:2020qdt} & \multirow{2}{*}{1} \\
        & $1.34^{+0.36}_{-0.33}$ & \cite{Sirunyan:2021ybb} & \\
        \hline
        \multirow{2}{*}{$\mu^{ZZ}_{\text{Vh}}$} 
        & $1.53^{+1.13}_{-0.92}$ & \cite{ATLAS:2020qdt} & \multirow{2}{*}{1} \\
        & $1.10^{+0.96}_{-0.74}$ & \cite{CMS:2020gsy} & \\
        \hline
        \multirow{2}{*}{$\mu^{bb}_{\text{Vh}}$ }
        & $1.02^{+0.18}_{-0.17}$ & \cite{ATLAS:2020qdt} & \multirow{2}{*}{1} \\
        & $1.03^{+0.28}_{-0.17}$ & \cite{Aad:2020jym} & \\
        \hline
        \multirow{3}{*}{$\mu^{\gamma\gamma}_{\text{tth}}$} 
        & $0.90^{+0.41}_{-0.35}$ & \cite{ATLAS:2020qdt} & \multirow{3}{*}{1} \\
        & $1.62^{+0.52}_{-0.43}$ & \cite{CMS:2020gsy} & \\
        & $1.35^{+0.34}_{-0.28}$ & \cite{Sirunyan:2021ybb} & \\
        \hline
        $\mu^{ZZ}_{\text{tth}}$ 
        & $0.25^{+1.03}_{-0.25}$ & \cite{CMS:2020gsy} & 1 \\
        \hline
        $\mu^{WW}_{\text{tth}}$
        & $0.93^{+0.48}_{-0.45}$ & \cite{CMS:2020gsy} & 1 \\
        \hline
        $\mu^{VV}_{\text{tth}}$ 
        & $1.72^{+0.56}_{-0.53}$ & \cite{ATLAS:2020qdt} & 1 \\
        \hline
        \multirow{2}{*}{$\mu^{\tau\tau}_{\text{tth}}$}
        & $1.20^{+1.07}_{-0.93}$ & \cite{ATLAS:2020qdt} & \multirow{2}{*}{1} \\
        & $0.81^{+0.74}_{-0.67}$ & \cite{CMS:2020gsy} & \\
        \hline
        \multirow{2}{*}{$\mu^{bb}_{\text{tth}}$} 
        & $0.79^{+0.60}_{-0.59}$ & \cite{ATLAS:2020qdt} &\multirow{2}{*}{1} \\
        & $1.13^{+0.33}_{-0.30}$ & \cite{CMS:2020gsy} & \\
        \hline
    \caption{Table of the Higgs signal strengths used in the analysis in 
    Section \ref{sec:signals}.}
    \label{tab:List-of-Higgs-signal-strengths}
    \end{longtable}
    }

\newpage
\renewcommand{\arraystretch}{1.25}
    { 
    \begin{longtable}{|C{3cm}|c|C{2cm}|c|}
        \hline 
        Observable & Experimental value & Source & SM prediction \\
        \hline \hline
        \multicolumn{4}{|c|}{\bf Oblique parameters}\\
        \hline
        S 
        & $0.02 \pm 0.10$ & \cite{Zyla:2020zbs} & 0 \\ 
        T 
        & $0.07 \pm 0.12$ & \cite{Zyla:2020zbs} & 0 \\ 
        U 
        & $0.00 \pm 0.09$ & \cite{Zyla:2020zbs} & 0 \\
        \hline

    \caption{List of the Oblique parameters used in the analysis in Section~\ref{sec:STU}.}
    \label{tab:List-of-oblique-parameters}
    \end{longtable}
    }

\setlength\LTleft{-18pt}
\renewcommand{\arraystretch}{1.25}
    { 
    \begin{longtable}{|l|c|C{1.5cm}|c|}
        \hline 
        Observable & Experimental value & Source & SM prediction  \\
        \hline \hline
        \multicolumn{4}{|c|}{\bf Tree-level leptonic decays} \\
        \hline
        $\Gamma(\pi^+\to\mu^+\nu_\mu)$ 
        & $(2.528\pm0.005)\times10^{-17}\,$eV 
        & \cite{Zyla:2020zbs}
        & $(2.521\pm0.003)\times10^{-17}\,$eV \\ 
        ${\cal B}(K^+\to\mu^+\nu_\mu)$ 
        & $(63.56 \pm 0.11)\times10^{-2}$ 
        & \cite{Zyla:2020zbs}
        & $(63.91\pm0.93)\times10^{-2}$ \\ 
        ${\cal B} (B^+\to\mu^+\nu_\mu)$ 
        & $(5.3\pm2.2)\times10^{-7}$ 
        & \cite{Zyla:2020zbs}
        & $(4.15\pm0.51)\times10^{-7}$ \\ 
        ${\cal B} (B^+\to\tau^+\nu_\tau)$ 
        & $(1.09 \pm 0.24)\times10^{-4}$
        & \cite{Zyla:2020zbs}
        & $(9.24\pm1.13)\times10^{-5}$ \\ 
        ${\cal B} (D^+\to\mu^+\nu_\mu)$ 
        & $(3.74 \pm 0.17)\times 10^{-4}$
        & \cite{Zyla:2020zbs}
        & $(3.97\pm0.04)\times10^{-4}$ \\ 
        ${\cal B} (D^+ \to \tau^+ \nu_\tau)$ 
        & $(1.20 \pm 0.27)\times 10^{-3}$
        & \cite{Zyla:2020zbs}
        & $(1.07\pm0.01) \times10^{-3}$ \\ 
        ${\cal B} (D_s^+\to\mu^+\nu_\mu)$ 
        & $(5.49 \pm 0.16)\times10^{-3}$
        & \cite{Zyla:2020zbs}
        & $(5.31\pm0.04)\times10^{-3}$ \\ 
        ${\cal B} (D_s^+\to\tau^+\nu_\tau)$ 
        & $(5.48\pm0.23)\times10^{-2}$ 
        & \cite{Zyla:2020zbs}
        & $(5.22\pm0.04)\times10^{-2}$ \\
        ${\cal B}(\tau^+\to K^+\nu_\tau)$ 
        & $(6.96\pm0.10)\times10^{-3}$ 
        & \cite{Zyla:2020zbs}
        & $(7.07\pm0.10)\times10^{-3}$ \\ 
        ${\cal B}(\tau^+\to\pi^+\nu_\tau)$ 
        & $(10.82\pm0.05)\times10^{-2}$ 
        & \cite{Zyla:2020zbs}
        & $(10.85\pm0.13)\times10^{-2}$ \\[2mm]
        \hline \hline
        \multicolumn{4}{|c|}{\bf Tree-level semi-leptonic decays}\\
        \hline
        ${\cal B} (B^0 \to D^- \ell^+ \nu_\ell)$
        & $(2.31 \pm 0.10) \times 10^{-2}$
        & \cite{Amhis:2019ckw}
        & $(2.26 \pm 0.18) \times 10^{-2}$
        \\
        ${\cal B} (B^+ \to D^0 \ell^+ \nu_\ell)$
        & $(2.35 \pm 0.10) \times 10^{-2}$
        & \cite{Amhis:2019ckw} 
        & $(2.45 \pm 0.19) \times 10^{-2}$
        \\
        ${\cal B} (B^0 \to D^{*-} \ell^+ \nu_\ell)$
        & $(5.06 \pm 0.12) \times 10^{-2}$
        & \cite{Amhis:2019ckw} 
        & $(4.82 \pm 0.44) \times 10^{-2}$
        \\
        ${\cal B} (B^+ \to D^{*0} \ell^+ \nu_\ell)$
        & $(5.66 \pm 0.22) \times 10^{-2}$
        & \cite{Amhis:2019ckw}  
        & $(5.20 \pm 0.48) \times 10^{-2}$
        \\
        ${\cal B} (B^+ \to \pi^0 \ell^+ \nu_\ell)$
        & $(7.80 \pm 0.27) \times 10^{-5}$ 
        & \cite{Zyla:2020zbs}
        & $(6.37 \pm 1.03) \times 10^{-5}$ 
        \\
        ${\cal B} (B^0 \to \pi^- \ell^+ \nu_\ell)$
        & $(1.50 \pm 0.06) \times 10^{-4}$
        & \cite{Zyla:2020zbs}
        & $(1.18 \pm 0.19) \times 10^{-4}$
        \\
        ${\cal B} (B^+ \to \rho^0 \ell^+ \nu_\ell)$
        & $(1.58 \pm 0.11) \times 10^{-4}$
        & \cite{Zyla:2020zbs}
        & $(2.19 \pm 0.44) \times 10^{-4}$
        \\
        ${\cal B} (B^0 \to \rho^- \ell^+ \nu_\ell)$
        & $(2.94 \pm 0.21) \times 10^{-4}$
        & \cite{Zyla:2020zbs}
        & $(4.06 \pm 0.82) \times 10^{-4}$
        \\
        ${\cal B} (B^0_s \to D^-_s \mu^+ \nu_\mu)$
        & $(2.40 \pm 0.12 \pm 0.15 \pm 0.06 \pm 0.10) \times 10^{-2}$
        & \cite{Aaij:2020hsi,Aaij:2021nyr}
        & $(2.36 \pm 0.18) \times 10^{-2} $
        {\red *}  \cite{McLean:2019qcx}
        \\
        ${\cal B} (B^0_s \to D^{*-}_s \mu^+ \nu_\mu)$
        & $(5.19 \pm 0.24 \pm 0.47 \pm 0.13 \pm 0.14) \times 10^{-2}$ 
        & \cite{Aaij:2020hsi,Aaij:2021nyr}
        & $(4.99 \pm 0.45) \times 10^{-2} $
        {\red *}  \cite{McLean:2019qcx}
        \\
        ${\cal B} (D^+ \to \bar K^0 e^+ \nu_e)$
        & $(8.73 \pm 0.10) \times 10^{-2}$
        & \cite{Zyla:2020zbs}
        & $(9.08 \pm 0.61) \times 10^{-2}$
        \\
        ${\cal B} (D^+ \to \bar K^0 \mu^+ \nu_\mu)$
        & $(8.76 \pm 0.19) \times 10^{-2}$
        & \cite{Zyla:2020zbs}
        & $(9.04 \pm 0.61) \times 10^{-2}$
        \\
        ${\cal B} (D^+ \to \pi^0 e^+ \nu_e)$
        & $(3.72 \pm 0.17) \times 10^{-3}$
        & \cite{Zyla:2020zbs}
        & $(3.47 \pm 0.20) \times 10^{-3}$
        \\
        ${\cal B} (D^+ \to \pi^0 \mu^+ \nu_\mu)$
        & $(3.50 \pm 0.15) \times 10^{-3}$
        & \cite{Zyla:2020zbs}
        & $(3.46 \pm 0.19) \times 10^{-3}$
        \\
        ${\cal B} (D^0 \to K^- e^+ \nu_e)$
        & $(3.542 \pm 0.035) \times 10^{-2}$
        & \cite{Zyla:2020zbs}
        & $(3.55 \pm 0.24) \times 10^{-2}$
        \\
        ${\cal B} (D^0 \to K^- \mu^+ \nu_\mu)$
        & $(3.41 \pm 0.04) \times 10^{-2}$
        & \cite{Zyla:2020zbs}
        & $(3.54 \pm 0.24) \times 10^{-2}$
        \\
        ${\cal B} (D^0 \to \pi^- e^+ \nu_e)$
        & $(2.91 \pm 0.04) \times 10^{-3}$
        & \cite{Zyla:2020zbs}
        & $(2.68 \pm 0.15) \times 10^{-3}$
        \\
        ${\cal B} (D^0 \to \pi^- \mu^+ \nu_\mu)$
        & $(2.67 \pm 0.12) \times 10^{-3}$
        & \cite{Zyla:2020zbs}
        & $(2.67 \pm 0.15) \times 10^{-3}$
        \\
        ${\cal B} (K^+ \to \pi^0 e^+ \nu_e)$
        & $(5.07 \pm 0.04) \times 10^{-2}$
        & \cite{Zyla:2020zbs}
        & $(5.12 \pm 0.06) \times 10^{-2}$
        \\
        ${\cal B} (K^+ \to \pi^0 \mu^+ \nu_\mu)$
        & $(3.352 \pm 0.033) \times 10^{-2}$
        & \cite{Zyla:2020zbs}
        & $(3.380 \pm 0.043) \times 10^{-2}$
        \\
        ${\cal B} (K^0_L \to \pi^+ e^- \bar \nu_e)$
        & $(40.55 \pm 0.11) \times 10^{-2}$
        & \cite{Zyla:2020zbs}
        & $(40.82 \pm 0.43) \times 10^{-2}$
        \\
        ${\cal B} (K^0_L \to \pi^+ \mu^- \bar \nu_\mu)$
        & $(27.04 \pm 0.07) \times 10^{-2}$
        & \cite{Zyla:2020zbs}
        & $(27.04 \pm 0.30) \times 10^{-2}$
        \\
        ${\cal B} (K^0_S \to \pi^+ e^- \bar \nu_e)$
        & $(7.04 \pm 0.08) \times 10^{-4}$
        & \cite{Zyla:2020zbs}
        & $(7.15 \pm 0.07) \times 10^{-4}$
        \\
        ${\cal B} (K^0_S \to \pi^+ \mu^- \bar \nu_\mu)$
        & $(4.69 \pm 0.05) \times 10^{-4}$
        & \cite{Zyla:2020zbs}
        & $(4.73 \pm 0.05) \times 10^{-4}$
        \\
        \hline \hline
        \multicolumn{4}{|c|}{\bf Tree-level LFU Ratios}\\
        \hline
        $\mathcal{R}(D)$ 
        & $(0.340 \pm 0.030)$ 
        & \cite{Amhis:2019ckw} 
        & $(0.303 \pm 0.006)$  \\ 
        $\mathcal{R}(D^*)$ 
        & $(0.295 \pm 0.014)$ 
        & \cite{Amhis:2019ckw}
        & $(0.255 \pm 0.003)$  \\
        \hline \hline 
        \multicolumn{4}{|c|}{\bf Radiative decays}\\
        \hline
        ${\cal B} (\bar{B} \to X_s \gamma)$ 
        & $(3.32 \pm 0.15) \times10^{-4}$
        & \cite{Amhis:2019ckw}
        & $(3.40 \pm 0.17) \times10^{-4}$ ${\red *}$ \cite{Misiak:2020vlo} \\
        \hline \hline
        \multicolumn{4}{|c|}{\bf FCNC leptonic decays}\\
        \hline 
        ${\cal B} (B_d\to\mu^+\mu^-)$ 
        & $(0.56 \pm 0.70)\times10^{-10}$ 
        & \cite{Altmannshofer:2021qrr}
        & $(1.08\pm0.13)\times10^{-10}$  \\ 
        ${\cal B} (B_s \to \mu^+ \mu^-)$ 
        & $(2.93 \pm 0.35)\times10^{-9}$ 
        & \cite{Altmannshofer:2021qrr}
        & $(3.48\pm0.26)\times10^{-9}$  \\
        \hline \hline 
        \multicolumn{4}{|c|}{\bf \boldmath $B$-Mixing}\\
        \hline \hline
        $\Delta m_d$ 
        & $(0.5065\pm0.0019)\,$ps$^{-1}$ 
        & \cite{Amhis:2019ckw}
        & $(0.533^{+0.022}_{-0.036})\,$ps$^{-1}$ {\red *}  \cite{DiLuzio:2019jyq}\\ 
        $\Delta m_s$ 
        & $(17.757\pm0.021)\,$ps$^{-1}$ 
        & \cite{Amhis:2019ckw}
        & $(18.4^{+0.7}_{-1.2})\,$ps$^{-1}$ {\red *} \cite{DiLuzio:2019jyq}\\ 
        \hline \hline
        \multicolumn{4}{|c|}{\bf Anomalous magnetic moments}\\
        \hline
        $a_\mu$ 
        & $116592091(54)(33)\times10^{-11}$ 
        & \cite{Zyla:2020zbs}
        & $116591810(43)\times10^{-11}$ {\red *} \cite{Aoyama:2020ynm}\\
        & & 
        & $116591954(55)\times10^{-11}$ {\red *} \cite{Borsanyi:2020mff}\\
        \hline
    \caption{Table of flavour observables 
    (except $b\to s \ell^+ \ell^-$) used in the analysis. 
    The stated SM predictions are our calculations through \textbf{flavio}, where we have modified some appropriately in order to closely replicate the most recent averages or predictions, for example in $B$-mixing and $B\to X_s\gamma$ (marked by ${\red *}$ in the fourth column).}
    \label{tab:List-of-flavour-observables}
    \end{longtable}}

\newpage
\setlength\LTleft{-40pt}
\setlength\LTright{0pt}
\begin{longtable}{|l|c|c|c|c|c|}
\hline
Mode & Observable & Source & Bin & Experimental value & SM prediction \\
\hline
$B^+ \to K^+ \mu^+ \mu^-$ 
& $10^{\, 8} \, {\rm GeV}^2 \times \frac{\Delta \cal B}{\Delta q^2} $ 
& LHCb \cite{Aaij:2014pli} 
& $[0.1, 0.98]$ 
& $3.32 \pm 0.18 \pm 0.17 $  
& $3.34 \pm 0.63 $ 
\\
& & 
& $[1.1, 2]$
& $2.33 \pm 0.15 \pm 0.12 $
& $3.33 \pm 0.63 $ 
\\
& & 
& $[2, 3]$
& $2.82 \pm 0.16 \pm 0.14 $ 
& $3.32 \pm 0.62 $
\\
& & 
& $[3, 4]$
& $2.54 \pm 0.15 \pm 0.13 $ 
& $3.30 \pm 0.62 $ 
\\
& & 
& $[4, 5]$
& $2.21 \pm 0.14 \pm 0.11 $ 
& $3.28 \pm 0.62 $ 
\\
& & 
& $[5, 6]$
& $2.31 \pm 0.14 \pm 0.12 $ 
& $3.26 \pm 0.61 $
\\
& & 
& $[15, 16]$
& $1.61 \pm 0.10 \pm 0.08 $ 
& $2.25 \pm 0.33 $
\\
& & 
& $[16, 17]$
& $1.64 \pm 0.10 \pm 0.08 $ 
& $2.03 \pm 0.28 $
\\
& & 
& $[17, 18]$
& $2.06 \pm 0.11 \pm 0.10 $ 
& $1.77 \pm 0.23 $ 
\\
& & 
& $[18, 19]$
& $1.37 \pm 0.10 \pm 0.07 $ 
& $1.49 \pm 0.19 $
\\
& & 
& $[19, 20]$
& $0.74 \pm 0.08 \pm 0.04 $ 
& $1.17 \pm 0.14 $ 
\\
& & 
& $[20, 21]$
& $0.59 \pm 0.07 \pm 0.03 $ 
& $0.82 \pm 0.10 $
\\
& & 
& $[21, 22]$
& $0.43 \pm 0.07 \pm 0.02 $ 
& $0.44 \pm 0.05 $ 
\\
\hline
  $B^0 \to K^0 \mu^+ \mu^-$ 
& $10^{\, 8} \, {\rm GeV}^2 \times \frac{\Delta \cal B}{\Delta q^2} $ 
& LHCb \cite{Aaij:2014pli} 
& $[0.1, 2]$ 
& $1.22^{+0.59}_{-0.52} \pm 0.06 $ 
& $3.10 \pm 0.58 $ 
\\
& &
& $[2, 4]$ 
& $1.87^{+0.55}_{-0.49} \pm 0.09 $ 
& $3.07 \pm 0.58 $
\\
& &
& $[4, 6]$ 
& $1.73^{+0.53}_{-0.48} \pm 0.09 $ 
& $3.03 \pm 0.57 $
\\
& &
& $[15, 17]$ 
& $1.43^{+0.35}_{-0.32} \pm 0.07 $ 
& $1.98 \pm 0.28 $ 
\\
& &
& $[17, 19]$ 
& $0.78^{+0.17}_{-0.15} \pm 0.04 $ 
& $1.50 \pm 0.19 $
\\
\hline
  $ B^+ \to K^{*+} \mu^+ \mu^- $ 
& $10^{\, 8} \, {\rm GeV}^2 \times \frac{\Delta \cal B}{\Delta q^2}$ 
& LHCb \cite{Aaij:2014pli} 
& $[0.1,2]$
& $5.92^{+1.44}_{-1.30} \pm 0.40 $ 
& $7.53 \pm 1.11 $ \\
& & 
& $[2,4]$
& $5.59^{+1.59}_{-1.44} \pm 0.38 $ 
& $4.60 \pm 0.63 $ \\
& & 
& $[4,6]$
& $2.49^{+1.10}_{-0.96} \pm 0.17 $ 
& $5.13 \pm 0.74 $ \\
& & 
& $[15,17]$
& $6.44^{+1.29}_{-1.15} \pm 0.44 $ 
& $7.19 \pm 0.89 $ \\
& &
& $[17,19]$
& $1.16^{+0.91}_{-0.76} \pm 0.08 $ 
& $4.96 \pm 0.68 $ \\
\hline 
$B^+ \to K^{*+} \mu^+ \mu^-$ 
& $F_L$ 
& LHCb \cite{Aaij:2020ruw} 
& $[0.1, 0.98]$ 
& $0.34^{+0.10}_{-0.10} \pm 0.06 $ 
& $0.303 \pm 0.052$    \\
& &
& $[1.1, 2.5]$ 
& $0.54^{+0.18}_{-0.19} \pm 0.03 $ 
& $0.767 \pm 0.043$    \\
& &
& $[2.5, 4]$ 
& $0.17^{+0.24}_{-0.14} \pm 0.04 $ 
& $0.798 \pm 0.035$    \\
& &
& $[4, 6]$ 
& $0.67^{+0.11}_{-0.14} \pm 0.03 $ 
& $0.712 \pm 0.048$    \\
& &
& $[15, 17]$ 
& $0.41^{+0.18}_{-0.14} \pm 0.02 $ 
& $0.346 \pm 0.022$    \\
& &
& $[17, 19]$ 
& $0.34^{+0.11}_{-0.12} \pm 0.04 $ 
& $0.326 \pm 0.016$    \\
\hline
$B^+ \to K^{*+} \mu^+ \mu^-$ 
& $P_1$ 
& LHCb \cite{Aaij:2020ruw} 
& $[0.1, 0.98]$ 
& $\phantom{-}0.44^{+0.38}_{-0.40} \pm 0.11 $ 
& $\phantom{-}0.045 \pm 0.027$    \\
& &
& $[1.1, 2.5]$ 
& $\phantom{-}1.60^{+4.92}_{-1.75} \pm 0.32 $ 
& $\phantom{-}0.022 \pm 0.047$    \\
& &
& $[2.5, 4]$ 
& $-0.29^{+1.43}_{-1.04} \pm 0.22 $ 
& $-0.116 \pm 0.036$    \\
& &
& $[4, 6]$ 
& $-1.24^{+0.99}_{-1.17} \pm 0.29 $ 
& $-0.176 \pm 0.046$    \\
& &
& $[15, 17]$ 
& $-0.88^{+0.41}_{-0.67} \pm 0.07 $ 
& $-0.531 \pm 0.050$    \\
& &
& $[17, 19]$ 
& $-0.40^{+0.58}_{-0.57} \pm 0.09 $ 
& $-0.746 \pm 0.036$    \\
\hline
$B^+ \to K^{*+} \mu^+ \mu^-$ 
& $P_2$ 
& LHCb \cite{Aaij:2020ruw} 
& $[0.1, 0.98]$ 
& $-0.05^{+0.12}_{-0.12} \pm 0.03 $ 
& $-0.136 \pm 0.008$    \\
& &
& $[1.1, 2.5]$ 
& $-0.28^{+0.24}_{-0.42} \pm 0.15 $ 
& $-0.451 \pm 0.017$    \\
& &
& $[2.5, 4]$ 
& $\phantom{-}0.03^{+0.26}_{-0.25} \pm 0.11 $ 
& $-0.054 \pm 0.114$    \\
& &
& $[4, 6]$ 
& $-0.15^{+0.19}_{-0.20} \pm 0.06 $ 
& $\phantom{-}0.294 \pm 0.081$    \\
& &
& $[15, 17]$ 
& $\phantom{-}0.45^{+0.03}_{-0.07} \pm 0.03 $ 
& $\phantom{-}0.414 \pm 0.016$    \\
& &
& $[17, 19]$ 
& $\phantom{-}0.14^{+0.10}_{-0.10} \pm 0.04 $ 
& $\phantom{-}0.320 \pm 0.020$    \\
\hline
$B^+ \to K^{*+} \mu^+ \mu^-$ 
& $P_3$ 
& LHCb \cite{Aaij:2020ruw} 
& $[0.1, 0.98]$ 
& $-0.42^{+0.20}_{-0.21} \pm 0.05 $ 
& $\phantom{-}0.0015 \pm 0.0134$    \\
& &
& $[1.1, 2.5]$ 
& $-0.09^{+0.70}_{-0.99} \pm 0.18 $ 
& $\phantom{-}0.0040 \pm 0.0242$    \\
& &
& $[2.5, 4]$ 
& $-0.45^{+0.50}_{-0.62} \pm 0.20 $ 
& $\phantom{-}0.0039 \pm 0.0103$    \\
& &
& $[4, 6]$ 
& $\phantom{-}0.52^{+0.82}_{-0.62} \pm 0.15 $ 
& $\phantom{-}0.0025 \pm 0.0154$    \\
& &
& $[15, 17]$ 
& $-0.23^{+0.16}_{-0.20} \pm 0.02 $ 
& $-0.0004 \pm 0.0003$    \\
& &
& $[17, 19]$ 
& $\phantom{-}0.12^{+0.21}_{-0.21} \pm 0.02 $ 
& $-0.0003 \pm 0.0002$    \\
\hline
$B^+ \to K^{*+} \mu^+ \mu^-$ 
& $P_4^\prime$ 
& LHCb \cite{Aaij:2020ruw} 
& $[0.1, 0.98]$ 
& $-0.09^{+0.36}_{-0.35} \pm 0.12 $ 
& $\phantom{-}0.234 \pm 0.012$    \\
& &
& $[1.1, 2.5]$ 
& $\phantom{-}0.58^{+0.62}_{-0.56} \pm 0.11 $ 
& $-0.066 \pm 0.047$    \\
& &
& $[2.5, 4]$ 
& $-0.81^{+1.09}_{-0.84} \pm 0.14 $ 
& $-0.392 \pm 0.048$    \\
& &
& $[4, 6]$ 
& $-0.79^{+0.47}_{-0.28} \pm 0.09 $ 
& $-0.502 \pm 0.028$    \\
& &
& $[15, 17]$ 
& $-0.32^{+0.23}_{-0.22} \pm 0.08 $ 
& $-0.618 \pm 0.010$    \\
& &
& $[17, 19]$ 
& $-0.57^{+0.29}_{-0.36} \pm 0.13 $ 
& $-0.660 \pm 0.007$    \\
\hline
$B^+ \to K^{*+} \mu^+ \mu^-$ 
& $P_5^\prime$ 
& LHCb \cite{Aaij:2020ruw} 
& $[0.1, 0.98]$ 
& $\phantom{-}0.51^{+0.30}_{-0.28} \pm 0.12 $ 
& $\phantom{-}0.684 \pm 0.024$    \\
& &
& $[1.1, 2.5]$ 
& $\phantom{-}0.88^{+0.70}_{-0.71} \pm 0.10 $ 
& $\phantom{-}0.112 \pm 0.083$    \\
& &
& $[2.5, 4]$ 
& $-0.87^{+1.00}_{-1.68} \pm 0.09 $ 
& $-0.517 \pm 0.107$    \\
& &
& $[4, 6]$ 
& $-0.25^{+0.32}_{-0.40} \pm 0.09 $ 
& $-0.763 \pm 0.077$    \\
& &
& $[15, 17]$ 
& $-0.14^{+0.21}_{-0.20} \pm 0.06 $ 
& $-0.673 \pm 0.036$    \\
& &
& $[17, 19]$ 
& $-0.66^{+0.36}_{-0.80} \pm 0.13 $ 
& $-0.487 \pm 0.035$    \\
\hline
$B^+ \to K^{*+} \mu^+ \mu^-$ 
& $P_6^\prime$ 
& LHCb \cite{Aaij:2020ruw} 
& $[0.1, 0.98]$ 
& $-0.02^{+0.40}_{-0.34} \pm 0.06 $ 
& $-0.0476 \pm 0.0376$    \\
& &
& $[1.1, 2.5]$ 
& $\phantom{-}0.25^{+1.22}_{-1.32} \pm 0.08 $ 
& $-0.0539 \pm 0.0784$    \\
& &
& $[2.5, 4]$ 
& $-0.37^{+1.59}_{-3.91} \pm 0.05 $ 
& $-0.0439 \pm 0.1071$    \\
& &
& $[4, 6]$ 
& $-0.09^{+0.40}_{-0.41} \pm 0.05 $ 
& $-0.0285 \pm 0.1159$    \\
& &
& $[15, 17]$ 
& $-0.48^{+0.21}_{-0.21} \pm 0.02 $ 
& $-0.0029 \pm 0.0015$    \\
& &
& $[17, 19]$ 
& $\phantom{-}0.12^{+0.33}_{-0.33} \pm 0.04 $ 
& $-0.0012 \pm 0.0007$    \\
\hline
$B^+ \to K^{*+} \mu^+ \mu^-$ 
& $P_8^\prime$ 
& LHCb \cite{Aaij:2020ruw} 
& $[0.1, 0.98]$ 
& $\phantom{-}0.45^{+0.50}_{-0.39} \pm 0.09 $ 
& $-0.0326 \pm 0.0225$    \\
& &
& $[1.1, 2.5]$ 
& $\phantom{-}0.12^{+0.75}_{-0.76} \pm 0.05 $ 
& $-0.0271 \pm 0.0354$    \\
& &
& $[2.5, 4]$ 
& $\phantom{-}0.12^{+7.89}_{-4.95} \pm 0.07 $ 
& $-0.0182 \pm 0.0394$    \\
& &
& $[4, 6]$ 
& $-0.15^{+0.44}_{-0.48} \pm 0.05 $ 
& $-0.0114 \pm 0.0377$    \\
& &
& $[15, 17]$ 
& $-0.34^{+0.23}_{-0.22} \pm 0.04 $ 
& $\phantom{-}0.0007 \pm 0.0006$    \\
& &
& $[17, 19]$ 
& $\phantom{-}0.36^{+0.37}_{-0.33} \pm 0.07 $ 
& $\phantom{-}0.0003 \pm 0.0002$    \\
\hline
$B^0 \to K^{*0} \mu^+ \mu^-$ 
& $F_L$ 
& LHCb \cite{Aaij:2020nrf} 
& $[0.1, 0.98]$ 
& $0.255 \pm 0.032 \pm 0.007 $ 
& $0.295 \pm 0.050$    \\
& &
& $[1.1, 2.5]$ 
& $0.655 \pm 0.046 \pm 0.017 $ 
& $0.760 \pm 0.044$    \\
& &
& $[2.5, 4]$ 
& $0.756 \pm 0.047 \pm 0.023 $ 
& $0.796 \pm 0.036$    \\
& &
& $[4, 6]$ 
& $0.684 \pm 0.035 \pm 0.015 $ 
& $0.711 \pm 0.048$    \\
& &
& $[15, 17]$ 
& $0.352 \pm 0.026 \pm 0.009 $ 
& $0.348 \pm 0.022$    \\
& &
& $[17, 19]$ 
& $0.344 \pm 0.032 \pm 0.025 $ 
& $0.328 \pm 0.016$    \\
\hline
$B^0 \to K^{*0} \mu^+ \mu^-$ 
& $P_1$ 
& LHCb \cite{Aaij:2020nrf}
& $[0.1, 0.98]$ 
& $\phantom{-}0.090 \pm 0.119 \pm 0.009 $ 
& $\phantom{-}0.0441 \pm 0.0465$    \\
& &
& $[1.1, 2.5]$ 
& $-0.617 \pm 0.296 \pm 0.023 $
& $\phantom{-}0.0234 \pm 0.0463$    \\
& &
& $[2.5, 4]$ 
& $\phantom{-}0.168 \pm 0.371 \pm 0.043 $
& $-0.116 \pm 0.037$    \\
& &
& $[4, 6]$ 
& $\phantom{-}0.088 \pm 0.235 \pm 0.029 $
& $-0.178 \pm 0.047$    \\
& &
& $[15, 17]$ 
& $-0.511 \pm 0.096 \pm 0.020 $ 
& $-0.534 \pm 0.050$    \\
& &
& $[17, 19]$ 
& $-0.763 \pm 0.152 \pm 0.094 $ 
& $-0.751 \pm 0.035$    \\
\hline
$B^0 \to K^{*0} \mu^+ \mu^-$ 
& $P_2$ 
& LHCb \cite{Aaij:2020nrf} 
& $[0.1, 0.98]$ 
& $-0.003 \pm 0.038 \pm 0.003 $
& $-0.132 \pm 0.008$    \\
& &
& $[1.1, 2.5]$ 
& $-0.443 \pm 0.100 \pm 0.027 $ 
& $-0.451 \pm 0.014$    \\
& &
& $[2.5, 4]$ 
& $-0.191 \pm 0.116 \pm 0.043 $ 
& $-0.063 \pm 0.113$    \\
& &
& $[4, 6]$ 
& $\phantom{-}0.105 \pm 0.068 \pm 0.009 $ 
& $\phantom{-}0.292 \pm 0.081$    \\
& &
& $[15, 17]$ 
& $\phantom{-}0.396 \pm 0.022 \pm 0.004 $ 
& $\phantom{-}0.413 \pm 0.016$    \\
& &
& $[17, 19]$ 
& $\phantom{-}0.328 \pm 0.032 \pm 0.017 $
& $\phantom{-}0.317 \pm 0.020$    \\
\hline
$B^0 \to K^{*0} \mu^+ \mu^-$ 
& $P_3$ 
& LHCb \cite{Aaij:2020nrf} 
& $[0.1, 0.98]$ 
& $\phantom{-}0.073 \pm 0.057 \pm 0.003 $ 
& $\phantom{-}0.0014 \pm 0.0135$    \\
& &
& $[1.1, 2.5]$ 
& $\phantom{-}0.324 \pm 0.147 \pm 0.014 $ 
& $\phantom{-}0.0039 \pm 0.0245$    \\
& &
& $[2.5, 4]$ 
& $\phantom{-}0.049 \pm 0.195 \pm 0.014 $ 
& $\phantom{-}0.0040 \pm 0.0104$    \\
& &
& $[4, 6]$ 
& $-0.090 \pm 0.139 \pm 0.006 $ 
& $\phantom{-}0.0026 \pm 0.0155$    \\
& &
& $[15, 17]$ 
& $-0.000 \pm 0.056 \pm 0.003 $ 
& $-0.0004 \pm 0.0003$    \\
& &
& $[17, 19]$ 
& $\phantom{-}0.085 \pm 0.068 \pm 0.004 $
& $-0.0003 \pm 0.0002$    \\
\hline
$B^0 \to K^{*0} \mu^+ \mu^-$ 
& $P_4^\prime$ 
& LHCb \cite{Aaij:2020nrf}
& $[0.1, 0.98]$ 
& $\phantom{-}0.135 \pm 0.118 \pm 0.010 $ 
& $\phantom{-}0.251 \pm 0.017$    \\
& &
& $[1.1, 2.5]$ 
& $-0.080 \pm 0.142 \pm 0.019 $ 
& $-0.064 \pm 0.045$    \\
& &
& $[2.5, 4]$ 
& $-0.435 \pm 0.169 \pm 0.035 $ 
& $-0.393 \pm 0.046$    \\
& &
& $[4, 6]$ 
& $-0.312 \pm 0.115 \pm 0.013 $ 
& $-0.504 \pm 0.028$    \\
& &
& $[15, 17]$ 
& $-0.626 \pm 0.069 \pm 0.018 $ 
& $-0.619 \pm 0.010$    \\
& &
& $[17, 19]$ 
& $-0.647 \pm 0.086 \pm 0.057 $ 
& $-0.661 \pm 0.007$    \\
\hline
$B^0 \to K^{*0} \mu^+ \mu^-$ 
& $P_5^\prime$ 
& LHCb \cite{Aaij:2020nrf} 
& $[0.1, 0.98]$ 
& $\phantom{-}0.521 \pm 0.095 \pm 0.024 $ 
& $\phantom{-}0.688 \pm 0.024$    \\
& &
& $[1.1, 2.5]$ 
& $\phantom{-}0.365 \pm 0.122 \pm 0.013 $
& $\phantom{-}0.140 \pm 0.081$    \\
& &
& $[2.5, 4]$ 
& $-0.150 \pm 0.144 \pm 0.032 $
& $-0.500 \pm 0.102$    \\
& &
& $[4, 6]$ 
& $-0.439 \pm 0.111 \pm 0.036 $ 
& $-0.757 \pm 0.079$    \\
& &
& $[15, 17]$ 
& $-0.714 \pm 0.074 \pm 0.021 $ 
& $-0.670 \pm 0.036$    \\
& &
& $[17, 19]$ 
& $-0.590 \pm 0.084 \pm 0.059 $
& $-0.483 \pm 0.035$    \\
\hline
$B^0 \to K^{*0} \mu^+ \mu^-$ 
& $P_6^\prime$ 
& LHCb \cite{Aaij:2020nrf} 
& $[0.1, 0.98]$ 
& $\phantom{-}0.015 \pm 0.094 \pm 0.007 $ 
& $-0.055 \pm 0.036$    \\
& &
& $[1.1, 2.5]$ 
& $-0.226 \pm 0.128 \pm 0.005 $
& $-0.070 \pm 0.077$    \\
& &
& $[2.5, 4]$ 
& $-0.155 \pm 0.148 \pm 0.024 $ 
& $-0.052 \pm 0.105$    \\
& &
& $[4, 6]$ 
& $-0.293 \pm 0.117 \pm 0.004 $ 
& $-0.030 \pm 0.117$    \\
& &
& $[15, 17]$ 
& $\phantom{-}0.061 \pm 0.085 \pm 0.003 $
& $-0.003 \pm 0.002$    \\
& &
& $[17, 19]$ 
& $\phantom{-}0.103 \pm 0.105 \pm 0.016 $
& $-0.001 \pm 0.001$    \\
\hline
$B^0 \to K^{*0} \mu^+ \mu^-$ 
& $P_8^\prime$ 
& LHCb \cite{Aaij:2020nrf}
& $[0.1, 0.98]$ 
& $-0.007 \pm 0.122 \pm 0.002 $
& $-0.0056 \pm 0.0222$    \\
& &
& $[1.1, 2.5]$ 
& $-0.366 \pm 0.158 \pm 0.005 $ 
& $-0.0177 \pm 0.0350$    \\
& &
& $[2.5, 4]$ 
& $\phantom{-}0.037 \pm 0.169 \pm 0.007 $
& $-0.0173 \pm 0.0370$    \\
& &
& $[4, 6]$ 
& $\phantom{-}0.166 \pm 0.127 \pm 0.004 $ 
& $-0.0115 \pm 0.0376$    \\
& &
& $[15, 17]$ 
& $\phantom{-}0.007 \pm 0.086 \pm 0.002 $ 
& $\phantom{-}0.0007 \pm 0.0006$    \\
& &
& $[17, 19]$ 
& $-0.055 \pm 0.099 \pm 0.006 $
& $\phantom{-}0.0003 \pm 0.0002$    \\
\hline
$B^0 \to K^{*0} \mu^+ \mu^-$ 
& $F_L$ 
& ATLAS \cite{Aaboud:2018krd}
& $[0.04, 2.0]$ 
& $0.44 \pm 0.08 \pm 0.07 $
& $0.388 \pm 0.057$ \\
& &
& $[2, 4]$ 
& $0.64 \pm 0.11 \pm 0.05 $ 
& $0.799 \pm 0.035$ \\
& &
& $[4, 6]$ 
& $0.42 \pm 0.13 \pm 0.12 $
& $0.711 \pm 0.048$ \\
\hline
$B^0 \to K^{*0} \mu^+ \mu^-$ 
& $P_1$ 
& ATLAS \cite{Aaboud:2018krd}
& $[0.04, 2.0]$ 
& $-0.05 \pm 0.30 \pm 0.08 $
& $\phantom{-}0.0435 \pm 0.0280$ \\
& &
& $[2, 4]$ 
& $-0.78 \pm 0.51 \pm 0.34 $ 
& $-0.095 \pm 0.038$ \\
& &
& $[4, 6]$ 
& $\phantom{-}0.14 \pm 0.43 \pm 0.26 $
& $-0.178 \pm 0.047$ \\
\hline
$B^0 \to K^{*0} \mu^+ \mu^-$ 
& $P_4^\prime$ 
& ATLAS \cite{Aaboud:2018krd}
& $[0.04, 2.0]$ 
& $\phantom{-}0.31 \pm 0.40 \pm 0.20 $
& $\phantom{-}0.150 \pm 0.018$ \\
& &
& $[2, 4]$ 
& $-0.76 \pm 0.31 \pm 0.21 $ 
& $-0.349 \pm 0.050$ \\
& &
& $[4, 6]$ 
& $\phantom{-}0.64 \pm 0.33 \pm 0.18 $
& $-0.504 \pm 0.028$ \\
\hline
$B^0 \to K^{*0} \mu^+ \mu^-$ 
& $P_5^\prime$ 
& ATLAS \cite{Aaboud:2018krd}
& $[0.04, 2.0]$ 
& $\phantom{-}0.67 \pm 0.26 \pm 0.16 $
& $\phantom{-}0.513 \pm 0.033$ \\
& &
& $[2, 4]$ 
& $-0.33 \pm 0.31 \pm 0.13 $ 
& $-0.409 \pm 0.110$ \\
& &
& $[4, 6]$ 
& $\phantom{-}0.26 \pm 0.35 \pm 0.18 $
& $-0.757 \pm 0.079$ \\
\hline
$B^0 \to K^{*0} \mu^+ \mu^-$ 
& $P_6^\prime$ 
& ATLAS \cite{Aaboud:2018krd}
& $[0.04, 2.0]$ 
& $-0.18 \pm 0.21 \pm 0.04 $
& $-0.056 \pm 0.042$ \\
& &
& $[2, 4]$ 
& $\phantom{-}0.31 \pm 0.28 \pm 0.19 $ 
& $-0.056 \pm 0.042$ \\
& &
& $[4, 6]$ 
& $\phantom{-}0.06 \pm 0.27 \pm 0.13 $
& $-0.030 \pm 0.117$ \\
\hline
$B^0 \to K^{*0} \mu^+ \mu^-$ 
& $P_8^\prime$ 
& ATLAS \cite{Aaboud:2018krd}
& $[0.04, 2.0]$ 
& $-0.29 \pm 0.48 \pm 0.18 $
& $-0.0084 \pm 0.0233$ \\
& &
& $[2, 4]$ 
& $\phantom{-}1.07 \pm 0.41 \pm 0.39 $ 
& $-0.0179 \pm 0.0388$ \\
& &
& $[4, 6]$ 
& $-0.24 \pm 0.42 \pm 0.09 $
& $-0.0115 \pm 0.0376$ \\
\hline
$B^0 \to K^{*0} \mu^+ \mu^-$ 
& $10^{\, 8} \, {\rm GeV}^2 \times \frac{\Delta \cal B}{\Delta q^2}$ 
& CMS \cite{Khachatryan:2015isa}
& $[1, 2]$ 
& $4.6 \pm 0.7 \pm 0.3 $
& $4.59 \pm 0.61$ \\
& &
& $[2, 4.3]$ 
& $3.3 \pm 0.5 \pm 0.2 $
& $4.25 \pm 0.59$ \\
& &
& $[4.3, 6]$ 
& $3.4 \pm 0.5 \pm 0.3 $
& $4.79 \pm 0.71$ \\
& &
& $[14.18, 16]$ 
& $6.7 \pm 0.6 \pm 0.5 $
& $7.06 \pm 0.85$ \\
& &
& $[16, 19]$ 
& $4.2 \pm 0.3 \pm 0.3 $
& $5.16 \pm 0.68$ \\
\hline
$B^0 \to K^{*0} \mu^+ \mu^-$ 
& $F_L$ 
& CMS \cite{Khachatryan:2015isa}
& $[1, 2]$ 
& $0.64^{+0.10}_{-0.09} \pm 0.07 $
& $0.724 \pm 0.050$ \\
& &
& $[2, 4.3]$ 
& $0.80^{+0.08}_{-0.08} \pm 0.06 $ 
& $0.793 \pm 0.036$ \\
& &
& $[4.3, 6]$ 
& $0.62^{+0.10}_{-0.09} \pm 0.07 $
& $0.703 \pm 0.049$ \\
& &
& $[14.18, 16]$ 
& $0.48^{+0.05}_{-0.06} \pm 0.04 $
& $0.362 \pm 0.027$ \\
& &
& $[16, 19]$ 
& $0.38^{+0.05}_{-0.06} \pm 0.04 $
& $0.333 \pm 0.020$ \\
\hline
$B^0 \to K^{*0} \mu^+ \mu^-$ 
& $A_{\rm FB}$ 
& CMS \cite{Khachatryan:2015isa}
& $[1, 2]$ 
& $-0.27^{+0.17}_{-0.40} \pm 0.07 $
& $-0.156 \pm 0.031$ \\
& &
& $[2, 4.3]$ 
& $-0.12^{+0.15}_{-0.17}\pm 0.05 $ 
& $-0.026 \pm 0.033$ \\
& &
& $[4.3, 6]$ 
& $\phantom{-}0.01^{+0.15}_{-0.15} \pm 0.03 $
& $\phantom{-}0.132 \pm 0.045$ \\
& &
& $[14.18, 16]$ 
& $\phantom{-}0.39^{+0.04}_{-0.06} \pm 0.01 $
& $\phantom{-}0.412 \pm 0.018$ \\
& &
& $[16, 19]$ 
& $\phantom{-}0.35^{+0.07}_{-0.07} \pm 0.01 $
& $\phantom{-}0.349 \pm 0.021$ \\
\hline
$B^0 \to K^{*0} \mu^+ \mu^-$ 
& $P_{1}$ 
& CMS \cite{Sirunyan:2017dhj}
& $[1, 2]$ 
& $-0.12^{+0.46}_{-0.47} \pm 0.10 $
& $\phantom{-}0.044 \pm 0.047$ \\
& &
& $[2, 4.3]$ 
& $-0.69^{+0.58}_{-0.27}\pm 0.23 $ 
& $-0.106 \pm 0.038$ \\
& &
& $[4.3, 6]$ 
& $\phantom{-}0.53^{+0.24}_{-0.33} \pm 0.19 $
& $-0.180 \pm 0.049$ \\
& &
& $[14.18, 16]$ 
& $-0.33^{+0.24}_{-0.23} \pm 0.20 $
& $-0.465 \pm 0.051$ \\
& &
& $[16, 19]$ 
& $-0.53^{+0.19}_{-0.19} \pm 0.16 $
& $-0.680 \pm 0.041$ \\
\hline
$B^0 \to K^{*0} \mu^+ \mu^-$ 
& $P_{5}^\prime$ 
& CMS \cite{Sirunyan:2017dhj}
& $[1, 2]$ 
& $\phantom{-}0.10^{+0.32}_{-0.31} \pm 0.07 $
& $\phantom{-}0.289 \pm 0.063$ \\
& &
& $[2, 4.3]$ 
& $-0.57^{+0.34}_{-0.31}\pm 0.18 $ 
& $-0.449 \pm 0.104$ \\
& &
& $[4.3, 6]$ 
& $-0.96^{+0.22}_{-0.21} \pm 0.25 $
& $-0.769 \pm 0.073$ \\
& &
& $[14.18, 16]$ 
& $-0.66^{+0.13}_{-0.20} \pm 0.18 $
& $-0.719 \pm 0.035$ \\
& &
& $[16, 19]$ 
& $-0.56^{+0.12}_{-0.12} \pm 0.07 $
& $-0.547 \pm 0.036$ \\
\hline
$B^+ \to K^+ \mu^+ \mu^-$ 
& $10^{\, 8} \, {\rm GeV}^2 \times \frac{\Delta \cal B}{\Delta q^2} $
& Belle \cite{Abdesselam:2019lab} 
& $[0.1,4]$
& $4.51^{+1.05}_{-0.94}\pm 0.20$
& $3.33 \pm 0.60$ \\
& & 
& $[4,8.12]$ 
& $3.00^{+0.68}_{-0.61} \pm 0.07$ 
& $3.24 \pm 0.61$ \\
& & 
& $[1, 6]$ 
& $4.60^{+0.82}_{-0.76} \pm 0.10 $ 
& $3.30 \pm 0.60 $ \\
\hline
$B^+ \to K^{*+} \mu^+ \mu^-$ 
& $10^{\, 8} \, {\rm GeV}^2 \times \frac{\Delta \cal B}{\Delta q^2} $ 
& Belle \cite{Abdesselam:2019wac} 
& $[1.1, 6]$ 
& $2.45^{+1.84}_{-1.43} \pm 0.41 $ 
& $4.87 \pm 0.77$ \\
& &
& $[15, 19]$ 
& $7.25^{+2.50}_{-2.00} \pm 0.75 $ 
& $6.08 \pm 0.73$ \\
\hline
$B^0 \to K^{*0} \mu^+ \mu^-$ 
& $10^{\, 8} \, {\rm GeV}^2 \times \frac{\Delta \cal B}{\Delta q^2} $ 
& Belle \cite{Abdesselam:2019wac} 
& $[1.1, 6]$ 
& $3.88^{+1.22}_{-1.02} \pm 0.61 $ 
& $4.48 \pm 0.71$ \\
& &
& $[15, 19]$ 
& $5.50^{+1.25}_{-1.00} \pm 0.50 $ 
& $5.60 \pm 0.67 $ \\
\hline
$B_s^0 \to \phi^0 \mu^+ \mu^-$ 
& $10^{\, 8} \, {\rm GeV}^2 \times \frac{\Delta \cal B}{\Delta q^2} $ 
& LHCb \cite{Aaij:2015esa,Aaij:2021nyr} 
& $[0.1, 2]$ 
& $5.54^{+0.69}_{-0.65} \pm 0.13 \pm 0.27 $  
& $7.92 \pm 0.98$ \\
& &
& $[2, 5]$
& $2.42^{+0.40}_{-0.38} \pm 0.06 \pm 0.12 $  
& $4.94 \pm 0.69$ \\
& &
& $[15, 17]$
& $4.28^{+0.54}_{-0.51} \pm 0.11 \pm 0.21 $  
& $6.60 \pm 0.65$ \\
& &
& $[17, 19]$
& $3.75^{+0.54}_{-0.51} \pm 0.13 \pm 0.18 $  
& $3.93 \pm 0.41$ \\
\hline
$B_s^0 \to \phi^0 \mu^+ \mu^-$ 
& $F_L$ 
& LHCb \cite{Aaij:2015esa} 
& $[0.1, 2]$ 
& $0.20^{+0.08}_{-0.09} \pm 0.02 $  
& $0.496 \pm 0.037$ \\
& &
& $[2, 5]$
& $0.68^{+0.16}_{-0.13} \pm 0.03 $  
& $0.811 \pm 0.017$ \\
& &
& $[15, 17]$
& $0.23^{+0.09}_{-0.08} \pm 0.02 $
& $0.349 \pm 0.015$ \\
& &
& $[17, 19]$
& $0.40^{+0.13}_{-0.15} \pm 0.02 $  
& $0.328 \pm 0.014$ \\
\hline 
$B_s^0 \to \phi^0 \mu^+ \mu^-$ 
& $S_3$ 
& LHCb \cite{Aaij:2015esa} 
& $[0.1, 2]$ 
& $-0.05^{+0.13}_{-0.13} \pm 0.01 $  
& $\phantom{-}0.020 \pm 0.006$ \\
& &
& $[2, 5]$
& $-0.06^{+0.19}_{-0.23} \pm 0.01 $  
& $-0.009 \pm 0.004$ \\
& &
& $[15, 17]$
& $-0.06^{+0.16}_{-0.19} \pm 0.01 $
& $-0.180 \pm 0.007$ \\
& &
& $[17, 19]$
& $-0.07^{+0.23}_{-0.27} \pm 0.02 $  
& $-0.262 \pm 0.006$ \\
\hline
$B_s^0 \to \phi^0 \mu^+ \mu^-$ 
& $S_4$ 
& LHCb \cite{Aaij:2015esa} 
& $[0.1, 2]$ 
& $\phantom{-}0.27^{+0.28}_{-0.18} \pm 0.01 $  
& $\phantom{-}0.060 \pm 0.007$ \\
& &
& $[2, 5]$
& $-0.47^{+0.30}_{-0.44} \pm 0.01 $  
& $-0.148 \pm 0.017$ \\
& &
& $[15, 17]$
& $-0.03^{+0.15}_{-0.15} \pm 0.01 $
& $-0.296 \pm 0.004$ \\
& &
& $[17, 19]$
& $-0.39^{+0.25}_{-0.34} \pm 0.02 $  
& $-0.313 \pm 0.004$ \\
\hline
$B_s^0 \to \phi^0 \mu^+ \mu^-$ 
& $S_7$ 
& LHCb \cite{Aaij:2015esa} 
& $[0.1, 2]$ 
& $\phantom{-}0.04^{+0.12}_{-0.12} \pm 0.00 $  
& $-0.025 \pm 0.018$ \\
& &
& $[2, 5]$
& $-0.03^{+0.18}_{-0.23} \pm 0.01 $  
& $-0.020 \pm 0.038$ \\
& &
& $[15, 17]$
& $\phantom{-}0.12^{+0.16}_{-0.15} \pm 0.01 $
& $-0.0013 \pm 0.0006$ \\
& &
& $[17, 19]$
& $\phantom{-}0.20^{+0.29}_{-0.22} \pm 0.01 $  
& $-0.0005 \pm 0.0003$ \\
\hline
$B \to X_s \, \mu^+\mu^-$ $\! \! \phantom{B}^{\red *}$ 
& $ 10^{6} \times \Delta {\cal B} $
& BaBar \cite{Lees:2013nxa}
& $[1,6]$
& $0.66^{+0.87}_{-0.80} \pm 0.07$
& $1.68 \pm 0.16$ \\
&
& 
& $[14.2,25]$
& $0.60^{+0.31}_{-0.29} \pm 0.00$
& $0.35 \pm 0.04$ \\
\hline
$B \to X_s \, e^+ e^-$ $\! \! \phantom{B}^{\red *}$ 
& $ 10^{6} \times \Delta {\cal B} $
& BaBar \cite{Lees:2013nxa}
& $[0.1,2]$
& $3.05^{+0.60}_{-0.53} \pm 0.35$
& $1.24 \pm 0.12$ \\
&
&
& $[2.0,4.3]$
& $0.69^{+0.33}_{-0.29} \pm 0.07$
& $0.79 \pm 0.08$ \\
&
&
& $[4.3,6.8]$
& $0.69^{+0.34}_{-0.31} \pm 0.05$
& $0.74 \pm 0.07$ \\
&
&
& $[14.2,25]$
& $0.56^{+0.19}_{-0.18} \pm 0.00$
& $0.30 \pm 0.03$ \\
\hline
$B \to K^* \mu^+ \mu^-$ $\! \! \phantom{B}^{\red *}$ 
& $P_4^\prime$ 
& Belle \cite{Wehle:2016yoi} 
& $[1, 6]$
& $-0.22^{+0.35}_{-0.34} \pm 0.15 $  
& $-0.342 \pm 0.041$ \\
& &
& $[14.18, 19]$
& $-0.10^{+0.39}_{-0.39} \pm 0.07 $
& $-0.627 \pm 0.010$ \\
\hline
$B \to K^* \mu^+ \mu^-$ $\! \! \phantom{B}^{\red *}$ 
& $P_5^\prime$ 
& Belle \cite{Wehle:2016yoi} 
& $[1, 6]$
& $\phantom{-}0.43^{+0.26}_{-0.28} \pm 0.10 $    
& $-0.431 \pm 0.096$ \\
& &
& $[14.18, 19]$
& $-0.13^{+0.39}_{-0.35} \pm 0.06 $
& $-0.626 \pm 0.036$ \\
\hline
$B \to K^* e^+ e^-$ $\! \! \phantom{B}^{\red *}$ 
& $P_4^\prime$ 
& Belle \cite{Wehle:2016yoi} 
& $[1, 6]$
& $-0.72^{+0.40}_{-0.39} \pm 0.06 $  
& $-0.339 \pm 0.041$ \\
& &
& $[14.18, 19]$
& $-0.15^{+0.41}_{-0.40} \pm 0.04 $
& $-0.627 \pm 0.010$ \\
\hline
$B \to K^* e^+ e^-$ $\! \! \phantom{B}^{\red *}$ 
& $P_5^\prime$ 
& Belle \cite{Wehle:2016yoi} 
& $[1, 6]$
& $-0.22^{+0.39}_{-0.41} \pm 0.03 $  
& $-0.424 \pm 0.095$ \\
& &
& $[14.18, 19]$
& $-0.91^{+0.36}_{-0.30} \pm 0.03 $
& $-0.626 \pm 0.036$ \\
\hline
$B^0 \to K^{*0} e^+ e^-$ 
& $F_L$
& LHCb \cite{Aaij:2015dea}
& $[0.002,1.12]$
& $0.16 \pm 0.06 \pm 0.03$
& $0.18 \pm 0.04$ \\
\hline
$B^+ \to K^+ \mu^+ \mu^-$ 
& $A_{\rm FB}$ 
& CMS \cite{Sirunyan:2018jll} 
& $[1, 6]$
& $-0.14^{+0.07}_{-0.06} \pm 0.03 $  
& $0.0 \pm 0.0 $ \\
& &
& $[16,18]$
& $\phantom{-}0.04^{+0.05}_{-0.04} \pm 0.03 $
& $0.0 \pm 0.0$ \\
& &
& $[18,22]$
& $\phantom{-}0.05^{+0.05}_{-0.04} \pm 0.02 $
& $0.0 \pm 0.0$ \\
\hline
$B^+ \to K^+ \mu^+ \mu^-$ 
& $F_{H}$ 
& CMS \cite{Sirunyan:2018jll} 
& $[1, 6]$
& $0.38^{+0.17}_{-0.21} \pm 0.09 $  
& $0.0247 \pm 0.0003$ \\
& &
& $[16,18]$
& $0.07^{+0.06}_{-0.07} \pm 0.07 $
& $0.0064 \pm 0.0004$ \\
& &
& $[18,22]$
& $0.10^{+0.06}_{-0.10} \pm 0.09 $
& $0.0079 \pm 0.0007$ \\
\hline
$\Lambda_b^0 \to \Lambda \mu^+ \mu^-$ 
& $10^{\, 8} \, {\rm GeV}^2 \times \frac{\Delta \cal B}{\Delta q^2} $
& LHCb \cite{Aaij:2015xza} 
& $[1.1, 6]$ 
& $\phantom{1}0.9^{+0.6 \, +0.1}_{-0.5 \, -0.1} \pm 0.2 $ 
& $0.98 \pm 0.57$ \\
& &
& $[15, 16]$ 
& $11.2^{+1.9 \, +0.5}_{-1.8 \, - 0.5} \pm 2.3 $ 
& $6.75 \pm 0.90$ \\
& &
& $[16, 18]$ 
& $12.2^{+1.4 \, +0.3}_{-1.4 \, - 0.6} \pm 2.5 $ 
& $7.28 \pm 0.88$ \\
& &
& $[18, 20]$ 
& $12.4^{+1.4 \, +0.6}_{-1.4 \, - 0.5} \pm 2.6 $ 
& $6.12 \pm 0.76$ \\
\hline
$\Lambda_b^0 \to \Lambda \mu^+ \mu^-$ 
& $A_{\rm FB}^\ell $
& LHCb \cite{Aaij:2018gwm}
& $[15, 20]$
& $ -0.39 \pm 0.04 \pm 0.01$
& $-0.353 \pm 0.018$ \\
& $A_{\rm FB}^h $
& 
& $[15, 20]$
& $ -0.30 \pm 0.05 \pm 0.02$
& $-0.318 \pm 0.011$ \\
& $A_{\rm FB}^{\ell h} $
& 
& $[15, 20]$
& $\phantom{-}0.25 \pm 0.04 \pm 0.01$
& $\phantom{-}0.164 \pm 0.009$ \\
\hline
$B_s \to \mu^+ \mu^-$ 
& $ 10^{9} \times {\cal B} $
& Comb. \cite{Altmannshofer:2021qrr}
&
& $ 2.93 \pm 0.35$ 
& $ 3.48 \pm 0.26$  \\
\hline
\caption{\label{tab:bsll} Table of $b\to s \ell^+ \ell^-$ observables (except $R_{K^{(*)}}$) used in the analysis. We include correlations between measurements when provided by the sources. The stated SM predictions are our calculations through \textbf{flavio}.
The modes marked by {\red *} assume averaging of $B^+$ and $B^0$ decays.}
\end{longtable}

\newpage
\begin{longtable}[c]{|l|c|c|c|c|}
\hline
Observable & Source & Bin & Experimental value & SM prediction \\
\hline
$ R_{K^+} $ 
& LHCb \cite{Aaij:2021vac} 
& $[1.1,6] $
& $0.846^{+0.042 \, + 0.013}_{-0.039 \, -0.012} $
& $1.0008 \pm 0.0003$     \\
\hline
$ R_{K^{*0}} $ 
& LHCb \cite{Aaij:2017vbb} 
& $[0.045, 1.1] $
& $0.66^{+0.11}_{-0.07} \pm 0.03 $
& $0.926 \pm 0.004$    \\
&
& $[1.1, 6.0] $
& $0.69^{+0.11}_{-0.07} \pm 0.05 $
& $0.9964 \pm 0.0006$    \\
\hline
$ R_{K} $ 
& BaBar \cite{Lees:2012tva}
& $[0.10, 8.12]$ 
& $0.74^{+0.40}_{-0.31} \pm 0.06 $
& $1.0007 \pm 0.0003$ \\
\hline
$ R_{K^*} $ 
& BaBar \cite{Lees:2012tva}
& $[0.10, 8.12]$ 
& $1.06^{+0.48}_{-0.33} \pm 0.08 $
& $0.9949 \pm 0.0015$ \\
\hline
$ R_{K^+} $ 
& Belle \cite{Abdesselam:2019lab} 
& $[1,6] $
& $1.39^{+0.36}_{-0.33} \pm 0.02 $
& $1.0008 \pm 0.0003$    \\
\hline
$ R_{K^{*+}} $ 
& Belle \cite{Abdesselam:2019wac} 
& $[0.045, 1.1] $
& $0.62^{+0.60}_{-0.36} \pm 0.09 $
& $0.928 \pm 0.005$    \\
& 
& $[1.1, 6] $
& $0.72^{+0.99}_{-0.44} \pm 0.15 $
& $0.997 \pm 0.001$    \\
& 
& $[15, 19]$
& $1.40^{+1.99}_{-0.68} \pm 0.12 $
& $0.99807 \pm 0.00004$ \\
\hline
$ R_{K^{*0}} $ 
& Belle \cite{Abdesselam:2019wac} 
& $[0.045, 1.1] $
& $0.46^{+0.55}_{-0.27} \pm 0.13 $
& $0.926 \pm 0.004$     \\
& 
& $[1.1, 6] $
& $1.06^{+0.63}_{-0.38} \pm 0.14 $
& $0.996 \pm 0.001$     \\
& 
& $[15, 19]$
& $1.12^{+0.61}_{-0.36} \pm 0.10 $
& $0.99807 \pm 0.0004$  \\
\hline
\caption{Table of $R_{K^{(*)}}$ observables used in the analysis. The stated SM predictions are our calculations through \textbf{flavio}.}
\label{tab:RKtab}
\end{longtable}

\section{Calculation of $\delta$ from $\lambda_i$ and $\tan \beta$}
\label{delta_formula}
We start from Eq.~\eqref{eq:tan-alpha} and rearrange to give an expression for $m_{12}^2$.
% In order to implement Eq.~\eqref{eq:tan-alpha} in our theory scans, 
% we rearrange Eq.~\eqref{eq:mh0} and substitute it in to Eq.~\eqref{eq:tan-alpha}. 
Using small angle approximations on $\delta =\beta-\alpha-\pi/2$, we expand
to second order, and find that
\begin{equation}
     m_{12}^2 \, \delta^2 =\GMt + \GMo \delta + \Gz \, \delta^2 + \odc
     \label{eq:m12sq-exp}
\end{equation}
with
\begin{align}
    \GMt&=\Sb \, \Cb\left(m_{h^0}^2-\lt{1}\Cb^4-\lt{2}\Sb^4-\frac{\lt{345}\,\Stb^2}{2}\right),\\
    \GMo&=\Sb \, \Cb\left(-2\lt{1}\Cb^3\Sb+2\lt{2}\Sb^3\Cb+\lt{345} \,\Stb\Ctb\right),\\
    \Gz&=\Sb \, \Cb\left(\lt{1}(\Cb^4-\Sb^2\Cb^2)+\lt{2}(\Sb^4-\Cb^2\Sb^2)+\lt{345}\, \Stb^2\right),
\end{align}
and where we have used as shorthand:
\begin{equation}
    s_\theta=\sin(\theta),\,\, c_\theta=\cos(\theta),\,\, \lambda_{345}=\lambda_3+\lambda_4+\lambda_5,\,\,\text{and}\,\,\lt{i}=v^2\lambda_{i}. 
\end{equation}
Then inserting Eq.~\eqref{eq:m12sq-exp} into Eq.~\eqref{eq:tan-alpha}, we find an equation for $\delta$ of the form 
\begin{equation}
    c+b\, \delta+a\, \delta^2+\odc=0,
\end{equation}
where 
\newcommand{\dtb}{\Delta t_\beta}
\begin{eqnarray}
    a & = & 2\Stb\dtb\,\GMt+2\Ctb\dtb\,\GMo-2\Stb\dtb\,\Gz-\Stb\lt{1}\Cb^2+\Stb\lt{2}\Sb^2
    \nonumber \\
    & & - 2\Ctb\,\Gz+\lt{345} \, \Ctb\Stb-4\Stb \, \GMo + 4\Ctb \, \GMt, 
    \\
    b & = & 2\Ctb\dtb\,\GMt-\Stb\dtb\,\GMo-2\Ctb\,\GMo-4\Stb\,\GMt, 
    \\
    c & = &  -\Stb\dtb\,\GMt-2\Ctb\,\GMt,
    \label{eq:c}
\end{eqnarray}
having defined
\begin{equation}
    \dtb = \tan(\beta)-\frac{1}{\tan(\beta)}.
\end{equation}
By trigonometric manipulation we find that $c=0$. We are therefore left with the two solutions 
\begin{equation}
    \delta\rightarrow 0\quad \text{and}\quad \delta=-\frac{b}{a},
\end{equation}
where the first solution corresponds to the alignment limit. 
Note that the second solution is only valid for small $\delta$ as a second-order Taylor expansion was used in its derivation.

\section{Formulas for the Oblique Parameters in the 2HDM}
\label{sec:Oblique_formula}
The formulae for the oblique parameters in a general 2HDM as derived from \cite{Grimus:2008nb} are
\begin{align}
    \begin{split}
        T^{\text{2HDM}}&=\frac{G_F}{8\sqrt{2} \, \alpha\pi^2}\Big\{ F(m_{H^\pm}^2, m_{A^0}^2)
        +\sin^2(\theta) F(m_{H^\pm}^2,m_{H^0}^2)
        +\cos^2(\theta) F(m_{H^\pm}^2,m_{h^0}^2)
        \\
         &\qquad\qquad -\sin^2(\theta)F(m_{A^0}^2,m_{H^0}^2)-\cos^2(\theta)F(m_{A^0}^2,m_{h^0}^2) \\
         &\qquad\qquad +3\cos^2(\theta)[F(m_Z^2,m_{H^0}^2)-F(m_W^2,m_{H^0}^2)] \\
         &\qquad\qquad -3\cos^2(\theta)[F(m_{Z}^2,m_{h^0}^2)-F(m_W^2,m_{h^0}^2)]\Big\},
        \label{T2HDM}
    \end{split}\\[2mm]
    \begin{split}
        S^{\text{2HDM}} & 
        = \frac{G_F m_W^2 \sin^2(\theta_W)}{12 \sqrt{2} 
        \, \alpha \pi^2} \Big\{ (2 \sin^2(\theta_W)-1)^2 \, G(m_{H^\pm}^2,m_{H^\pm}^2,m_Z^2)
        \\ 
        & \qquad\qquad +\sin^2(\theta)\, G(m_{A^0}^2, m_{H^0}^2, m_Z^2) \\
        & \qquad\qquad +\cos^2(\theta)\, G(m_{A^0}^2, m_{h^0}^2, m_Z^2) +2\ln\left(\frac{m_{A^0} \, m_{H^0}}{m_{H^\pm}^2} \right) 
        \\
        & \qquad\qquad + 
        \cos^2(\theta)  
        \left[\hat{G}(m_{H^0}^2,m_Z^2)-\hat{G} (m_{h^0}^2,m_Z^2)\right]\Big\},
        \label{S2HDM}
    \end{split}\\[2mm]
    \begin{split}
        U^{\text{2HDM}}&=\frac{G_F m_W^2}{48\sqrt{2}\, \alpha\pi^2}\Big\{ G(m_{H^\pm}^2,m_{A^0}^2,m_W^2) +\sin^2(\theta) \, G(m_{H^\pm}^2,m_{H^0}^2,m_W^2) 
        \\
        &\qquad\qquad +\cos^2(\theta) \, G(m_{H^\pm}^2,m_{h^0}^2,m_W^2)
         -[2\sin^2(\theta_W)-1]^2 G(m_{H^\pm}^2,m_{H^\pm}^2,m_Z^2) 
         \\
        &\qquad\qquad -\sin^2(\theta) \, G(m_{A^0}^2,m_{H^0}^2,m_Z^2)-\cos^2(\theta) \, G(m_{A^0}^2,m_{h^0}^2,m_Z^2) \\
        &\qquad\qquad +\cos^2(\theta)[\hat{G}(m_{H^0}^2,m_W^2)-\hat{G}(m_{H^0}^2,m_Z^2)] \\
        &\qquad\qquad -\cos^2(\theta)[\hat{G}(m_{h^0}^2,m_W^2)-\hat{G}(m_{h^0}^2,m_Z^2)] \Big \},
        \label{U2HDM}
    \end{split}
\end{align}
where $\theta = \beta-\alpha$. The loop functions used above are defined as
\begin{align}
    F(x,y)&=
    \begin{cases}
        \displaystyle\frac{x+y}{2}-\displaystyle\frac{xy}{x-y}\ln\left(\frac{x}{y}\right), & \text{for }x\neq y, \\
        0, & \text{for }x=y,
    \end{cases} \\
        G(x,y,z)&=-\frac{16}{3}+\frac{5(x+y)}{z}-\frac{2(x-y)^2}{z^2} \nonumber \\[2mm]
        &\qquad +\frac{3}{z}\left[\frac{x^2+y^2}{x-y}-\frac{x^2-y^2}{z}+\frac{(x-y)^3}{3z^2}\right] \ln\left(\frac{x}{y}\right)+\frac{r}{z^3}f(t,r),
        \\[2mm]
    \tilde{G}(x,y,z)
    & = -2 + \left(\frac{x-y}{z} - \frac{x+y}{x-y}\right)\ln\left(\frac{x}{y}\right)+\frac{f(t,r)}{z},
    \\[2mm]
    \hat{G}(x,z)&=G(x,z,z)+12 \, \tilde{G}(x,z,z).
\end{align}
These functions depend on the sub-function,
\begin{align}
    f(t,r)&=
    \begin{cases}
        r^{1/2}\displaystyle\ln\left|\frac{t-r^{1/2}}{t+r^{1/2}}\right| & \text{for } r>0, 
        \\
        0 & \text{for } r=0,
          \\
        2(-r)^{1/2}\displaystyle \tan^{-1}\left(\frac{(-r)^{1/2}}{t}\right) & \text{for } r<0,
    \end{cases} 
\end{align}
where
\begin{equation}
\begin{aligned}
    t &= x+y-z, \\
    r &= z^2 - 2z(x+y) + (x-y)^2.
\end{aligned}
\end{equation}

\bibliographystyle{JHEP}
\bibliography{References}

\end{document}